\documentclass[graybox,twoside]{svmult}
\usepackage{mathptmx}       
\usepackage{helvet}         
\usepackage{courier}        
\usepackage{microtype}

\usepackage{makeidx}         
\usepackage{graphicx}        
\usepackage{multicol}        
\usepackage[bottom]{footmisc}
\usepackage{psfrag}          
\usepackage{dcolumn}         
\usepackage[numbers]{natbib}
\usepackage{hyperref}
\usepackage{afterpage}       
\usepackage{booktabs}        

\usepackage{amssymb,latexsym,amsmath,amsbsy}

\graphicspath{{./figures/pdf/}}

\newcommand{\choosebib}[2]{#1}

\newcommand{\abs}[1]{\lvert{#1}\rvert}            
\newcommand{\Abs}[1]{\left\lvert{#1}\right\rvert} 
\newcommand{\norm}[1]{\lVert{#1}\rVert}           
\newcommand{\mat}[1]{\boldsymbol{\mathbf{#1}}}    

\DeclareMathOperator{\order}{O}                   
\DeclareMathOperator{\tr}{Tr}                     
\DeclareMathOperator{\diag}{diag}                 

\newcommand{\pdiff}[3][{}]{\frac{\partial^{#1}{#2}}{\partial{#3}^{#1}}}

\newcommand{\bra}[1]{\mathinner{\langle{#1}|}}
\newcommand{\ket}[1]{\mathinner{|{#1}\rangle}}

\newcommand{\braket}[1]{\mathinner{\langle{#1}\rangle}}{\catcode`\|=\active
  \gdef\Braket#1{\left<\mathcode`\|"8000\let|\bravert {#1}\right>}}
\newcommand{\bravert}{\egroup\,\vrule\,\bgroup}

\newcommand{\op}[1]{\pmb{#1}}

\newcommand{\Xint}[1]{\mathchoice
   {\XXint\displaystyle\textstyle{#1}}%
   {\XXint\textstyle\scriptstyle{#1}}%
   {\XXint\scriptstyle\scriptscriptstyle{#1}}%
   {\XXint\scriptscriptstyle\scriptscriptstyle{#1}}%
   \!\int}
\newcommand{\XXint}[3]{{\setbox0=\hbox{$#1{#2#3}{\int}$}
     \vcenter{\hbox{$#2#3$}}\kern-.5\wd0}}

\newcommand{\dashint}{\Xint-}

\newcommand{\BdG}{\textsc{b}d\textsc{g}}

\newcommand{\UV}{\textsc{uv}}
\newcommand{\IR}{\textsc{ir}}
\newcommand{\LDA}{\textsc{lda}}
\newcommand{\Lda}{L\textsc{da}}
\newcommand{\SLDA}{\textsc{slda}}
\newcommand{\Slda}{S\textsc{lda}}
\newcommand{\ASLDA}{\textsc{aslda}}
\newcommand{\Aslda}{A\textsc{slda}}
\newcommand{\LOFF}{\textsc{loff}}
\newcommand{\Loff}{L\textsc{off}}
\newcommand{\FFLO}{\textsc{fflo}}
\newcommand{\Fflo}{F\textsc{flo}}
\newcommand{\LO}{\textsc{lo}}
\newcommand{\MIT}{\textsc{mit}}
\newcommand{\DFT}{\textsc{dft}}
\newcommand{\MC}{\textsc{mc}}
\newcommand{\QMC}{\textsc{qmc}}
\newcommand{\AFQMC}{\textsc{afqmc}}
\newcommand{\QCD}{\textsc{qcd}}
\newcommand{\SVD}{\textsc{svd}}
\newcommand{\FFT}{\textsc{fft}}
\newcommand{\TD}{\textsc{td}}
\newcommand{\RG}{\textsc{rg}}
\newcommand{\TDDFT}{\textsc{tddft}}
\newcommand{\GFMC}{\textsc{gfmc}}
\newcommand{\FNDMC}{\textsc{fn}-\textsc{dmc}}
\newcommand{\BCS}{\textsc{bcs}}
\newcommand{\BEC}{\textsc{bec}}
\newcommand{\AdS}{\textsc{a}d\textsc{s}}
\newcommand{\CFT}{\textsc{cft}}
\newcommand{\LGGP}{\textsc{lg}/\textsc{gp}}

\newcommand\bookchapter[1]{
  \setcounter{chapter}{#1}
  \addtocounter{chapter}{-1}
}

\makeatletter
\def\tableofcontents{
  \newpage
  \section*{\contentsname}
  \def\authcount##1{\setcounter{auco}{##1}\setcounter{@auth}{1}}
  \def\lastand{\ifnum\value{auco}=2\relax
                 \unskip{} \andname\
              \else
                 \unskip \lastandname\
              \fi}%
  \def\and{\stepcounter{@auth}\relax
          \ifnum\value{@auth}=\value{auco}%
             \lastand
          \else
             \unskip,
          \fi}%
 \@starttoc{toc}\if@restonecol\twocolumn\fi}

\renewcommand\paragraph{\@startsection{paragraph}{4}{\z@}%
                       {-24\p@}
                       {12\p@}
                       {\normalfont\normalsize\scshape
                        \rightskip=\z@ \@plus 8em\pretolerance=10000 }}
\makeatother

\usepackage{fancyhdr}
 \usepackage[overload]{textcase}

\begin{document}

\author{Aurel Bulgac, Michael McNeil Forbes and Piotr Magierski}

\institute{
  Aurel Bulgac 
  \at Department of Physics, University of Washington, Seattle, WA, USA,\\
  \email{bulgac@uw.edu}
  \and Michael McNeil Forbes
  \at Institute for Nuclear Theory and Department of Physics,\\
  University of Washington, Seattle, WA, USA,\\
  T-2, Los Alamos National Laboratory, Los Alamos, NM, USA\\
  \email{mforbes@uw.edu}
  \and Piotr Magierski
  \at Faculty of Physics, Warsaw University of Technology, Warsaw, Poland,\\
  \email{Piotr.Magierski@if.pw.edu.pl}}

\title{The Unitary Fermi Gas:\\
  From Monte Carlo to Density Functionals}

\bookchapter{9}
\maketitle
\abstract{In this chapter, we describe three related studies of the universal
physics of two-component unitary Fermi gases with resonant
short-ranged interactions.  First we discuss an ab initio auxiliary
field quantum Monte Carlo technique for calculating thermodynamic
properties of the unitary gas from first principles.  We then describe
in detail a Density Functional Theory (\DFT) fit to these
thermodynamic properties: the Superfluid Local Density Approximation
(\SLDA) and its Asymmetric (\ASLDA) generalization.  We present
several applications, including vortex structure, trapped systems, and
a supersolid Larkin-Ovchinnikov (\FFLO/\LOFF) state.  Finally, we
discuss the time-dependent extension to the density functional
(\TDDFT) which can describe quantum dynamics in these systems,
including non-adiabatic evolution, superfluid to normal transitions
and other modes not accessible in traditional frameworks such as a
Landau-Ginzburg, Gross-Pitaevskii, or quantum hydrodynamics.

}
\tableofcontents
\section{Introduction}
\label{cha:introduction}

The question of how pairing correlations between two types of fermions
develop with interaction strength has fascinated physicists for
decades, beginning with the papers of Eagles~\cite{Eagles:1969-10} and
Leggett~\cite{Leggett:1980uq}, and followed by many
others~\cite{Nozieres:1985, Sa-de-Melo:1993, PhysRevB.55.15153,
  Randeria:1995}.  These initial studies focused on the inter-species
pairing gap at various temperatures as the pairing interaction varied
throughout the entire \BCS--\BEC\ crossover from weak to strong
attraction.

Eagles and Leggett~\cite{Eagles:1969-10, Leggett:1980uq} solved the
Bardeen-Cooper-Schrieffer (\BCS) mean-field equations only in the
particle-particle (pairing) channel: The prevailing attitude
(influenced by electronic systems) was that the pairing gap is much
smaller than the self-energy (exponentially suppressed in
weak-coupling), and that the presence or absence of pairing
correlations was a tiny effect compared to the background density
which determined the self-energy.  Subsequent improvements to the
theory focused only on a more accurate description of the pairing
channel~\cite{Nozieres:1985, Sa-de-Melo:1993, PhysRevB.55.15153,
  Randeria:1995, Haussmann:1993, Haussmann:1994, Pistolesi:1994,
  Pistolesi:1996, Pieri:2000}, neglecting the so called
``Hartree-Fock'' contributions to the total energy of such a system.

However, even in the weak coupling limit ($a<0$ and $k_F\abs{a}\ll 1$
where the Fermi energy $\varepsilon_{F} = p_{F}^2/2m$, the Fermi
momentum $p_F = \hbar k_{F} = \hbar (3\pi^2 n)^{1/3}$, $n$ is the
total density, and $a$ is the two-body $s$-wave scattering
length)---which was rather thoroughly studied in the
1950's~\cite{AGD:1975, GM-B:1961}---it was evident that the
``Hartree-Fock'' and higher order particle-hole contributions dominate
in the total energy.  These contributions can be described
perturbatively in terms of the small parameter $k_F a$ (in both \BCS\
and \BEC\ limits).  In the \BCS\ limit, for example, the leading
contributions enter at linear order $\propto \varepsilon_F k_Fa$ while
the particle-particle pairing contributions are exponentially
suppressed $\propto\varepsilon_F \exp (\pi/k_Fa)$.

Despite neglecting the dominant particle-hole contributions, these
mean-field studies correctly captured many of the \emph{qualitative}
features of the \BCS--\BEC\ crossover. This can be partially
attributed to the fact that the particle-particle channel correctly
accounts for the two-body bound state that dominates in the extreme
\BEC\ limit at strong attraction (however, higher order
effects---describing the dimer-dimer interaction for example---are not
correct~\cite{Petrov:2004a, BBF:2003, Brodsky:2006, Levinsen:2006}).

At unitarity, the majority of the interaction energy is due to the
particle-hole channel: see~\cite{CCPS:2003} where the energy of the
normal state at $T=0$ was evaluated for the first time and the
discussion in Sec.~\ref{sec:results:-energy-entr}. In
particular---above the critical temperature $T_c$, for example---the
total energy of the normal phase exceeds the ground state energy by
only about 20\% or so~\cite{Bulgac:2006}: This means that the
condensation energy gained by the particle-particle pairing
interaction is a relatively small contribution to the total
interaction energy.  A quantitative description of unitary physics
must thus include these ``Hartree-Fock'' contributions and go beyond
the simple mean-field models used initially to study the crossover.

In 1999, G.~F.~Bertsch~\cite{Bertsch:1999:mbx} emphasized the special
role played by the problem of a two-species Fermi gas at unitarity
with large scattering length.  In the crust of neutron stars one can
find a situation where the scattering length $a$ of the interaction is
anomalously large compared to the other length scales, the average
interparticle separation $n^{-1/3}$, and the range $r_{0}$ of the
interaction: $r_0\ll n^{-1/3} \ll \abs{a}$. Since the Fermi momentum
is small ($k_Fr_0\ll 1)$, the neutrons effectively interact only in
the relative $s$-partial wave, and the ground state energy should be
some function of the physical parameters defining the system
$E_{gs}=f(N,V,r_0,a,\hbar,m)$, where $N$ is the particle number
contained in a volume $V$ of the system.  In the formal limit of
$k_Fr_0\rightarrow 0$ and $ 1/k_Fa \rightarrow 0$ this function
simplifies:
\begin{equation}
  E_{gs} = f(N,V,\hbar,m)=\frac{3}{5}\varepsilon_FN \xi,
\end{equation}
and all the non-perturbative effects are described by a single
dimensionless constant: $\xi$ (often referred to as the Bertsch
parameter).  At finite temperatures the total energy of the system
becomes a slightly more complicated function, since now it depends
also on the temperature $T$:
\begin{equation}
  E(T) = f(N,V,T,\hbar,m) 
       = \frac{3}{5}\varepsilon_F N \xi\left(\frac{T}{\varepsilon_F}\right).
\end{equation}
The Bertsch parameter (along with all other thermodynamic
properties) becomes a ``universal'' function of the dimensionless
variable $T/\varepsilon_F$~\cite{PhysRevLett.92.090402}.

In 1999 it was not yet clear whether this limit existed: One might
expect such a system to collapse, since the na\"\i{}ve coupling
constant $g=4\pi\hbar^2a/m$ is infinite at unitarity.
Baker~\cite{Baker:1999:PhysRevC.60.054311, baker00:_mbx_chall_compet,
  Heiselberg:2001} provided the first clue that this system was
actually stable.  Carlson and collaborators~\cite{CCPS:2003}
subsequently calculated the energy of this system, proving that it was
stable, and showing that the superfluid paring gap was very
large. Meanwhile, using a Feshbach resonance to induce an extremely
large scattering length, J.~E.~Thomas and his
collaborators~\cite{OHara:2002} produce for the first time a quantum
degenerate unitary gas of cold-atoms in a trap, thus providing
experimental evidence that this system is indeed stable.

There has since been an explosion in both theoretical and experimental
studies of resonant Fermi gases near the unitary regime (see for
example the reviews~\cite{giorgini-2007, Bloch;Dalibard;Zwerger:2008,
  KZ:2008_full, Luo:2009, Grimm:2007}). On one hand, cold-atom
experiments can simulate other systems of interest; for example,
dilute superfluid neutron matter which can only exist  in the
crust of neutron stars, various condensed matter systems (the unitary
gas exhibits a pseudogap that might shed light on the pseudogap in
high-temperature superconductors), and quantum systems with extremely
low viscosity similar to quark-gluon plasmas observed in
ultra-relativistic heavy-ion collisions. On the other hand, the
simplicity of the system provides an excellent vehicle through which
the plethora of many-body techniques can be put to rigorous test,
including both traditional approaches, as well as modern developments
such as the $\epsilon$-expansion and \AdS/\CFT\
correspondence~\cite{Nishida:2006, Son:2008}.

We shall not provide a cursory review of current theoretical
techniques, but will instead focus on a couple of theoretical methods
that have produced a large reliable set of information about the
properties of unitary Fermi gases. The first approach is an ab initio
Quantum Monte Carlo (\QMC) method that has accurately evaluated many
properties of these systems, and has been confirmed by
experiments. The second approach is Density Functional Theory (\DFT),
which is in principle, an exact approach commonly used for describing
``normal'' systems (no superfluidity). We show how to extend the \DFT\
to describe both superfluid systems and time-dependent phenomena, and
how the \DFT\ allows us to address phenomena that are essentially
impossible to describe within a \QMC\ approach.


\section{The Quantum Monte Carlo Approach}
\label{cha:quantum-monte-carlo}

\subsection{From the Physical Problem to the Lattice Formulation}
\label{sec:from-phys-probl}

Atomic collisions in a trap occur at very low relative velocities (due
to the diluteness of the gas) and this fact allows us to restrict the
description to using the lowest partial waves only.  In practice, the
$s$-wave scattering phase shift fully determines the properties of a
unitary Fermi gas, for which $r_0\ll n^{-1/3} \ll \abs{a}$.  The
detailed physics of the collision is more complicated since atoms are
not point-like objects and can appear in various configurations.
Roughly speaking, these can be associated with various valence
electronic configurations.  For example, atoms with a single valence
electron (such as $^{6}$Li) can form two possible electronic
configurations in a binary system: a singlet and a triplet
configuration.  The inter-atomic potential describing a singlet
configuration corresponds to the symmetric spatial wave function.  It
admits the existence of a bound state and corresponds to the closed
(inaccessible) scattering channel.  The triplet channel, on the other
hand, is open and shallow: due to large (mainly electronic) magnetic
moment, its energy can be easily tuned with respect to the closed
singlet channel threshold by adjusting an external magnetic
field. This allows experimentalists to use a Feshbach resonance to
tune the effective interaction in the open channel to virtually any
value: in particular, experiments with dilute clouds of cold atoms can
directly probe the unitary regime.

A typical Hamiltonian describing the two channel atom-atom collision
is of the form~\cite{Stoof:1988, Tiesinga:1991, Tiesinga:1992,
  Tiesinga:1993, Moerdijk:1995}:
\begin{equation}
  \op{H}=\frac{p^2}{2M_{r}} + \sum_{i=1}^{2}(V^{hf}_{i} + V^{Z}_{i}) +
  V^{c} + V^{d},
\end{equation}
where $M_{r}$ is the reduced mass of two atoms,
$V^{hf}=a_{hf}\vec{S}^{e}\cdot\vec{S}^{n}/\hbar^2$ is a hyperfine
interaction term for each atom (with hyperfine constant $a_{hf}$), and
$\vec{S}^{e}$ and $\vec{S}^{n}$ are the total electron spin and the
total nuclear spin respectively.  The Zeeman term $V^{Z}=( \gamma_{e}
S^{e}_{z} + \gamma_{n} S^{p}_{z} ) B $ describes the interaction with
the external magnetic field $B$ which is assumed to be parallel to the
$z$-axis.  The terms $V^{c}$ and $V^{d}$ denote the Coulomb
interaction and dipole-dipole magnetic interaction, respectively.  The
dipole term contributes weakly to the interaction and can be
neglected. The Coulomb term distinguishes singlet and triplet channels
(due to different symmetry properties of electronic wave function) and
produces different interaction potentials in both channels.
Consequently, the continuum of the singlet channel lies above the
continuum of the incident triplet channel. At very low collision
energies, only the singlet channel is open.  However the hyperfine
interaction couples the singlet and triplet states and consequently,
resonant scattering may occur due to the bound state of the singlet
potential (see reviews~\cite{Kohler:2006,Duine:2004, Timmermans:1999}
and references therein).  An external magnetic field can thus be used
as an experimental knob to control the resonance position, effectively
altering the atom-atom collision cross-section.  In the limit of low
collisional energy, the effective scattering length for two colliding
atoms is well described by
\begin{equation}
  \label{eq:a_B}
  a(B) = a_{0} + \frac{C}{B - B_{\text{res}}},
\end{equation}
where $a_{0}$ is the triplet channel off-resonant background
scattering length, and $C>0$.  The second term results from the
coupling to the closed channel, and $B_{\text{res}}$ is the value of
the magnetic field where the Feshbach resonance occurs.  In this way,
experiments may realized the unitary Fermi gas by considering dilute
systems ($r_0 \ll n^{-1/3}$) and tuning the scattering length
(\ref{eq:a_B}) near the resonance ($n^{-1/3} \ll \abs{a}$).

To determine the thermodynamic properties of an ensemble of fermionic
atoms in a non-perturbative manner, we consider the system on a three
dimensional (3D) cubic spatial lattice with periodic boundary
conditions. The system consists of two species of fermions that we
shall denote ``$a$'' and ``$b$''.  In dilute neutron matter these
would correspond to the two spin states of the neutrons, while in cold
atom experiments these are the two populated hyperfine states.
Although there are physical processes that can convert one species to
another, for the purposes of the experiments we shall describe, these
transitions are highly suppressed and one can consider each species to
be independently conserved.

The lattice spacing $l$ and size $L = N_s l$ introduce natural
ultraviolet (\UV) and infrared (\IR) momentum cut-offs given by $\hbar
k_c=\pi\hbar/l$ and $\hbar\Lambda_0=2\pi\hbar/L$, respectively.  The
momentum space has the shape of a cubic lattice, but in order to
simplify the analysis, we place a spherically symmetric \UV\ cut-off,
including only momenta satisfying $k \le k_c \le \pi/l$.  In order to
minimize the discretization errors, the absolute value of scattering
length must be much larger than the lattice spacing: $a \gg l$.

\subsection{Effective Hamiltonian}
\label{sec:effect-hamilt}

As discussed in the introduction, it has by now been well established
that the unitary regime exists and is stable.  Hence, any sufficiently
short-ranged interaction with large scattering length will exhibit the
same universal physics.  Here we use a contact (zero-range)
interaction $V(\vec{r}_1 - \vec{r}_2) = -g \delta(\vec{r}_1 -
\vec{r}_2)$ regularized by the lattice, which defines a momentum
cut-off $\hbar k_c$.  (We require all two-body matrix elements to
vanish if the relative momentum of the incoming particles exceeds this
cutoff.)  The second quantized Hamiltonian of this system is 
\begin{equation}
  \label{eq:Hamiltonian}
  \hat{H}=
  \int \D^{3}r \left (
    -\sum_{\sigma=a,b}
    \hat{\psi}_{\sigma}^{+}(\vec{r})
    \frac{\hbar^2\nabla^2}{2m}\hat{\psi}_{\sigma}(\vec{r})
    + g \hat{n}_a (\vec{r}) \hat{n}_b (\vec{r}) \right ),
\end{equation}
where $\hat{n}_\sigma
(\vec{r})=\hat{\psi}_{\sigma}^{+}(\vec{r})\hat{\psi}_{\sigma}
(\vec{r})$. Once the cutoff is imposed, the value of the bare coupling
$g$ can be tuned to fix the value of the renormalized physical
coupling---in this case, the $s$-wave scattering length $a$.  The
relation between $a$ and the coupling constant $g$ can be obtained
from $T$ matrix describing two-particle scattering induced by the
interaction (\ref{eq:Hamiltonian}) with the $s$-wave phase shift:
\begin{equation}
  k\cot \delta = -\frac{4\pi\hbar^2}{gm} -\frac{2}{\pi}k_c
  -\frac{k}{\pi}\ln \Abs{\frac{k_c-k}{k_c+k}}.
\end{equation}
The low-momentum expansion of the scattering amplitude reads:
\begin{equation}
  f(k)\approx
  \left[{-\I k + \frac{4\pi\hbar^{2}}{gm} - \frac{2k_{c}}{\pi}+\frac{2k^{2}}{\pi
        k_{c}} + \order(k^{3})}\right]^{-1}.
\end{equation}
At low momentum we have $f(k) = [-\I k - 1/{a} +
r_\text{eff}k^2/2 + \order(k^3)]^{-1}$, which gives the relation between
the bare coupling constant $g$ and the scattering length $a$ at a
given momentum cutoff $\hbar k_c$:
\begin{equation}
  \label{eq:RenormalizedCoupling}
  \frac{1}{g} = \frac{m}{4 \pi \hbar^2 a}  - \frac{k_c m}{2 \pi^2
    \hbar^2}
  = \frac{m}{4 \pi \hbar^2 a} \left(1
     - \frac{2 k_c a}{\pi}\right).
\end{equation}
One has to remember, however, that the value of the coupling constant
$g$ has been determined for the two body system in its center of mass
frame.  On the other hand the Hamiltonian (\ref{eq:Hamiltonian}) is
supposed to describe an ensemble of fermions in the box.
Consequently, only a fraction of interacting pairs have their
center of mass at rest with respect to the box. Most of the
interaction processes will occur for pairs for which the center of
mass velocity is nonzero.  It implies that their mutual interaction
will be characterized by a slightly different scattering length than
(\ref{eq:RenormalizedCoupling}).  Consequently, the Hamiltonian will
generate a systematic error in the description of interacting
fermions.  This error will scale as $k_{F}/k_{c}$ and in order to
minimize its influence one should keep the particle density as small
as possible.  Another source of systematic error is related to the
nonzero effective range, which is generated by the interaction and is
independent of the coupling constant $r_\text{eff}=4/(\pi k_{c})$.
Note however that the choice of $k_c$ described above implies that
$r_\text{eff}<l$.

\subsection{The Hubbard-Stratonovich Transformation}
\label{sec:hubb-strat-transf}

Since we are interested in the finite temperature thermodynamic
properties of the system, it is natural to use the grand canonical
ensemble to evaluate physical quantities.  This is equivalent to
considering a small portion of volume $V=L^3$ in thermal and chemical
equilibrium with the larger system.  Consequently we allow for energy
and particle exchange between our subsystem and the larger system,
fixing only the average values of these quantities in the box.  The
thermodynamic variables are thus the temperature $T$, the chemical
potential $\mu$, and the volume $V$.  The partition function and
average of an observable $\hat{O}$ are calculated according to
\begin{align}
  Z(\beta,\mu,V) &=  {\mathrm{Tr}} \left \{ \exp [-\beta (\hat{H}-\mu
    \hat{N})] \right \} , \nonumber \\
  O(\beta,\mu,V) &=
  \frac{{\mathrm{Tr}} \; \left \{ \hat{O}\exp [-\beta (\hat{H}-\mu
      \hat{N})] \right \}}{Z(\beta,\mu,V)} ,
\end{align}
where $\beta = 1/T$ (in this work we will take Boltzmann's constant to
be $k_B = 1$ so that temperature is expressed in units of energy).  In
order to be able to calculate these quantities we first
factorize the statistical weight using the Trotter formula:
\begin{equation}
  \exp [-\beta (\hat{H}-\mu \hat{N})] = \prod_{j=1}^{N_\tau}\exp
  [-\tau (\hat{H}-\mu \hat{N})]
\end{equation}
where $\beta = N_{\tau} \tau$.  The next step is to decompose the
exponentials on the right hand side into exponentials that depend
separately on the kinetic and potential energy operators. The second
order expansion is (higher orders require more effort,
see~\cite{Forbert:2001, Suzuki:1990, Yoshida:1990, Creutz:1989}):
\begin{multline}
  \label{eq:fact-Exp}
  \exp[ -\tau(\hat{H}-\mu \hat{N})]\\
  =\exp \left [ -\frac{\tau (\hat{K}-\mu \hat{N})}{2}\right ]
  \exp(-\tau \hat{V} )
  \exp \left [ -\frac{\tau (\hat{K}-\mu \hat{N})}{2}\right ]
  +\order(\tau^3),
\end{multline}
where $\hat{K}$ is the kinetic energy operator, whose dispersion
relation, for momenta smaller than the cut-off, is given by
$\varepsilon_{\vec{k}}=\hbar^2 k^2/2m$. Since $\tau$ has the dimension
of inverse energy, the above approximate representation makes sense
only if $\tau_{ \max}\norm{\hat{V}} \ll 1$ and $\tau_{ \max} \norm{\hat{K}-\mu
  \hat{N}}\ll 1$.  Since both the interaction and kinetic energies are
extensive quantities, this restriction might appear as very
strict. However, after performing a Hubbard-Stratonovich
transformation (see below), this restriction is considerably eased
and both the kinetic and the interaction energies in these
inequalities are replaced by the corresponding intensive energies per
particle.  It is important to note that, because we have used the
expansion up to $\order(\tau^{3})$, when calculating the partition
function the error becomes $\order(\tau^{2})$. Indeed, the statistical
weight involves a product of $N_\tau$ factors and is given by the
following expression:
\begin{multline}
  \label{eq:GibbsWeight}
  \exp [-\beta (\hat{H}-\mu \hat{N})] =
  \exp \left [-\frac{\tau
      (\hat{K}-\mu \hat{N})}{2}\right ]\\
  \times
  \left ( \prod_{j=1}^{N_\tau}\exp [-\tau \hat{V} ] \exp [-\tau
    (\hat{K}-\mu \hat{N})] \right )
  \exp \left [+\frac{\tau (\hat{K}-\mu \hat{N})}{2}\right ]
  +
  \order(\tau^2)
\end{multline}
Note also that this approach does not depend on the choice of
dispersion relation in the kinetic energy term. However various
choices of representation of derivatives on the lattice may lead to
different discretization errors~\cite{BDM:2008}.  In our case we shall
consider the kinetic energy operator in momentum space, $\epsilon(k) =
\hbar^2 k^2/2m$, which minimizes the discretization errors.

In order to efficiently evaluate the term containing the interaction,
one has to replace it by the sum (or integral) of one body terms.
This can be done with the Hubbard-Stratonovich
transformation~\cite{Negele:1998}. The transformation is not
unique, and we take advantage of this freedom to ensure an
efficient summation (or integration) scheme.  In our case, due to the
simplicity of the interaction term, a discrete Hubbard-Stratonovich
transformation can be applied, similar to that in~\cite{Hirsch:1983}:
\begin{equation}
  \exp [ -g \tau  \hat{n}_a (\vec{r}) \hat{n}_b (\vec{r}) ]=
  \frac{1}{2}\sum _{\sigma (\vec{r},\tau_j)=\pm 1}
  [1 + A \sigma(\vec{r},\tau_j)\hat{n}_a   (\vec{r}) ]
  [1 + A \sigma(\vec{r},\tau_j)\hat{n}_b (\vec{r}) ],
\end{equation}
where $A=\sqrt{ \exp(-g\tau)-1}$, $\tau_j$ labels the location on the
imaginary time axis, $j=1,\dotsc,N_{\tau}$, and
$\sigma(\vec{r},\tau_j)$ is a field that can take values $\pm 1$ at
each point on the space-time lattice.  This identity can be proved
simply by evaluating both sides at $\hat{n}_{\{a,b \}} (\vec{r}) =
0,1$. This discrete Hubbard-Stratonovich transformation is sensible
only for $A<1$, which means that the imaginary time step cannot exceed
$\abs{g}^{-1}\log 2$. The advantages of this transform is discussed,
for example, in~\cite{Hirsch:1983, BDM:2008}.

Taking all this into account, the grand canonical partition function
becomes
\begin{equation}
  Z(\beta,\mu,V) = {\mathrm{Tr}} \left \{ \exp [-\beta (\hat{H}-\mu
    \hat{N})] \right \} =
  \int\prod_{\vec{r},\tau_j}\mathcal{D}\sigma(\vec{r},\tau_j){\mathrm{Tr}}\; 
  \hat{\mathcal{U}}(\{ \sigma \}),
\end{equation}
where we define
\begin{equation}
  \label{eq:FactorizedU}
  \hat{\mathcal{U}}(\{ \sigma \})=\prod_{j=1}^{N_{\tau}}
  \hat{\mathcal{W}}_{j}(\{ \sigma \})
\end{equation}
and
\begin{multline}
  \label{eq:FactorizedW}
  \hat{\mathcal{W}}_{j}(\{ \sigma \})=\exp \left [-\frac{\tau
      (\hat{K}-\mu \hat{N})}{2}\right ]\\
  \times\left (
    \prod_{\bf i} [1 + A \sigma(\vec{r},\tau_j)\hat{n}_a
    (\vec{r}) ]
    [1 + A \sigma(\vec{r},\tau_j)\hat{n}_b (\vec{r}) ] 
  \right )
  \exp \left [-\frac{\tau (\hat{K}-\mu \hat{N})}{2}\right ].
\end{multline}
Since $\sigma(\vec{r}, \tau)$ is discrete, the integration is in fact
a summation:
\begin{equation}
  \int \prod_{\vec{r},\tau_j}\mathcal{D}\sigma(\vec{r},\tau_j) 
  \equiv \sum_{\{\sigma\}}\frac{1}{2^{N_{s}^{3}N_{\tau}}}
  \sum_{\{\sigma(\vec{r},\tau_1)\}=\pm 1}\sum_{\{\sigma(\vec{r},\tau_2)\}=\pm 1}\dotsi
  \sum_{\{\sigma(\vec{r},\tau_{N_{\tau}})\}=\pm 1},
\end{equation}
where
\begin{equation}
  \sum_{\{\sigma(\vec{r},\tau_j)\}=\pm 1}= \sum_{\sigma((1,0,0),\tau_j)=\pm 1}
  \sum_{\sigma((2,0,0),\tau_j)=\pm 1}
  \dots\sum_{\sigma((N_{s},N_{s},N_{s}),\tau_j) = \pm 1}.
\end{equation}
In a shorthand notation we will write
\begin{equation*}
  \hat{\mathcal{U}}(\{ \sigma \}) = {\mathrm{T}}_\tau\exp
  \left \{ -\int \D\tau[\hat{h}(\{\sigma\})-\mu \hat{N}]\right \},
\end{equation*}
where ${\mathrm{T}}_\tau$ stands for an imaginary time ordering operator and
$\hat{h}(\{\sigma\})$ is a resulting $\sigma$-dependent one-body Hamiltonian.
It is crucial to note that $\hat{\mathcal{U}}(\{ \sigma \})$ can be
expressed as a product of two operators which describe the imaginary
time evolution of two species of fermions:
\begin{subequations}
  \label{eq:decomp}
  \begin{align}
    \hat{\mathcal{U}}(\{ \sigma \})&=
    \hat{\mathcal{U}}_{b}(\{ \sigma
    \})\hat{\mathcal{U}}_{a}(\{ \sigma \}), \\
    \hat{\mathcal{U}}_{b}(\{ \sigma \})&=\prod_{j=1}^{N_{\tau}}
    \hat{\mathcal{W}}_{jb}(\{ \sigma \}), &
    \hat{\mathcal{U}}_{a}(\{ \sigma \}) &= \prod_{j=1}^{N_{\tau}}
    \hat{\mathcal{W}}_{ja}(\{ \sigma \}).
  \end{align}
\end{subequations}
As we only consider unpolarized systems, for which $\mu_a = \mu_b = \mu$,
the operators for both species $a$ and $b$ are identical.

The expectation values of operators take the form:
\begin{multline}
  \label{eq:ham-T}
  O(\beta,\mu,V) =
  \frac{{\mathrm{Tr}} \; \left \{ \hat{O}\exp [-\beta (\hat{H}-\mu
      \hat{N})] \right \}}
  {Z(\beta,\mu,V)}=\\
  =\int \frac{\prod_{{\bf i}j}\mathcal{D}\sigma(\vec{r},\tau_j)
    {\mathrm{Tr}}\;\hat{\mathcal{U}}(\{ \sigma \})}{Z(\beta,\mu,V)}
  \; \frac{ {\mathrm{Tr}}\; \hat{O}\hat{\mathcal{U}}(\{ \sigma \})}
  {{\mathrm{Tr}}\; \hat{\mathcal{U}}(\{ \sigma \})},
\end{multline}
where we have introduced ${{\mathrm{Tr}}\; \hat{\mathcal{U}}(\{ \sigma
  \})}$ for convenience: in the numerator it represents the
probability measure used in our simulations (see below), and in the
denominator it serves the purpose of moderating the variations of
${\mathrm{Tr}}\; \hat{O}\hat{\mathcal{U}}(\{ \sigma \})$ as a function
of the auxiliary field $\sigma$.

All of the above traces over Fock space acquire very simple forms
\cite{Koonin:1997,Alhassid:2001}, and can be easily evaluated. In particular,
${\mathrm{Tr}}\;\hat{\mathcal{U}}(\{ \sigma \})$ can be written as
\begin{equation}
  {\mathrm{Tr}}\; \hat{\mathcal{U}}(\{ \sigma \})= \det [1 +
  \mathcal{U}(\{ \sigma \})] =
  \det [1 + \mathcal{U}_{b}(\{ \sigma \})] \det [1 +
  \mathcal{U}_{a}(\{ \sigma \})],
\end{equation}
where $\mathcal{U}$ (without the hat) is the representation of
$\hat{\mathcal{U}}$ in the single-particle Hilbert space.  The second
equality is a result of the decomposition (\ref{eq:decomp}) and is
easy to prove by expanding both sides.  For symmetric (unpolarized)
systems the chemical potentials $\mu_a = \mu_b$ are the same for
both species of fermion, so it follows that $\det [1 +
\mathcal{U}_{b}(\{ \sigma \})] = \det [1 + \mathcal{U}_{a}(\{ \sigma
\})]$.  This implies that ${\mathrm{Tr}}\; \hat{\mathcal{U}}(\{ \sigma
\})$ is positive, i.e., that there is no fermion sign problem. Indeed,
this allows to define a positive definite probability measure:
\begin{multline}
  \label{eq:measure}
  P(\{ \sigma \}) = \frac{{\mathrm{Tr}}\; \hat{\mathcal{U}}(\{ \sigma
    \})}{Z(\beta,\mu,V)} = \frac{\{\det [1 + \mathcal{U}_{a}(\{
    \sigma \})]\}^2}{Z(\beta,\mu,V)}\\
  =\frac{1}{Z(\beta,\mu,V)}
  \exp(2~\text{tr}\left(\log [1 + \mathcal{U}_{a}(\{ \sigma \})]\right))
\end{multline}
where the exponent in the last equation defines the negative of the
so-called effective action.  The positive definite probability measure
is crucial for Monte Carlo (\MC) treatment, allowing for statistical
sampling of the $\sigma$ space.  When considering the polarized
system, the sign problem inevitably occurs, making the Monte
Carlo procedure very difficult.  The sign problem appears also when
more complicated forms of interaction are applied.  In such a case one
can sometimes cure the problem by properly choosing the
Hubbard-Stratonovich transformation~\cite{Wlazowski:2009}.

The many-fermion problem is thus reduced to an Auxiliary Field Quantum
Monte Carlo problem (\AFQMC), to which the standard Metropolis
algorithm can be applied, using (\ref{eq:measure}) as a probability
measure.  Before moving on to the details of our Monte Carlo
algorithm, we briefly discuss the expressions used to compute a few
specific thermal averages.

Let us consider the one body operator
\begin{equation}
  \hat{O}=\sum_{s,t=b,a}
  \int \D^3 {\vec{r}_{1}} d^3 {\vec{r}_{2}}\hat{\psi}_{s}^{+}(\vec{r}_{1})
  O_{st}({\vec{r}_{1}},{\vec{r}_{2}})\hat{\psi}_{t}(\vec{r}_{2})
\end{equation}
From (\ref{eq:ham-T}) it follows that
\begin{equation}
  \langle \hat{O}\rangle  = \sum_{\{\sigma\}}P(\{\sigma\})\frac{
    {\mathrm{Tr}}\;
    \hat{O}\hat{\mathcal{U}}(\{ \sigma \})}
  {{\mathrm{Tr}}\; \hat{\mathcal{U}}(\{ \sigma \})}=
  \sum_{\{\sigma\}}P(\{\sigma\})\frac{ {\mathrm{Tr}}\;
    \hat{O}\hat{\mathcal{U}}(\{ \sigma \})}
  {\det [1+\mathcal{U}(\{ \sigma \})]}.
\end{equation}
The calculation of the last term requires the evaluation of
\begin{equation}
  {\mathrm{Tr}}\left [
    \hat{\psi}_{s}^{+}(
    \vec{r}_{1})\hat{\psi}_{t}(\vec{r}_{2})\hat{\mathcal{U}}(\{
    \sigma \}) \right ]
  = \delta_{st}\det[1+\mathcal{U}(\{ \sigma \})]^2
  n_{s}(\vec{r}_{1},\vec{r}_{2},\{ \sigma\})
\end{equation}
where $s$ and $t$ run over both species ($a$ or $b$), and
\begin{equation}
  n_{s}(\vec{r}_{1},\vec{r}_{2},\{ \sigma\})=
  \sum_{\vec{k}_1,\vec{k}_2\le k_c}
  \varphi_{\vec{k}_1}(\vec{r}_{1})
  \left [ \frac{  \mathcal{U}_{s}(\{ \sigma \})  }
    {  1+\mathcal{U}_{s}(\{ \sigma \})   }
  \right ] _{ \vec{k}_{1},\vec{k}_{2} } \varphi_{\vec{k}_{2}}^*(\vec{r}_{2})
\end{equation}
Here $\varphi_{\vec{k}}(\vec{r})=\exp(i\vec{k}\cdot\vec{r})/L^{3/2}$
are the single-particle orbitals on the lattice with periodic boundary
conditions, and hence quantized momenta $\vec{k} = 2\pi \vec{n}/L$.
This holds for any 1-body operator $\hat O$, if $\mathcal{U}$ is a product
of exponentials of 1-body operators, as is the case once the
Hubbard-Stratonovich transformation is performed.  It is then obvious
that the momentum representation of the one-body density matrix has
the form
\begin{equation}
  n_{s}(\vec{k}_{1},\vec{k}_{2},\{ \sigma\})=
  \left [ \frac{  \mathcal{U}_{s}(\{ \sigma \})  }{  1+\mathcal{U}_{s}(\{
      \sigma \})   }
  \right ] _{ \vec{k}_{1},\vec{k}_{2} }
\end{equation}
which, for a non-interacting homogeneous Fermi gas, is diagonal and
equal to the occupation number probability $1/(\exp [\beta
(\varepsilon_{\vec{k}}-\mu)] + 1)$ of a state with the energy
$\varepsilon_{\vec{k}}=\hbar^{2}k^{2}/(2m)$.

Summarizing, the expectation value of any one-body operator may be
calculated by summing over samples of the auxiliary field $\sigma(\vec{r},\tau_j)$:
\begin{equation}
  \langle \hat{O}\rangle  
  = \int\prod_{\vec{r},\tau_j}\mathcal{D}\sigma(\vec{r},\tau_j) P(\{\sigma \})
  \sum_{\vec{r}_{1},\vec{r}_{2}}\sum_{s=a,b}
  O_{ss}(\vec{r}_{1},\vec{r}_{2})n_{s}(\vec{r}_{1},\vec{r}_{2},\{ \sigma\})
\end{equation}
In particular, the kinetic energy can be calculated according to:
\begin{multline}
  \langle \hat{K}\rangle =
  \int \frac{\prod_{\vec{r},\tau_j}\mathcal{D}\sigma(\vec{r},\tau_j)
    {\mathrm{Tr}}\;\mathcal{U}(\{ \sigma \})}{Z(\beta,\mu,V)}
  \; \frac{ {\mathrm{Tr}}\; \hat{K}\mathcal{U}(\{ \sigma \})}
  {{\mathrm{Tr}}\; \mathcal{U}(\{ \sigma \})}\\
  =\int \prod_{\vec{r},\tau_j}\mathcal{D}\sigma(\vec{r},\tau_j)P(\{\sigma \})
  \sum_{\vec{k}}^{k\le k_{c}}\sum_{s=a,b}
  \left [ n_{s} (\vec{k},\vec{k},\{ \sigma\}) \frac{\hbar^{2}\vec{k}^2}{2m}
  \right ]
\end{multline}
Analogously, for a generic two-body operator:
\begin{equation}
  \hat{O}=\!\!\!\!\!\!\!\sum_{s,t,u,v=b,a}
  \int \D^3 {\vec{r}_{1}'} \D^3 {\vec{r}_{2}'} \D^3 {\vec{r}_{1}} \D^3 {\vec{r}_{2}}
  \hat{\psi}_{s}^{+}(\vec{r}_{1}')\hat{\psi}_{t}^{+}(\vec{r}_{2}')
  O_{stuv}({\vec{r}_{1}'},{\vec{r}_{2}'},{\vec{r}_{1}},{\vec{r}_{2}})
  \hat{\psi}_{v}(\vec{r}_{2})\hat{\psi}_{u}(\vec{r}_{1}).
\end{equation}
In order to calculate $\langle\hat{O}\rangle$ one needs to evaluate the expression
\begin{multline}
  {\mathrm{Tr}}\left [
    \hat{\psi}_{s}^{+}(\vec{r}_{1}')\hat{\psi}_{t}^{+}(\vec{r}_{2}')
    \hat{\psi}_{v}(\vec{r}_{2})\hat{\psi}_{u}(\vec{r}_{1})
    \mathcal{\hat{U}}(\{ \sigma \}) \right ]\\
  =\left(\det[1 + \mathcal{U}(\{\sigma\})]\right )^2
  \Biggl(
    \delta_{su}\delta_{tv}
    n_{s}(\vec{r}_{1}',\vec{r}_{1},\{ \sigma\}) n_{t}(\vec{r}_{2}',\vec{r}_{2},\{ \sigma\})\\
    - \delta_{sv}\delta_{tu}n_{s}(\vec{r}_{1}',\vec{r}_{2},\{ \sigma\})
    n_{t}(\vec{r}_{2}',\vec{r}_{1},\{ \sigma\})
  \Biggr).
\end{multline}
Hence, for the expectation value of the two body operator we get
\begin{multline}
  \langle\hat{O}\rangle=
  \int\prod_{\vec{r},\tau_j}\mathcal{D}\sigma(\vec{r},\tau_j) P(\{\sigma \})\\
  \times \sum_{\vec{r}_{1}',\vec{r}_{2}',\vec{r}_{1},\vec{r}_{2}}\sum_{s,t=a,b}
  \Bigg [
  O_{stst}({\vec{r}_{1}'},{\vec{r}_{2}'},{\vec{r}_{1}},{\vec{r}_{2}})
  n_{s}(\vec{r}_{1}',\vec{r}_{1},\{ \sigma\}) n_{t}(\vec{r}_{2}',\vec{r}_{2},\{ \sigma\})\\
  -O_{stts}({\vec{r}_{1}'},{\vec{r}_{2}'},{\vec{r}_{1}},{\vec{r}_{2}})
  n_{s}(\vec{r}_{1}',\vec{r}_{2},\{ \sigma\}) n_{t}(\vec{r}_{2}',\vec{r}_{1},\{ \sigma\})
  \Bigg ].
\end{multline}
In particular, the expectation value of the interaction energy reads:
\begin{equation}
  \langle\hat{V}\rangle=-g\int\prod_{\vec{r},\tau_j}\mathcal{D}\sigma(\vec{r},\tau_j) P(\{\sigma \})
  \sum_{\vec{r}}n_{a}(\vec{r},\vec{r},\{ \sigma\})
  n_{b}(\vec{r},\vec{r},\{ \sigma\})
\end{equation}
It should be noted that in the symmetric system
($\mu_{a}=\mu_{b}$)
\begin{equation}
  n_{a}(\vec{r},\vec{r}',\{ \sigma\}) = n_{b}(\vec{r},\vec{r}',\{ \sigma\}).\
\end{equation}
Hence,
\begin{equation}
  \langle\hat{V}\rangle=-g\int\prod_{\vec{r},\tau_j}\mathcal{D}\sigma(\vec{r},\tau_j) P(\{\sigma \})
  \sum_{\vec{r}} [ n_{a}(\vec{r},\vec{r},\{ \sigma\}) ]^{2}
\end{equation}
It is useful to introduce the correlation function
\begin{align}
  g_2(\vec{r}) =& \left (\frac{2}{N}\right )^2
  \int \D^3 {\vec{r}_{1}} \D^3 {\vec{r}_{2}}
  \langle\psi_a^\dagger(\vec{r}_{1} + \vec{r})
  \psi_b^\dagger(\vec{r}_{2}+\vec{r})\psi_b(\vec{r}_{2})
  \psi_a(\vec{r}_{1})\rangle \; \nonumber\\
  =& \left (\frac{2}{N}\right )^{2}
  \int\prod_{\vec{r},\tau_j}\mathcal{D}\sigma(\vec{r},\tau_j) P(\{\sigma \})
  \nonumber \\
   &\times \int \D^3 {\vec{r}_{1}} \D^3 {\vec{r}_{2}}
  n_{a}(\vec{r}_{1} + \vec{r},\vec{r}_{1},\{ \sigma\})
  n_{b}(\vec{r}_{2} + \vec{r},\vec{r}_{2},\{ \sigma\}),
  \label{TBDM}
\end{align}
(where $N$ is the average particle number) which is normalized in such
a way that for a non-interacting homogeneous Fermi gas $g_{2}
(\vec{r}) = 3 j_{1}(k_{F}r)/(k_{F}r)$ and $g_{2} (0)=1$.

\subsection{Stabilization of the Algorithm for Small Temperatures}

Once we have written the observables as in (\ref{eq:ham-T}), the next
step is to sum over all possible configurations of
$\sigma(\vec{r},\tau_j)$.  This is still an impossible task, as for
example, a lattice size $N_x^3 \times N_\tau$ (where typically $N_x =
8$ and $N_\tau \simeq 1000$), requires performing the sum over the
$2^{N_x^3 \times N_\tau}$ points in configuration space.  It is in
these cases that a Monte Carlo approach becomes essential.  By
generating $\mathcal{N}$ independent samples of the field
$\sigma(\vec{r},\tau_j)$ with probability given by (\ref{eq:measure}),
and adding up the values of the integrand at those samples, one can
estimate averages of observables with $\order(1/\sqrt{\mathcal{N}})$
accuracy.

The standard Metropolis algorithm is used to generate the samples.
Namely, at every \MC\ step, the sign of $\sigma$ is changed at random
locations of the space-time lattice (see~\cite{Bulgac:2006, BDM:2008,
  Bulgac:2006a} for details).  This procedure allows to probe the
sigma space, in order to collect the set of statistically uncorrelated
samples.

In order to compute the probability of a given $\sigma$ configuration,
it is necessary to find the matrix elements of $\mathcal{U}$, which entails
applying it to a complete set of single-particle wave-functions. For
the latter we chose plane waves (with momenta $\hbar k\le \hbar
k_c$). This choice is particularly convenient because one can compute
the overlap of any given function with the whole basis of plane waves
by performing a single Fast Fourier Transform (\FFT) on that
function~\cite{BDM:2008}.

The procedure described above requires many matrix multiplications to
calculate $\mathcal{U}$. In particular at low temperatures the number
of matrix multiplications grows rapidly and the matrices have elements
that vary over a large range of magnitudes. To avoid numerical
instabilities it is necessary to separate the scales when multiplying
the matrices, and a more costly but robust algorithm such as the
Singular Value Decomposition (\SVD) is required. In this section we
follow the same approach developed in~\cite{Koonin:1997} to introduce
the \SVD\ to our calculations.

Let us write the matrix $\mathcal{U}(\{ \sigma \})$ more explicitly:
\begin{equation}
  \mathcal{U}(\{ \sigma \})=
  \prod_{j=1}^{N_\tau} \mathcal{W}_j(\{ \sigma \}) =
  \mathcal{W}_{N_\tau} \mathcal{W}_{N_{\tau}-1} \dotsm \mathcal{W}_2 \mathcal{W}_{1} ,
\end{equation}
where the $\mathcal{W}_k(\{ \sigma \})$ are $N\times N$ matrices,
for a single-particle basis of dimension $N$. Let us then define
\begin{align}
  \mathcal{U}_{0} &= 1 \\
  \mathcal{U}_{1} &= \mathcal{W}_{1} \nonumber\\
  \mathcal{U}_{2} &= \mathcal{W}_{2}\mathcal{W}_{1} \nonumber\\
  &\hspace{0.5em}\vdots \nonumber\\
  \mathcal{U}_{n} &= \mathcal{W}_{n}\mathcal{W}_{n-1}\dotsm \mathcal{W}_{1} =
  \mathcal{W}_{n}\mathcal{U}_{n-1}. \nonumber
\end{align}

To separate the scales one decomposes the matrix $\mathcal{U}_{n-1}$
before multiplying it by $\mathcal{W}_{n}$ to get $\mathcal{U}_{n}$. This
process begins as follows
\begin{align}
  \mathcal{U}_{0} &= 1 \\
  \mathcal{U}_{1} &= \mathcal{W}_{1} =
  \mathcal{S}_{1}\mathcal{D}_{1}\mathcal{V}_{1} \nonumber\\
  \mathcal{U}_{2} &= \mathcal{W}_{2}\mathcal{W}_{1} =
  (\mathcal{W}_{2}\mathcal{S}_{1}\mathcal{D}_{1})\mathcal{V}_{1} =
  \mathcal{S}_{2}\mathcal{D}_{2}\mathcal{V}_{2}\mathcal{V}_{1} \nonumber
\end{align}
where $\mathcal{S}_{1}$ and $\mathcal{V}_{1}$ are orthogonal matrices (not
necessarily inverses of each other), and $\mathcal{D}_{1}$ is a diagonal
positive matrix containing the singular values of $\mathcal{U}_{1}$. The
idea is that the actual multiplication should be done by first
computing the factor in parenthesis in the last equation.  This factor
is then decomposed into $\mathcal{S}_{2}\mathcal{D}_{2}\mathcal{V}_{2}$, in
preparation for the multiplication by $\mathcal{W}_{3}$, and so on.  A
generic step in this process looks like:
\begin{equation}
  \mathcal{U}_{n} = \mathcal{W}_{n}\mathcal{U}_{n-1} =
  \mathcal{W}_{n}\mathcal{S}_{n-1}\mathcal{D}_{n-1}\mathcal{V}_{n-1}
  \mathcal{V}_{n-2}\dotsm\mathcal{V}_{1},
\end{equation}
so that in the end
\begin{equation}
  \mathcal{U}_{N_\tau} = \mathcal{U}(\{\sigma\})=
  \mathcal{S}_{N_\tau}\mathcal{D}_{N_\tau}\mathcal{V}_{N_\tau}
  \mathcal{V}_{N_\tau-1}\dotsm \mathcal{V}_{1} =
  \mathcal{S}\mathcal{D}\mathcal{V},
\end{equation}
where we have decomposed the full product in the last step.
Calculating the determinant, and therefore of the probability
measure, is straightforward if we perform one final more \SVD\ in the
following chain of identities:
\begin{multline}
  \det(1+ \mathcal{U}(\{\sigma\}))
  = \det(1+\mathcal{S}\mathcal{D}\mathcal{V}) 
  = \det(\mathcal{S}(\mathcal{S}^\dagger\mathcal{V}^\dagger+\mathcal{D})\mathcal{V})\\
  = \det(\mathcal{S}\;\tilde{\mathcal{S}} \tilde{\mathcal{D}}\tilde{\mathcal{V}}\;\mathcal{V})
  = \det(\mathcal{S}\;\tilde{\mathcal{S}}) 
    \det(\tilde{\mathcal{D}})
    \det(\tilde{\mathcal{V}}\mathcal{V})
\end{multline}
For equal densities (the symmetric case) we need this determinant
squared, so we only care about the factor in the middle of the last
expression: the other two factors have unit magnitude. Indeed, in that
case we can write the probability measure as
\begin{equation}
  P(\{\sigma\}) = \exp\left(\sum^M_{i=1} \log \tilde d_{i}\right)
\end{equation}
where $\tilde d_{i} > 0$ are the elements in the diagonal of
$\tilde{\mathcal D}$, and $M$ is the dimension of the single particle
Hilbert space.  The number of \SVD's required to stabilize the
calculation grows as we increase $\beta$.  In our calculations we have
made limited use of the \SVD, ranging from 2 decompositions at the highest
$T$ to 8 decompositions at low $T$'s.

\subsection{Finite Size Scaling}
\label{sec:finite-size-scaling}

The Monte Carlo calculations are performed in a box of finite size
with a finite average number of particles.  We are interested,
however, in the thermodynamic limit $N\rightarrow\infty, V
\rightarrow\infty$ and $N/V=\text{const}$, so we need to consider the finite
size scaling of the system so we can properly relate the values
calculated in the box to their thermodynamic counterparts.  This
becomes particularly important in the vicinity of phase transitions
where the correlation length $\xi_{\text{corr}}$ characterizing the
non-local degree of correlation of a system diverges:
\begin{equation}
  \xi_\text{corr} \propto \abs{t}^{-\nu} ,
\end{equation}
where $t =1 - T/T_c$, $T_c$ is the critical temperature, and $\nu$ is
a universal critical exponent. For the $U(1)$ universality class,
(which contains superfluid phase transitions), this exponent is
well-known: $\nu = 0.671$.

When dealing with systems that have a finite size $L^3$, the theory of
the renormalization group (\RG) predicts a very specific behavior for
the correlation functions close enough to the transition temperature
(see e.g.~\cite{Barber:1983}). In particular, the two-body
density matrix $K(L,T)$ that gives the order parameter for
off-diagonal long-range order, scales as
\begin{equation}
  R(L,T) = L^{1+\eta}K(L,T) = f(x)(1+c L^{-\omega}+\dotsb),
\end{equation}
where $\eta=0.038$ is another universal critical exponent, $f(x)$ is a
universal analytic function, $x = (L/\xi_{corr})^{1/\nu}$, and $c$ is
a non-universal constant, and $\omega \simeq 0.8$ is the critical
exponent of the leading irrelevant field. One should keep in mind that
typically one knows neither $c$ nor $T_c$, but is interested in
finding the latter.

In a typical Monte Carlo calculation $K(L,T)$ is computed for various
lengths $L_i$ and temperatures $T$. The procedure to locate the
critical point (characterized by scale invariance) involves finding
the ``crossing'' temperatures $T_{ij}$, for which $R(L_i,T_{ij}) =
R(L_j,T_{ij})$ at two given lengths $L_i$ and $L_j$.  Assuming that
one is close to the transition (so that the correlation length is
large compared to any other scale), one can expand $f(x(\abs{t})) =
f(0) + f'(0)L^{1/\nu}b\abs{t}$ (where we set $\xi_{corr} = b
\abs{t}^{-\nu}$), and derive the relation
\begin{equation}
  \abs{T_c -T_{ij}} = \kappa g(L_i,L_j),
\end{equation}
where
\begin{equation}
  g(L_i,L_j) = L_j^{-(\omega + 1/\nu)}
  \left[
    \frac{\left(\frac{L_j}{L_i}\right)^\omega -1}{1-\left(
        \frac{L_i}{L_j}\right)^{1/\nu}}
  \right]
\end{equation}
and $\kappa = c T_c f(0)/bf'(0)$. If there were no non-universal
corrections to scaling (i.e. if $c=0$), then $\kappa = 0$ and $T_c =
T_{ij}$, which means that, upon scaling by the appropriate factor (as
above) all the curves $K(L,T)$ corresponding to different $L$'s would
cross exactly at $T_c$. In general these corrections are present, and
it is therefore necessary to perform a linear fit of $T_{ij}$
vs. $g(L_i,L_j)$ and extrapolate to infinite $L$ in order to determine
the true $T_c$~\cite{BDM:2008}.

\subsection{Results: the Energy and the Entropy}
\label{sec:results:-energy-entr}

\begin{figure}[t]
  \begin{center}
    \includegraphics[width=0.75\textwidth]{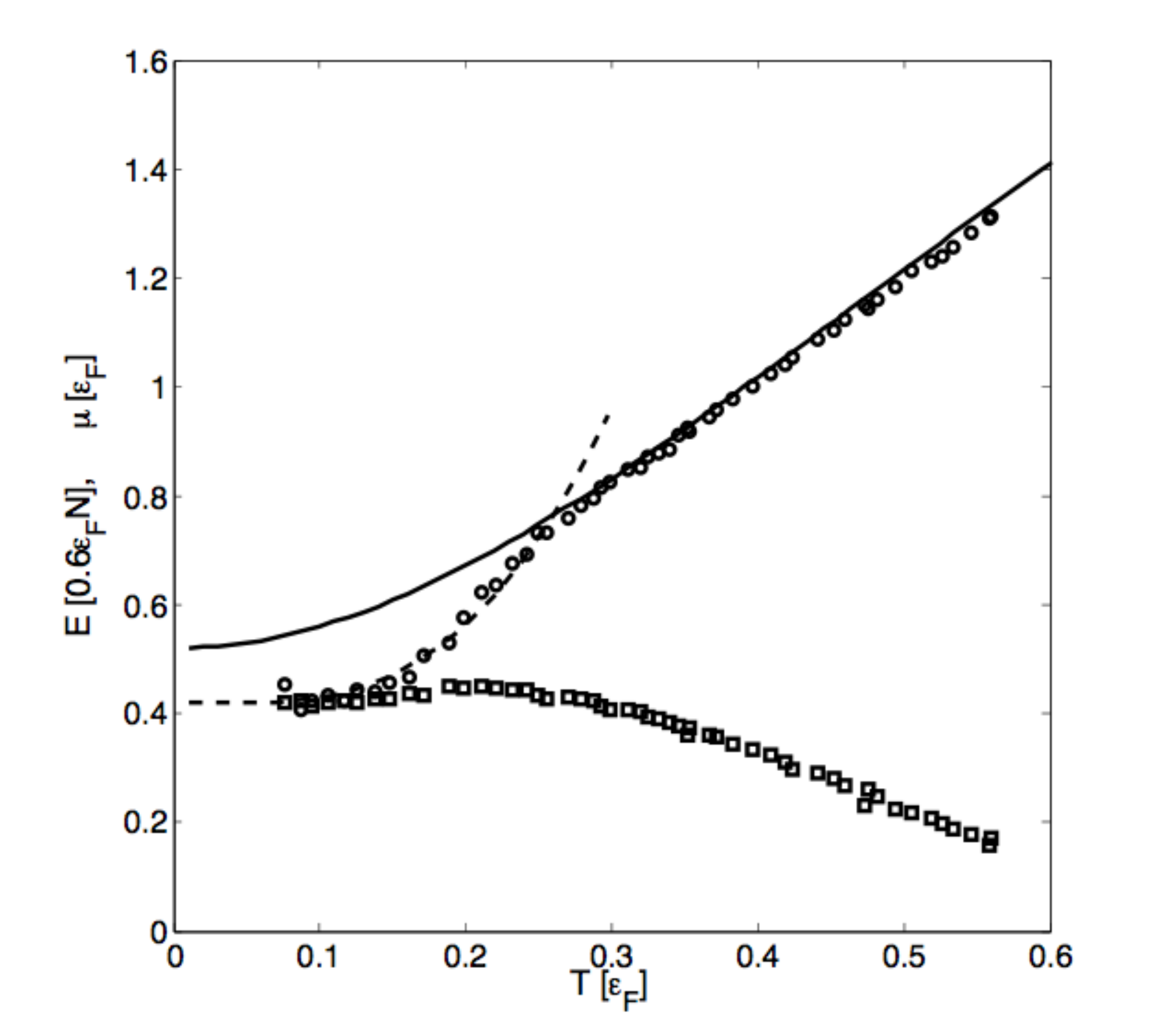}%
    \caption{
      \label{fig:energy} The total energy $E(T)$ with open circles,
      and the chemical potential $\mu(T)$ with squares, both for the
      case of an $8^3$ lattice. The combined Bogoliubov-Anderson
      phonon and fermion quasiparticle contributions $E_{\text{ph}+\text{qp}}(T)$
      (Eq.~(\ref{eq:phqp})) is shown as a dashed line. The solid line
      represents the energy of a free Fermi gas, with an offset (see
      text).  From~\cite{BDM:2008}.}
  \end{center}
\end{figure}

\begin{figure}[t]
  \begin{center}
    \includegraphics[width=0.85\textwidth]{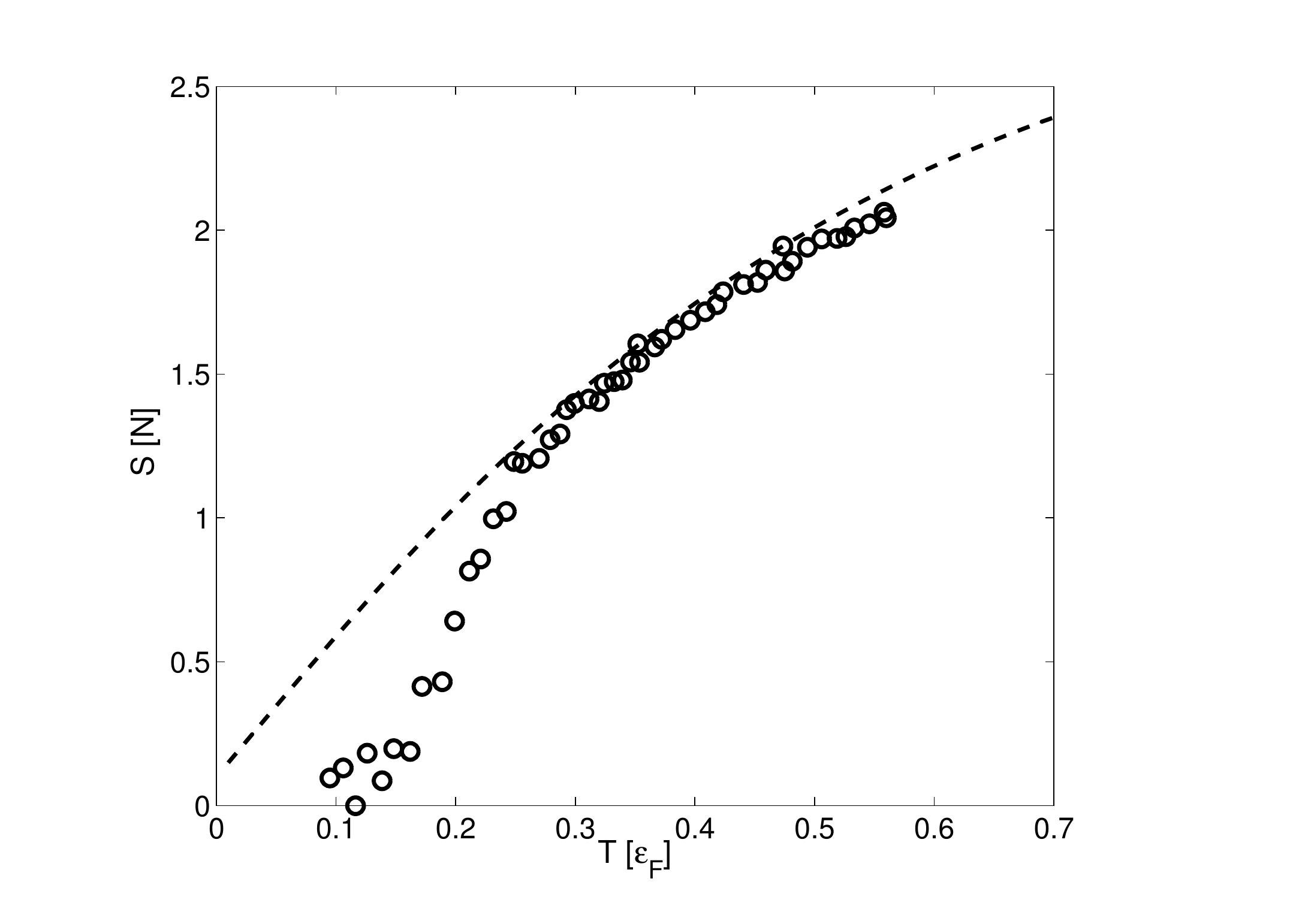}
    \caption{
      \label{fig:entropy} The entropy per particle with circles for
      $8^3$ lattice, and with a dashed line the entropy of a free
      Fermi gas with a slight vertical offset. The statistical errors
      are the size of the symbol or smaller. From~\cite{BDM:2008}.}
  \end{center}
\end{figure}

The results of our Monte Carlo simulations are shown in
Figs.~\ref{fig:energy} and \ref{fig:entropy}~\cite{Bulgac:2006,
  Bulgac:2006a,BDM:2008}.  The Monte Carlo autocorrelation length was
estimated (by computing the autocorrelation function of the total
energy) to be approximately $200$ Metropolis steps at $T\approx 0.2
\varepsilon_F$. Therefore, the statistical errors are of the order of
the size of the symbols in the figure. The chemical potential was
chosen so as to have a total of about $45$ particles for the $8^3$
lattice. We have also performed calculations for particle numbers
ranging from $30$ to $80$, for lattice sizes $8^3$ and $10^3$, and
various temperatures: in all cases, the results agree to within the
aforementioned errors.

According to the theory~\cite{Tan:2008uq, Braaten:2008} the asymptotic
behavior in the limit of large momenta $n(k)\propto C (k_F/k)^n$
should at all temperatures be governed by the same exponent, namely
$n=4$. This is consistent with a value of the exponent $n=4.5(5)$
extracted from the \MC\ data.  Both the energy~\ref{fig:energy} and
the entropy~\ref{fig:entropy} exhibit a definite transition between
low and a high temperature regimes separated by a characteristic
temperature $T_0$:
\begin{equation}
  T_{0} = 0.23(2)\varepsilon_F.
\end{equation}
We shall discuss the relation between $T_0$, the superfluid critical
temperature $T_c$, and the pair breaking temperature $T^{*}$ in
Sec.~\ref{sec:pair-gap-pseud}.  First we focus on the low temperature
limit.

At $T=0$, several interesting quantities describe the symmetric
unitary system: one is the energy as expressed through the Bertsch
parameter $\xi = E_{SF}/E_{FG}$; related is the somewhat fictitious
energy of the interacting normal state $\xi_{N} = E_{N}/E_{FG}$;
finally, there is the pairing gap $\Delta = \eta\varepsilon_{F}$.  The
$T=0$ value of these quantities have been obtained to high precision
by other groups using the variational fixed-node Monte Carlo
techniques~\cite{CCPS:2003, CPCS:2004, ABCG:2004, Carlson:2005kg}.
Unlike our approach, these $T=0$ techniques suffer from a sign problem
that is overcome by using a fixed-node constraint: This formally
provides only an upper bound on the energy.  Our result $\xi =
0.37(5)$ (see Table~\ref{table:results} agrees with these variational
bounds, $\xi=0.44(1)$~\cite{CCPS:2003, CPCS:2004},
$\xi=0.42(1)$~\cite{ABCG:2004, Carlson:2005kg}, and with more recently
quoted \AFQMC\ results $\xi=0.40(1)$~\cite{Gezerlis;Carlson:2008-03,
  Carlson:private_comm2}.  Although not as precise, our method is
truly ab initio and hence provide a non-trivial validation of these
variational results.

The quantity $\xi_{N}$ for the normal state---though not precisely
defined (since the normal state is not the ground state)---provides a
useful description of the physics.  For example, in the high
temperature regime $T>T_{0}$, the energy is described well by the
energy of a free Fermi gas shifted down by $1-\xi_{N}$ (shown as a
solid line in Fig.~\ref{fig:energy}), where $\xi_{N} = \xi +
\delta_{\xi} \approx 0.52$ can be found by determining what shift is
necessary to make the solid curve coincide with the high temperature
data (where the gas is expected to become normal).

Taking $\xi\approx 0.4$ this gives the condensation energy
$\delta_{\xi} \approx 0.12$ which is roughly consistent with the
estimate
\begin{equation}
  \label{eq:delta_xi}
  \delta_{\xi} =
  \frac{\delta E}{\frac{3}{5} \varepsilon_F N} =
  \frac{5}{8} \left (\frac{\Delta}{\varepsilon_F}\right)^2 \simeq 0.15
\end{equation}
based on the \BCS\ expression for $\delta E =
\frac{3}{8}\frac{\Delta^2}{\varepsilon_F} N$ (see~\cite{BY:2003}) and
the \QMC\ value of the pairing gap where $\Delta \simeq 0.50
\varepsilon_F$~\cite{CCPS:2003, Carlson:2005kg} and confirmed by us
in~\cite{Magierski:2009} (which turns out to be very close to the
weak-coupling prediction of Gorkov and
Melik-Barkhudarov~\cite{GM-B:1961, HPSV:2000}).  Our estimate should
also be compared with the results $\xi_{N} \approx 0.54$
of~\cite{CCPS:2003, Carlson:2003} and $\xi_{N} \approx 0.56$
of~\cite{LRGS:2006} obtained by considering only normal state nodal
constraints.  Finally, a similar result $\xi_{N} \approx 0.57(2)$ (see
Eq.~\eqref{eq:xi_N}) arises from fitting the \SLDA\ density functional
to be discussed in Sec.~\ref{sec:symm-superfl-state}.

At low temperatures, $T<T_0$, temperature dependence of the energy can
be accounted for by the elementary excitations present in the
superfluid phase: boson-like Bogoliubov-Anderson phonons and
fermion-like gapped Bogoliubov quasiparticles. Their contributions are
given by
\begin{align}
  \label{eq:phqp}
  E_{\text{ph}+\text{qp}}(T) &= \frac{3}{5}\varepsilon_F N \left [ \xi +
    \frac{\sqrt{3}\pi^4}{16\xi^{3/2}}
    \left(\frac{T}{\varepsilon_F}\right )^4  +
    \frac{5}{2}\sqrt{\frac{2\pi\Delta^3T}{\varepsilon_F^4}}
    \exp\left (-\frac{\Delta}{T}\right)\right ] ,\\
  \Delta &\approx \left (\frac{2}{e}\right )^{7/3} \!\!\!\!\!\!
  \varepsilon_F \exp \left (\frac{\pi}{2k_Fa}\right ),
\end{align}
The sum of the contributions from these excitations is plotted in
Fig.~\ref{fig:energy} as a dashed line: Both of these contributions
are comparable in magnitude over most of the temperature interval
$(T_0/2,T_0)$. Since the above expressions are only approximate for
$T\ll T_c$, the agreement with our numerical results may be
coincidental.

At $T>T_c$ the system is expected to become normal. If $T_0$ and $T_c$
are identified, then the fact that the specific heat is essentially
that of a normal Fermi liquid $E_F(T)$ above $T_0$ is somewhat of a
surprise: one would expect the presence of a large fraction of
non-condensed but unbroken pairs. Indeed, the pair-breaking temperature
has been estimated to be $T^* \simeq 0.55\varepsilon_F$, based on
fluctuations around the mean-field, see~\cite{Eagles:1969-10,
  Leggett:1980uq, Leggett:1980, Nozieres:1985, Sa-de-Melo:1993,
  Randeria:1995, Perali:2004}.  This implies that for $T_c < T < T^*$ there
should be a noticeable fraction of non-condensed pairs. In the next
sections we will show that this is indeed the case and that above the
superfluid critical temperature, the fermionic spectrum still contains
a gap, giving rise to the so-called pseudogap phase.

From the data for the energy $E$ and chemical potential $\mu$, one can
compute the entropy $S$ using the unitary relation $PV = \frac{2}{3}E$
(true of a free gas as well) which holds, where $P$ is the pressure,
$V$ is the volume and $E$ is the energy.  It is straightforward to
show that
\begin{equation}
  \frac{S}{N} = \frac{E + PV - \mu N}{N T} = \frac{\xi(x) - \zeta(x)}{x} ,
\end{equation}
where $\zeta(x) = \mu/\varepsilon_F$ and $x = T/\varepsilon_F$
determines the entropy per particle in terms of quantities extracted
from our simulation. As shown in Fig.~\ref{fig:entropy}, the entropy
also departs from the free gas behavior below $T_0$.

This data can be used to calibrate the temperature scale at
unitarity~\cite{Bulgac:2006, Bulgac:2006a}.  Indeed, extending the
suggestion of~\cite{Carr:2004}, from a known temperature in the \BCS\ limit, the
corresponding $S(T_{\BCS})$ can be determined. Then, by adiabatically
tuning the system to the unitary regime, one can uses $S(T_{\BCS}) =
S(T_{\text{unitary}})$ to determine $T$ at unitarity.  (In practice the
experimental procedure goes in the opposite direction, namely
measurements are performed at unitarity, and then the system is tuned
to the deep \BCS\ side, see~\cite{Luo:2007}.)

On the other hand, knowledge of the chemical potential as a function
of temperature allows for the construction of density profiles by
using of the Local Density Approximation (\LDA) (see the next
section).  In turn, this makes it possible to determine $S(E)$ for the
system in a trap, fixing the temperature scale via $\partial
S/\partial E = 1/T$. Direct comparison with experiment shows
remarkable agreement with our data (we discuss this later in
Fig.~\ref{fig:SofE})~\cite{BDM:2007}.

\begin{figure}[htb]
  \begin{center}
    \includegraphics[width=0.75\textwidth]{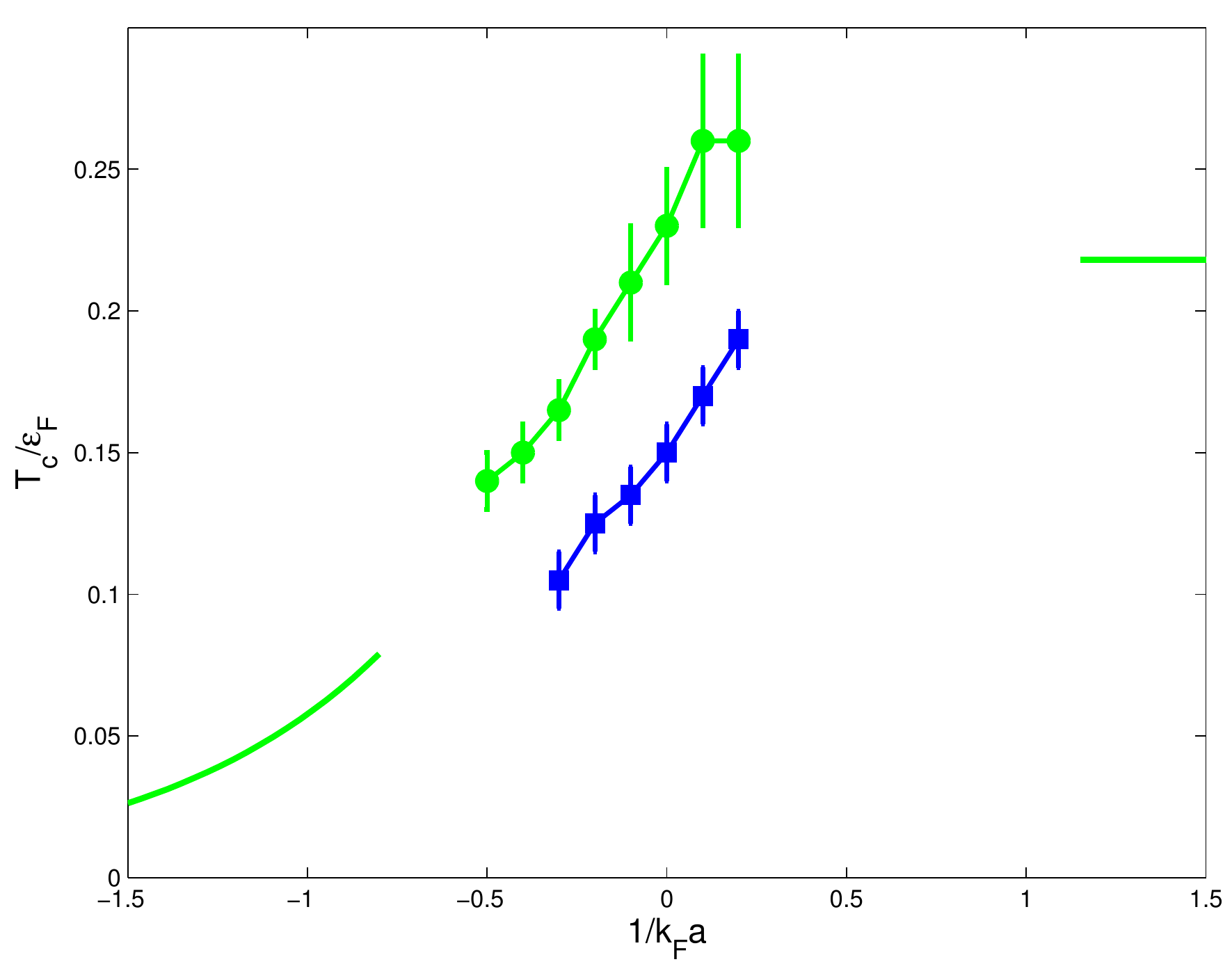}
    \caption{ \label{fig:crittemp}
      The critical temperature $T_c$ (squares
      either error bars) and the characteristic temperature $T_0$
      (circles with error bars) around the unitary point determined in
      \QMC\ and using finite size analysis. On the far left \BCS\ side
      of the critical point we show (solid green line) the expected
      \BCS\ critical temperature, including the corrections due to
      induced interactions \cite{GM-B:1961, HPSV:2000}, and on the far
      right side of the \BEC\ side of the unitary point we show (solid
      green line) the expect critical temperature in the \BEC\ limit.
      For more details see~\cite{BDM:2008}.}
  \end{center}
\end{figure}
  
In the following we present a brief summary of our results near
unitarity on both the \BCS\ $a<0$ and \BEC\ $a>0$ sides, (see
Fig.~\ref{fig:crittemp} and Table~\ref{table:results}).  The coupling
strength was varied in the range $-0.5 \leq 1/k_F a \leq 0.2$ (where
$k_F = (3 \pi^2 n)^{1/3}$), limited on the negative (\BCS) side by the
finite volume $V$ (which becomes comparable to the size of the Cooper
pairs), and on the positive (\BEC) side by the finite lattice spacing
$l$ (which becomes comparable to the size $a=\order(l)$ of the
localized dimers, manifesting as poor convergence of observables).

\begin{table}[ht]
  \begin{center}
    \begin{tabular}{rcclcccc}
      \toprule
      $1/k_Fa$   & $E(0)/E_F$    & $T_0$    &$\mu_0/\varepsilon_F$ & $E_0/E_F$ & $T_c < $ & $\mu_c/\varepsilon_F$ & $E_c/E_F$ \\
      \midrule
      -0.5 & 0.60(4)  &   0.14(1)     & 0.685(5)     & 0.77(2)    & --         & --       & --      \\
      -0.4 & 0.59(4)  &   0.15(1)     & 0.65(1)      & 0.75(1)    & --         & --       & --      \\
      -0.3 & 0.55(4)  &   0.165(10)   & 0.615(10)    & 0.735(10)  & 0.105(10)  & 0.61(1)  & 0.64(2) \\
      -0.2 & 0.51(4)  &   0.19(1)     & 0.565(10)    & 0.725(10)  & 0.125(10)  & 0.56(1)  & 0.61(2) \\
      -0.1 & 0.42(4)  &   0.21(2)     & 0.51(1)      & 0.71(2)    & 0.135(10)  & 0.50(1)  & 0.54(2) \\
      0    & 0.37(5)  &   0.23(2)     & 0.42(2)      & 0.68(5)    & 0.15(1)    & 0.43(1)  & 0.45(1) \\
      0.1  & 0.24(8)  &   0.26(3)     & 0.34(1)      & 0.56(8)    & 0.17(1)    & 0.35(1)  & 0.41(1) \\
      0.2  & 0.06(8)  &   0.26(3)     & 0.22(1)      & 0.39(8)    & 0.19(1)    & 0.21(1)  & 0.25(1) \\
      \bottomrule
    \end{tabular}
  \end{center}
  \caption{
    \label{table:results}
    Results for the ground state energy,
    the characteristic temperature $T_0$, and the corresponding
    chemical potential and energy, from the caloric curves  $E(T)$
    and the upper bounds on the critical
    temperature $T_c$ from finite size scaling and the corresponding
    chemical potentials and energies ~\cite{BDM:2008}.}
\end{table}

\subsection{Response to External Probes and the Spectral Function}

In order to get an insight into basic degrees of freedom which
contribute to the low energy excitations of the system one has to
investigate the response of the system to various external probes.
Here we will present the simplest possible probe: adding a particle to
the system and calculating the probability amplitude of finding it in
a given single particle state.  This requires calculating the one-body
finite temperature (Matsubara) Green's function~\cite{Fetter:1971fk}:
\begin{equation}
  \mathcal{G}(\vec{p},\tau)=
  \frac{1}{Z} \tr \{\exp[-(\beta-\tau) (H-\mu N)]\psi^\dagger(\vec{p})
  \exp[-\tau(H-\mu N)\psi(\vec{p})] \}, \label{eq:Gp}
\end{equation}
where $\beta = 1/T$ is the inverse temperature and $\tau >0$.  The
trace is performed over the Fock space, and $Z=\tr \{\exp[-\beta
(H-\mu N)]\}$. The spectral weight function $A(\vec{p},\omega)$ can be
extracted from the finite temperature Green's function using the
relation:
\begin{equation}
  \mathcal{G }(\vec{p},\tau)=-\frac{1}{2\pi}\int_{-\infty}^{\infty}
  \D\omega A(\vec{p},\omega)\frac{\exp(-\omega\tau)}{1+\exp(-\omega\beta)}.
  \label{eq:Ap}
\end{equation}
By definition, $A(\vec{p},\omega)$ fulfills the following constraints:
\begin{equation}
  A(\vec{p},\omega) \ge 0, \quad \quad
  \int_{-\infty}^{\infty}\frac{\D\omega}{2\pi} A(\vec{p},\omega) = 1
  \label{eq:Ap_con}.
\end{equation}
Since our study focuses on the symmetric (unpolarized) system and the
Hamiltonian is symmetric under $a \leftrightarrow b$, $\mathcal{G
}(\vec{p},\tau)$ is block diagonal and the species index is suppressed
in all formulae.  The numerical evaluation of the one-body temperature
propagator (\ref{eq:Gp}) is performed as described above, using a
Trotter expansion of $\exp[-\tau(H-\mu N)]$ followed by a
Hubbard-Stratonovich transformation and Metropolis importance
sampling.  Details can be found in~\cite{Magierski:2009}.

The numerical determination of $A(\vec{p},\omega)$ by inverting
(\ref{eq:Ap}) is an ill-posed problem that requires special
methods. We have used two, based on completely different
approaches. The first approach is the maximum entropy
method~\cite{Jaynes:1979, Silver:1990, Silver:1990a, White:1991},
which is based on Bayes' theorem.  Quantum Monte Carlo calculations
provide us with a discrete set of values
$\tilde{\mathcal{G}}(\vec{p},\tau_{i})$, where
$i=1,2,\dotsc,\mathcal{N}_{\tau}=50$. We treat them as normally
distributed random numbers around the true values
$\mathcal{G}(\vec{p},\tau_{i})$.  The Bayesian strategy consists in
maximizing the \emph{posterior probability}
\begin{equation}
  P(A|\tilde{G}) \propto P(\tilde{G}|A)P(A)
\end{equation}
of finding the right $A(\vec{p},\omega)$ under the condition that
$\mathcal{\tilde{G}}(\vec{p},\tau_{i})$ are known. Here,
\begin{equation}
  P(\tilde{G}|A)\propto\exp\left (-\frac{1}{2}\chi^{2}\right )
\end{equation}
is the \emph{likelihood function}, where
\begin{equation}
  \chi^{2}=\sum_{i=1}^{\mathcal{N}_{\tau}}
  \left [\mathcal{\tilde{G}}(\vec{p},\tau_{i}) - \mathcal{G}(\vec{p},\tau_{i})
  \right]^{2}/ \sigma^{2}.
\end{equation}
The quantity $\mathcal{G}(\vec{p},\tau_{i})$ is determined by the spectral
weight function in the discretized form of~(\ref{eq:Ap}) at
frequencies $\omega_k$.  The prior probability $P(A)$, describing our
ignorance about the spectral weight function, is defined as
$P(A)\propto\exp(\alpha S(\mathcal{M}))$, where $\alpha>0$ and $S(\mathcal{M})$ is 
the relative information entropy with respect to the assumed
model $\mathcal{M}$:
\begin{equation}
  S(\mathcal{M})= -\sum_{k}\Delta\omega \biggl [ A(\vec{p},\omega_{k})
  -
  \mathcal{M}(\omega_{k}) - A(\vec{p},\omega_{k})\ln
  \left ( \frac{A(\vec{p},\omega_{k})}{\mathcal{M}(\omega_{k})} \right ) \biggr ]
  . \label{eq:method}
\end{equation}
Hence the maximization of $P(A|\tilde{G})$ leads in practice to the
minimization of the quantity $\frac{1}{2}\chi^{2} - \alpha S(\mathcal{M})$ 
with respect to $A$~\cite{Magierski:2009}.

The second approach is based on the \SVD\ of the integral kernel
$\mathcal{K}$ of~(\ref{eq:Ap}), which can be rewritten in operator
form as
\begin{equation}
  \mathcal{G}(\vec{p},\tau_{i})=(\mathcal{K}{A})(\vec{p},\tau_{i}).
\end{equation}
The operator $\mathcal{K}$ possesses a singular subspace
\begin{equation}
  \mathcal{K}{u}_{i}=\lambda_{i}\vec{v}_{i}, \quad
  \mathcal{K}^{*}\vec{v}_{i}=\lambda_{i}{u}_{i},
\end{equation}
where $\mathcal{K}^{*}$ denotes the adjoint of $\mathcal{K}$,
$\lambda_{i}$ are the singular values, and ${u}_{i}$ and $\vec{v}_{i}$
are right-singular functions and left-singular vectors
respectively. The singular subspace forms a suitable basis for the
expansion of the spectral weight function~\cite{Bertero:1985,
  Bertero:1988, Creffield:1995}, which we can then write as
\begin{align}
  {A}(\vec{p},\omega)&=\sum_{i=1}^{r}b_{i}(\vec{p}){u}_{i}(\omega),
  \label{A:exp}  &
  b_{i}(\vec{p})&=\frac{1}{\lambda_{i}}( \vec{\mathcal{G}}(\vec{p})\cdot \vec{v}_{i}
  ),
\end{align}
where $(\_\cdot \_)$ is a scalar product and $r$ is the rank of the
operator $\mathcal{K}\mathcal{K}^{*}$. Since $\mathcal{G}(\vec{p},\tau_{i})$ is
affected by Monte Carlo errors $\sigma_i$, the coefficients $b_{i}$ carry
some uncertainty $\Delta b_{i}$. Each set of expansion coefficients
\mbox{$\tilde{b}_{i}\in (b_{i}-\Delta b_{i},b_{i}+\Delta b_{i})$}
reproduces $\mathcal{G}(\vec{p},\tau_{i})$ within its error bars. We use
this flexibility of choosing the expansion coefficients to produce a
solution satisfying constraints~(\ref{eq:Ap_con})~\cite{Villiers:1999}.
The relative advantages of each method will be discussed
elsewhere~\cite{MagierskyWlazlowski:inprep}.

\begin{figure}[p]
  \begin{minipage}{8cm}
    \includegraphics[width=8cm]{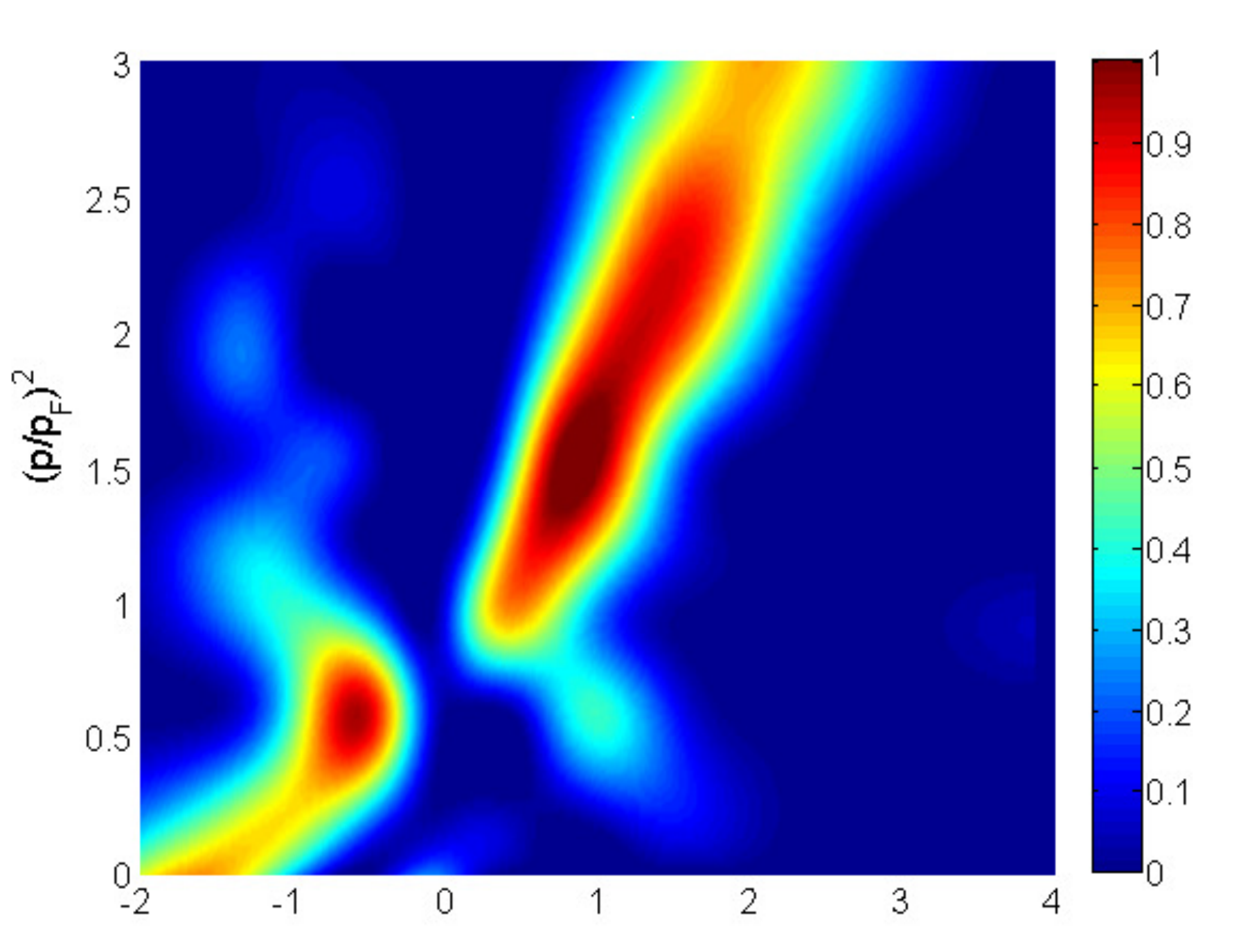}\\
    \includegraphics[width=8cm]{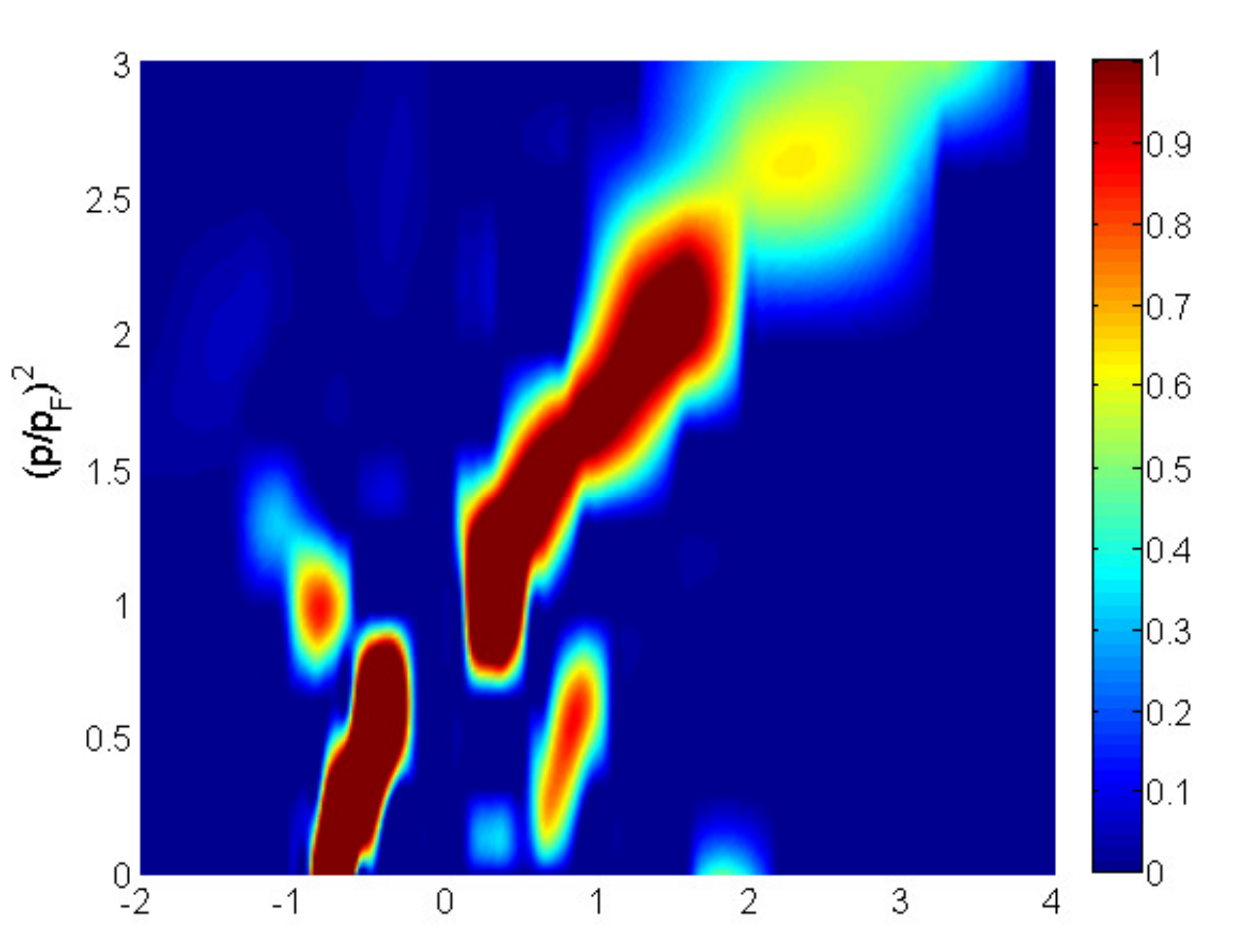}\\
    \includegraphics[width=8cm]{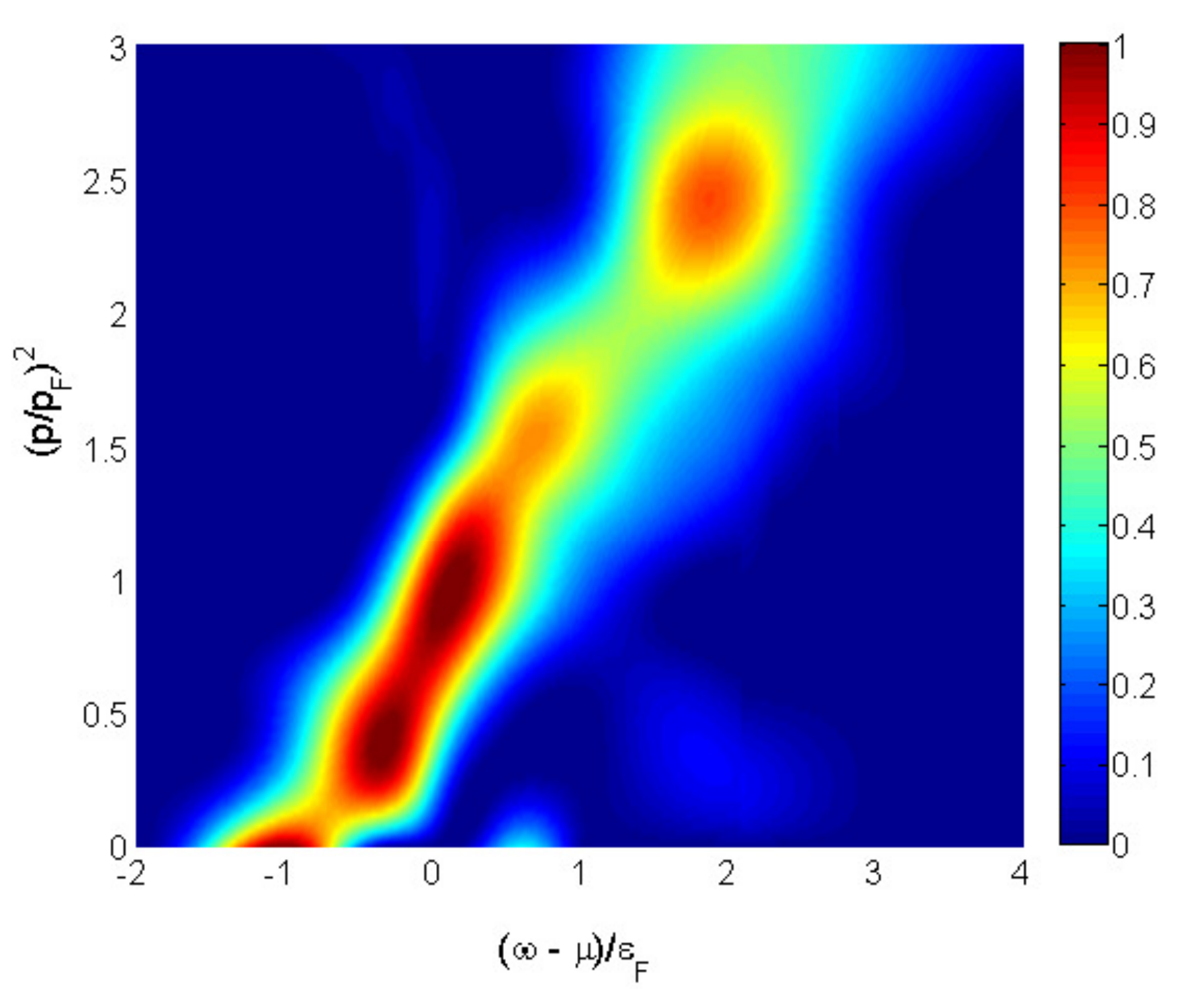}
  \end{minipage}
  \sidecaption
  \caption{
    \label{fig:A} 
    Spectral weight function $A(\vec{p},\omega)$ for three
    temperatures: $T=0.15\varepsilon_F\approx T_{c}$ (upper panel),
    $T=0.18\varepsilon_F\approx T_{c}$ (middle panel) and
    $T=0.20\varepsilon_{F}$ (lower panel).  The presence of a gap in
    clearly seen in the upper two panels. From~\cite{Magierski:2009}.
  }
\end{figure}

\begin{figure}[t]
  \begin{center}
    \includegraphics[width=0.52\textwidth]{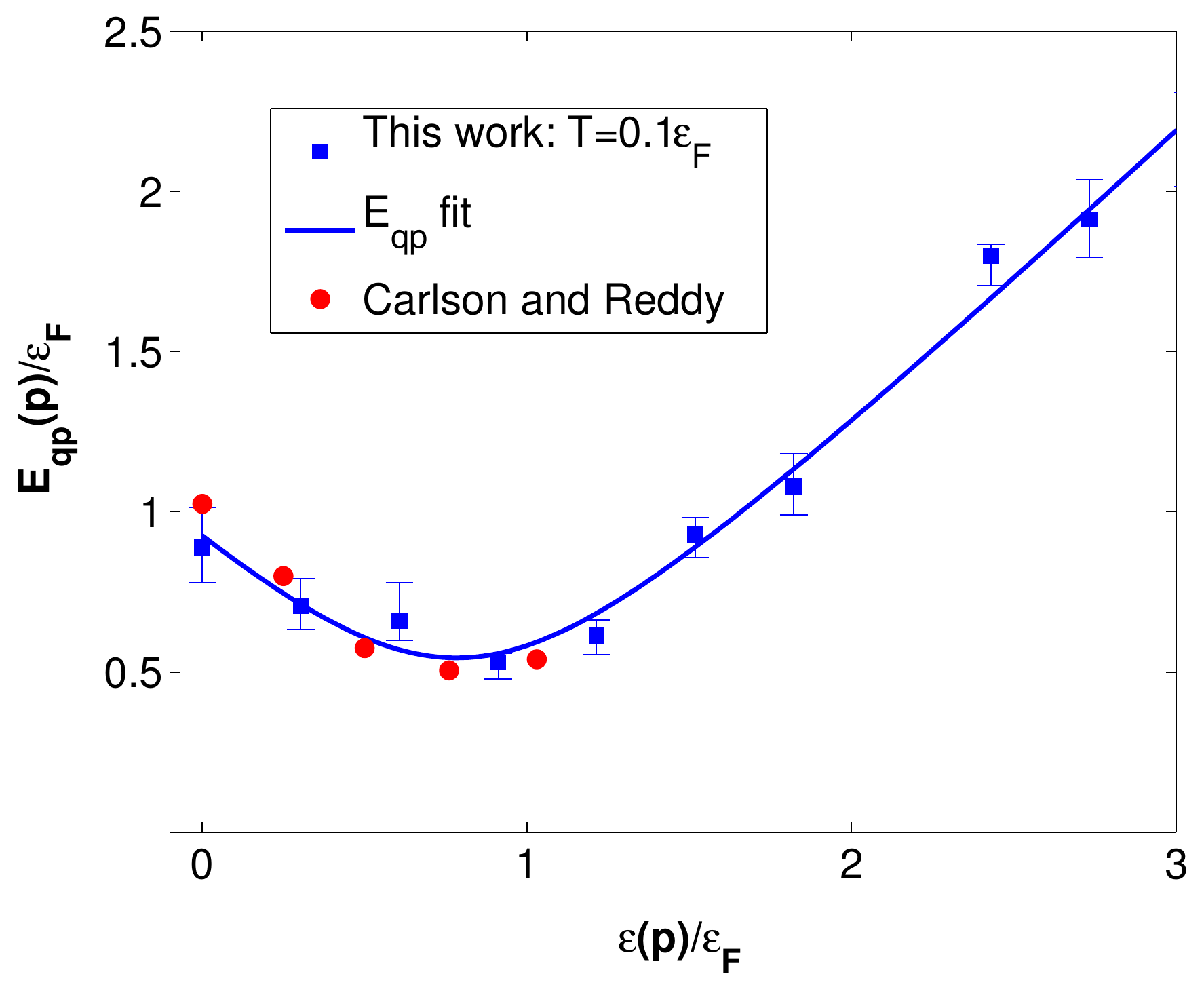}%
    \includegraphics[width=0.48\textwidth]{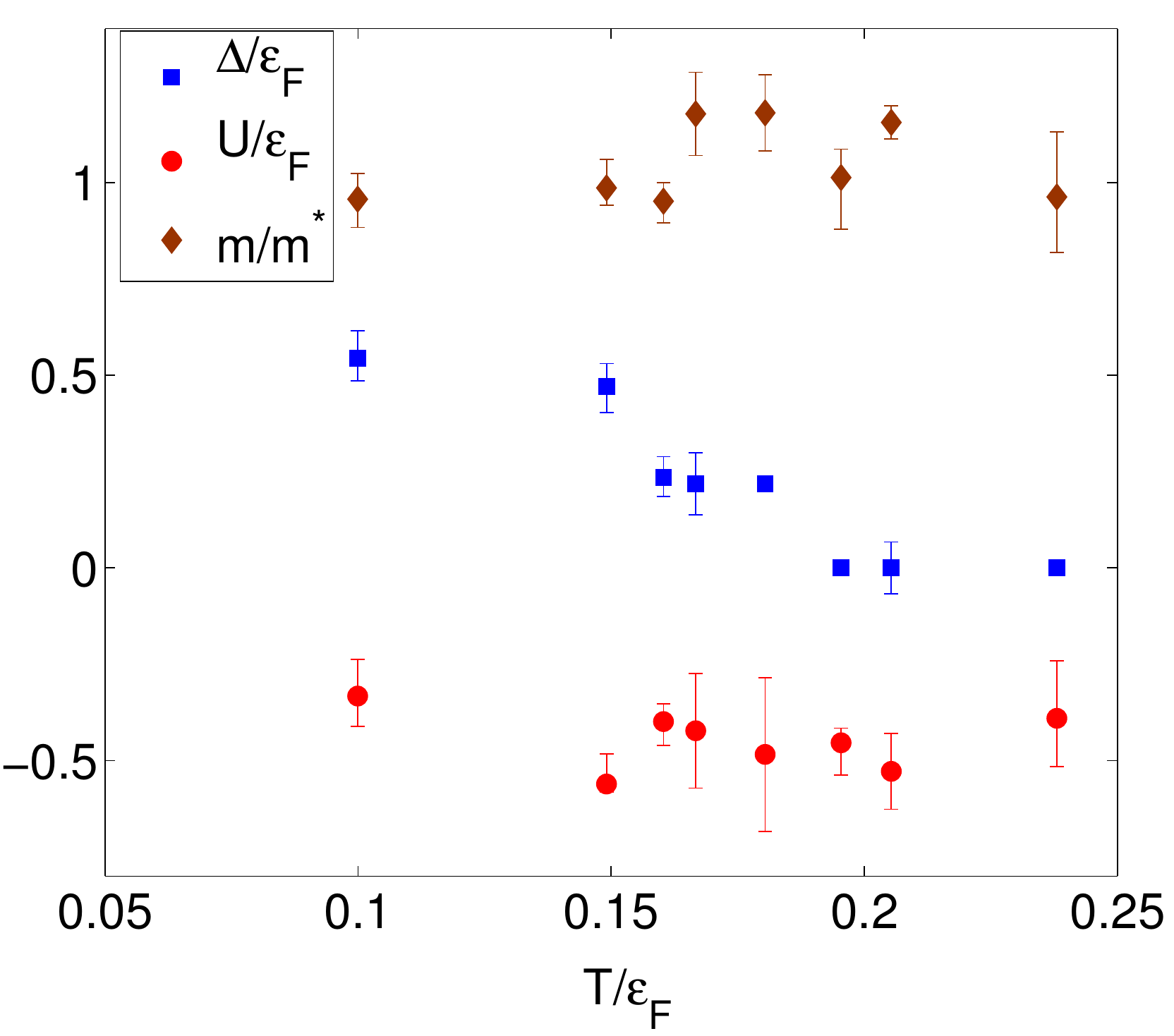}
    \caption{
      \label{fig:eqp_pmu}
      Quantities extracted from the spectral weight function
      $A(\vec{p},\omega)$ at $T=0.1\varepsilon_{F}$ at unitarity
      (from~\cite{Magierski:2009}).  Left: Quasiparticle energies
      $E(\vec{p})$ (squares).  The line corresponds to the fit to
      (\ref{eq:ek}). The circles are the results of Carlson and
      Reddy~\cite{Carlson:2005kg}. (See also Fig.~\ref{fig:qp} where
      the same data is used to fit the \SLDA\ density functional.)
      Right: The single-particle parameters.  One should note that
      while the effective mans and the self-energy show a very weak
      temperature dependence across the phase transition, the pairing
      gap halves in value at $T_c$ and vanishes around $T_0$.}
  \end{center}
\end{figure}

A sample of calculated spectral weight functions at unitarity are
shown in Fig.~\ref{fig:A}. In order to characterize the quasiparticle
excitation spectrum we have associated with the maximum of
$A(\vec{p},\omega)$ the quasiparticle energy $E(\vec{p})$:
\begin{equation}
  E(\vec{p}) = \pm \sqrt{ \left ( \frac{p^2}{2m^*}+U-\mu \right )^2+\Delta^2},
  \label{eq:ek}
\end{equation}
where the effective mass $m^{*}$, the effective potential $U$, and the
``pairing'' gap $\Delta$ depend on temperature, and $\mu$ is an input
parameter. In Fig.~\ref{fig:eqp_pmu} we compare the spectrum of elementary
fermionic excitations evaluated in~\cite{Carlson:2005kg}, with the one
extracted by us from our lowest temperature spectral weight function.

\subsection{The Pairing Gap, Pseudogap, and Critical Temperature}
\label{sec:pair-gap-pseud}

In order to find the critical temperature for the superfluid-normal
transition one has to perform the finite size analysis discussed in
the previous section.  Following this procedure, our data for the
condensate fraction of the unitary Fermi gas indicates that $T_c
\lesssim 0.15(1) \varepsilon_F$, considerably lower than the
characteristic temperature $T_0 = 0.23(2)$ found by studying the
behavior of the energy and the chemical potential (see
Fig.~\ref{fig:crittemp} and Table~\ref{table:results}).  Even though
this result for $T_c$ is close to estimates by other groups (see
e.g.~\cite{Burovski:2006, 
  BPST:2006, BKPST:2008}),
it should be pointed out that the experimental data of~\cite{Luo:2007}
shows a distinctive feature in the energy versus entropy curve at a
temperature close to $T_0$ (see~\cite{BDM:2007}).

It is notable that both methods (the maximum entropy method and the
\SVD\ method) admit a ``gapped'' spectral function above the critical
temperature $T_c$: a situation commonly called a pseudogap.  It
characterizes the range of temperatures where the system exists in an
exotic state which is neither normal, nor superfluid, and defies a
conventional \BCS\ description.  Therefore the onset of pairing and
superfluidity can occur at different temperatures.  On the other hand,
the pseudogap is easy to understand in the \BEC\ limit where stable
dimers exist well above the critical temperature.  This gives rise to
a pseudogap phase, where the system share a \BCS-like dispersion and a
partially gapped density of states, but does not exhibit
superfluidity.  Several groups have been advocating various aspects of
pseudogap physics in the unitary Fermi gas for the past few
years~\cite{Sa-de-Melo:1993, Perali:2004, Stajic:2004, Chen:2005,
  Levin:2006, He:2007}.

There have been several experimental attempts to extract the pairing
gap in ultra-cold dilute Fermi gases~\cite{Chin:2004,
  Greiner:2005, Schunck:2007} and a theoretical explanation of these
spectra was given in~\cite{Kinnunen:2004, He:2005}. It was later shown
in~\cite{Yu:2006, Baym:2007, Punk:2007, Perali:2008} that these
initial interpretations of the rf-spectra ignored the strong final
state interaction effects.  Recent experimental measurement of pair
condensation in momentum space and a measurement of the
single-particle spectral function using an analog to photo-emission
spectroscopy, directly probed the pseudogap phase and revealed its
existence for $1/(k_{F}a)\approx 0.15$~\cite{Gaebler:2010}.  Although
this lies on the \BEC\ side, there are indications that the pseudogap
persists well into the unitary regime~\cite{Stewart:2008,
  Kuhnle:2009}.

Our calculations show that the spectral function reveals the presence
of a gap in the spectrum up to about $T^*\approx 0.20\varepsilon_{F}$
(see Fig.~\ref{fig:eqp_pmu}), and a two peak structure around the
Fermi level at temperatures above
$T_{c}$~\cite{Magierski:2009,Magierski:2010}.  We note that $T^*$ is
close to $T_0$ (not surprising in hindsight), the temperature at which
the caloric curve $E(T)$ has a shoulder~\cite{Bulgac:2006,
  Bulgac:2006a} (called $T_0$ in~\cite{BDM:2008}).

\subsection{Describing Trapped Systems with Quantum Monte Carlo Results}

The Monte Carlo calculations presented above assume that the system is
uniform. In experiment, however, this condition is not fulfilled since
atoms are trapped in an external potential which induces inhomogeneity
of density distribution.  Most of the atomic trapping potentials used
in these experiments can be approximated rather well with harmonic
potential wells. Such potentials can be shown to satisfy the virial
theorem at unitarity, namely $E(T,N)=2 N\langle U\rangle =
3m\omega_z^2\langle z^2\rangle$~\cite{Thomas:2005}, and therefore simply
measuring the spatial shape of the cloud allows for a unique
determination of the unitary gas energy at any temperature. One of the
main goals is therefore to provide a link between the results of
experiment~\cite{Luo:2007} and the available finite temperature \QMC\
calculations.

At unitarity $(1/k_Fa=0)$ the pressure of a homogeneous unitary gas is
determined by a universal convex function $h_T(z)$:
\begin{align}
  \mathcal{P}(T, \mu )
  &= \frac{2}{5}\beta\left [ T h_T\left ( \frac{\mu}{T}\right )\right
  ]^{5/2}\!, &
  \beta &= \frac{1}{6\pi^2}\left(\frac{2m}{\hbar^2}\right)^{3/2},
\end{align}
where $T$ and $\mu$ are the temperature and the chemical potential,
respectively. $\mathcal{P}(T, \mu )$ is a convex function of its
arguments (second law of thermodynamics) if and only if $h_T(z)$ is
convex.  One can show~\cite{Bulgac:2006cv} that thermodynamic
stability implies positivity $h_T(z)\geq 0$, monotonicity $h_T'(z)\geq
0$, and convexity $h_T''(z)\geq 0$.  Remembering that the grand
canonical potential is $\Omega(V,T,\mu) = -V\mathcal{P}(T, \mu )$ one
can show that the energy of the system reads: $E=3\mathcal{P}V/2$,
where $V$ is the volume of the system. As it was mentioned before,
this relation between energy and pressure is identical in form to the
one corresponding to non-interacting particles.  In the
high-temperature limit $\mu \rightarrow -\infty$ and $\mathcal{P}(T,
\mu )$ tends from above to the free Fermi gas pressure. In the
low-temperature limit $\mathcal{P}(T, \mu )$ tends from above to
$\mathcal{P}(0, \xi \varepsilon_{F} )=4\beta
\varepsilon_F^{5/2}\xi/5$.

Standard manipulations show that all the thermodynamic potentials for
the unitary Fermi gas can be expressed in terms of a single function
of one variable, a property known as universality \cite{Bulgac:2006,
  PhysRevLett.92.090402, Bulgac:2006a, Burovski:2006, 
  BPST:2006}. This property was incorporated in our interpolation. At
high temperatures we notice that our results smoothly approach the
corresponding free Fermi gas results with some offsets for the energy,
chemical potential and entropy~\cite{Bulgac:2006, Bulgac:2006a}.

At this point we assume that the Local Density Approximation (\LDA)
can be used to describe the properties of an atomic cloud in a trap.
We will neglect the gradient corrections as one can show that for the
mostly-harmonic traps used in typical experiments the role of the
gradient corrections is relatively small~\cite{BDM:2007}, as the
average interparticle distance, and thus the Fermi wave length, is
much smaller than the harmonic oscillator length.

In this approach, the grand canonical thermodynamic potential for a
unitary Fermi gas confined by an external potential $U(\vec{r})$ is a
functional of the local density $n(\vec{r})$ given by
\begin{equation}
  \Omega = \int \mathrm{d}^3{\vec{r}}\left [
    \frac{3}{5}\varepsilon_F(\vec{r})\varphi (x)n(\vec{r}) + U(\vec{r})
    n(\vec{r})-\lambda n(\vec{r}) \right ],
\end{equation}
where
\begin{equation}
  \label{eq:XandEF}
  x = \frac{T}{\varepsilon_F(\vec{r})}, \quad
  \varepsilon_F(\vec{r}) = \frac{\hbar^2}{2m} [3 \pi^2  n(\vec{r})]^{2/3},
\end{equation}
and we have used the universal form for the free energy per particle
$F/N$ in the unitary regime:
\begin{equation}
  \frac{F}{N}=\frac{E-TS}{N} = \frac{3}{5}\varepsilon_F\varphi(x) =
  \frac{3}{5}\varepsilon_F [\xi(x)-x\sigma(x)],
\end{equation}
where for a homogeneous system $\xi (x)= 5E/3\varepsilon_FN$,
$\sigma(x)=S/N$ is the entropy per particle and $x = T/\varepsilon_F$ .  The
overall chemical potential $\lambda$ and the temperature $T$ are
constant throughout the system. The density profile will depend on the
shape of the trap as dictated by $\delta \Omega / \delta n(\vec{r}) =
0$, which results in:
\begin{equation}
  \label{eq:chempot}
  \frac{\delta \Omega}{\delta n(\vec{r})}=
  \frac{\delta(F-\lambda N)}{\delta n(\vec{r})}
  =\mu(x(\vec{r})) + U(\vec{r}) - \lambda = 0.
\end{equation}
At a given $T$ and $\lambda$, (\ref{eq:XandEF}) and (\ref{eq:chempot})
completely determine the density profile $n(\vec{r})$ (and
consequently both $E(T,N)$ and $S(T,N)$) in a given trap for a given
total particle number.  The only experimental input we have used is
the particle number, the trapping potential and the scattering length
at $B=1200$~G, taken from~\cite{Luo:2007}. The potential was assumed
to be an `isotropic' Gaussian, as suggested by the experimental group
\cite{Luo:2007}, although it is not entirely clear to us to what
extent this is accurate, especially in the axial direction. We have
approximated the properties of the atomic cloud at $B=840$~G with
those at unitarity ($B=834$~G), where we have \MC\ data. For $B=840$~G
and for the parameters of the Duke experiment \cite{Luo:2007} one
obtains $1/k_Fa = -0.06$, using data of~\cite{Bartenstein:2005uq}, if
the Fermi momentum corresponds to the central density of the cloud at
$T=0$.

\begin{figure}[p]
  \begin{center}
    \includegraphics[width=0.8\textwidth]{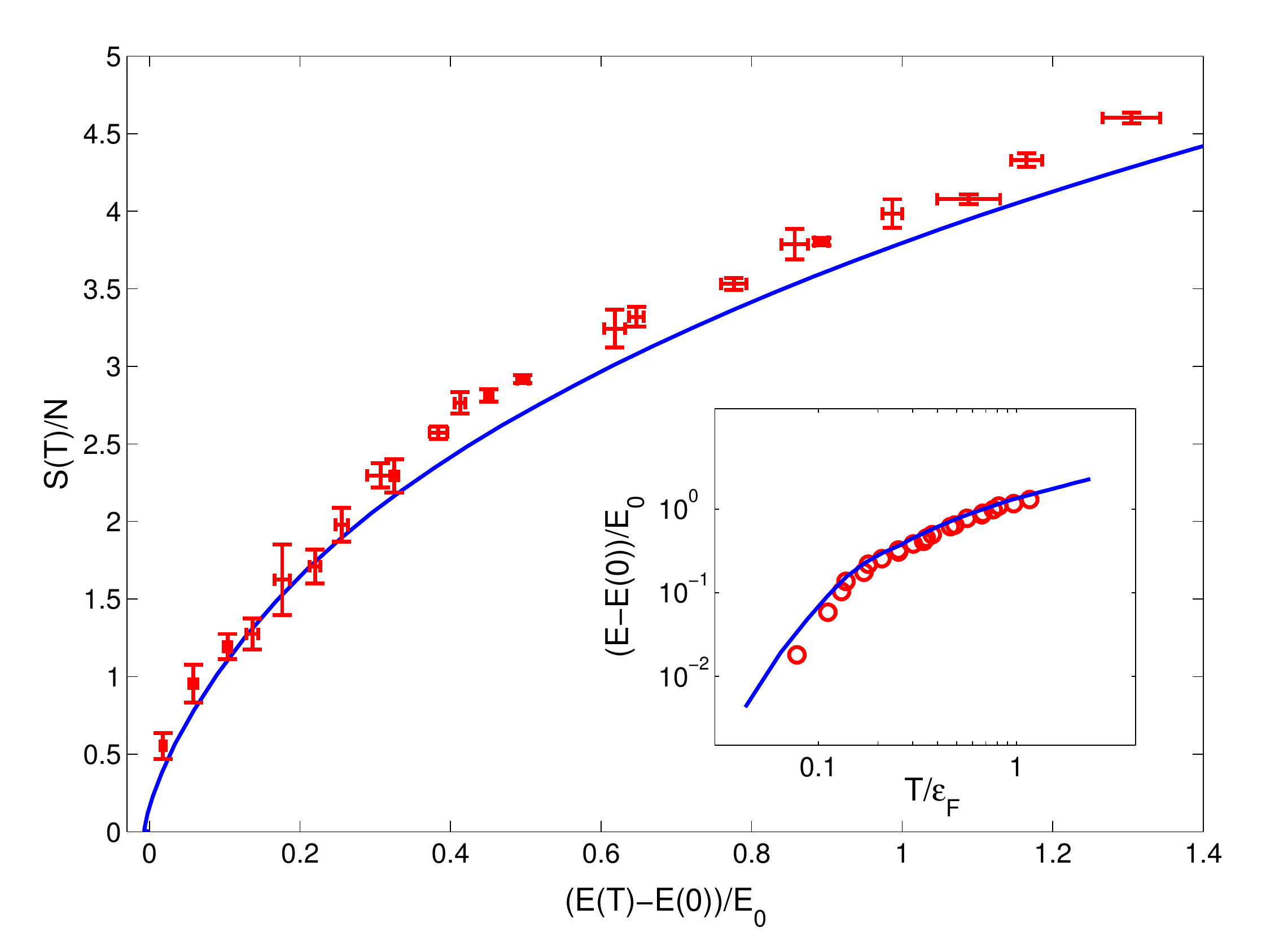}
    \caption{
      \label{fig:SofE}
      Entropy as a function of energy for the unitary Fermi gas in the
      Duke trap~\cite{Luo:2007}: experiment (points with error bars)
      and present work (solid curve), where
      $E_0=N\varepsilon_F^{HO}$. Inset: log-log plot of $E(T)$ as
      results from our calculations and as derived from experimental
      data~\cite{Luo:2007}. The temperature is units of the
      corresponding Fermi energy at the center of the trap:
      $\varepsilon_F(0)$. From~\cite{BDM:2007}.}
  \end{center}
\end{figure}
\begin{figure}[p]
  \begin{center}
    \includegraphics[width=0.9\textwidth]{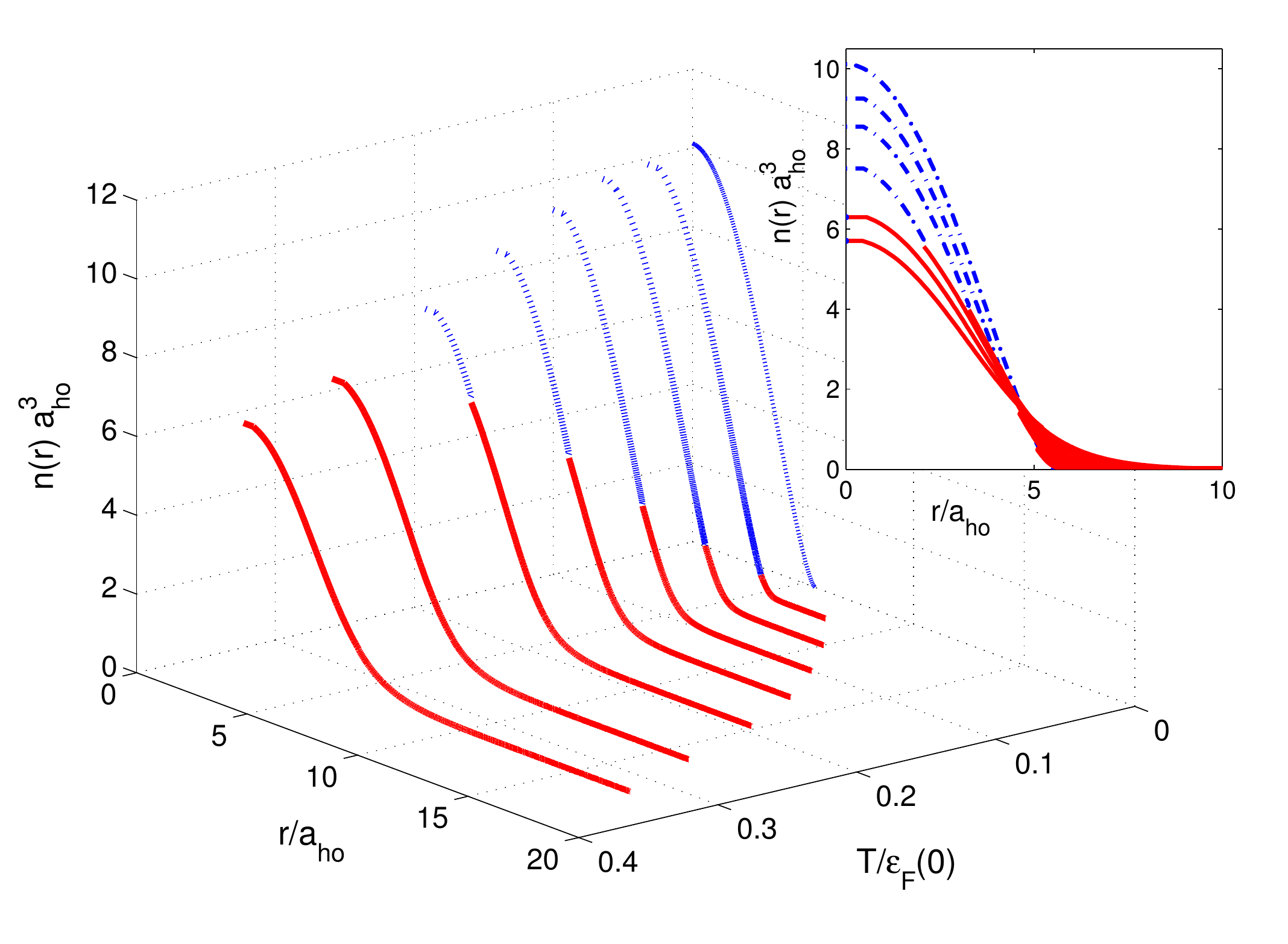}
    \caption{
      \label{fig:profiles}
      The radial (along shortest axis) density profiles of the Duke
      cloud at various temperatures, as determined theoretically using
      the \QMC\ results~\cite{Bulgac:2006, Bulgac:2006a}. The dotted
      blue line shows the superfluid part of the cloud, for which
      $x(\vec{r})=T/\varepsilon_F(\vec{r})\leq 0.23$. The solid red
      line shows the part of the system that is locally normal. Here
      $a_{ho}^2=\hbar /m\omega_{max}$. From~\cite{BDM:2007}.}
  \end{center}
\end{figure}

Our results for the entropy of the cloud and the density profiles for
several temperatures, are shown in Figs.~\ref{fig:SofE} and
\ref{fig:profiles}. In all the figures the temperature is measured in
natural units of $\varepsilon_F(0)$, corresponding to the actual
central density of the cloud at that specific
temperature. In~\cite{Luo:2007,KTTCSL:2005} the temperature is
expressed in units of the Fermi energy at $T=0$ in a harmonic trap:
$\varepsilon_F^{ho}=\hbar\Omega (3N)^{1/3}$. It is clear from
Fig.~\ref{fig:profiles} that the central density decreases with $T$
and that the superfluid core disappears at $T_c =
0.23(2)\varepsilon_F(0)$, which translates into $T_c =
0.27(3)\varepsilon_F^{ho}$ to be compared to $T_c =
0.29(2)\varepsilon_F^{ho}$ of~\cite{Luo:2007}. There is a noticeable
systematic difference between theory and experiment at high energies,
see Fig.~\ref{fig:SofE}. This discrepancy can be attributed to the
fact that the experiment was performed slightly off resonance, on the
\BCS\ side, where $1/k_Fa=-0.06$.

\begin{figure}[t]
  \begin{center}
    \includegraphics[width=\textwidth]{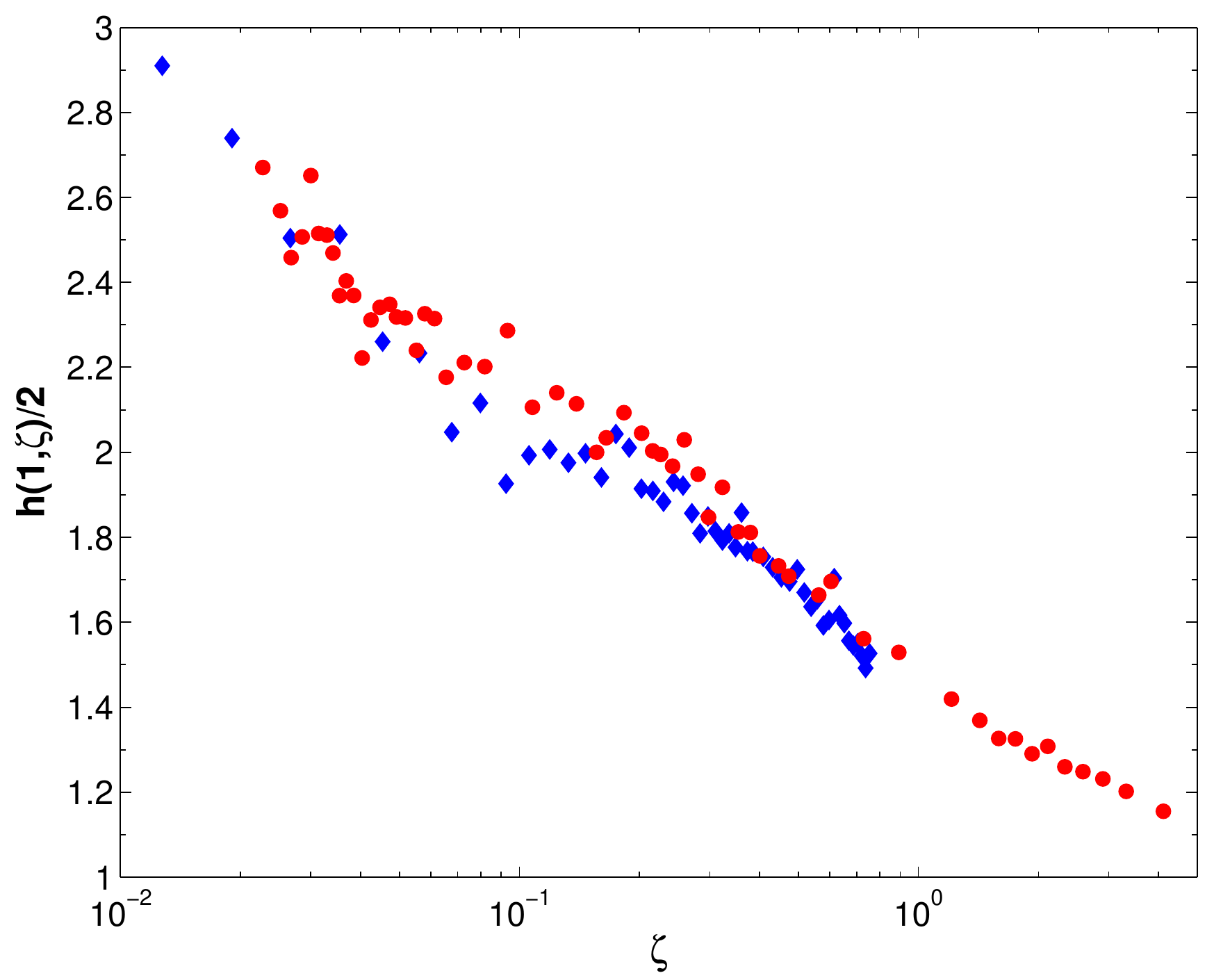}
    \caption{
      \label{fig:nascim}
      The comparison between the ration of the pressure versus
      fugacity of a unitary Fermi gas and the pressure of a free Fermi
      gas: measured in~\cite{Nascimbene:2010} (red filled circles) and
      calculated in \QMC\ in~\cite{BDM:2007} (blue filled diamonds). }
  \end{center}
\end{figure}

Recently a couple of new experiments have been published, one by the
Paris group~\cite{Sylvain-Nascimbene:2009lr} and another by the Tokyo
group~\cite{Horikoshi:2010}. Using new techniques these groups were
able to extract directly from cloud images the pressure as a function
of the fugacity.  While the Paris group has observed a very good
agreement with our \QMC\ results, see Fig.~\ref{fig:nascim}, they have
also noticed that the results of the Tokyo group show systematic
differences~\cite{Nascimbene:2010}.

One can summarize that so far the bulk of the theoretical predictions
obtained in ab initio \QMC\ have been confirmed experimentally with
impressive accuracy in most cases, often at a level of a few percent,
which is the accuracy of both theoretical calculations and of many
experimental results as well. The emergence of a pseudogap in the
unitary gas is a fascinating new feature, but still in its infancy both
theoretically and experimentally.


\renewcommand{\op}[1]{#1}

\section{Density Functional Theory for the Unitary Fermi Gas}
\label{cha:dens-funct-theory}

The idea of Density Functional Theory (\DFT) originated with Hohenberg
and Kohn~\cite{HK:1964} and Kohn and Sham~\cite{Kohn:1965fk} (see the
monographs~\cite{Dreizler:1990lr,Parr:1989uq} for an overview) where
they proved that the ground state energy and the density of a system
of interacting fermions in an \emph{arbitrary} external potential
$V_{\text{ext}}(\vec{r})$ may be found by minimizing a functional
\begin{equation}
  E[n(\vec{r})] + \int \D^{3}\vec{r}\;V_{\text{ext}}(\vec{r})n(\vec{r}).
\end{equation}
The utility of this approach is that the functional $E[n(\vec{r})]$
depends only on the interactions of the system and is independent of
the external potential.  Thus, if we were able to deduce
$E[n(\vec{r})]$ for the unitary Fermi gas, then by simply minimizing a
single functional, we could determine the ground state in any external
potential, including arbitrary trapping geometries and optical
lattices.

The challenge is that the Hohenberg-Kohn theorem is an existence
theorem.  The exact form of the functional $E[n(\vec{r})]$ is unknown,
and in general it may be extremely complicated and highly non-local.
In problems that are under perturbative control, the functional can be
formally derived (see~\cite{Drut:2010kx}), but in highly
non-perturbative problems such as the unitary gas, one must choose a
physically motivated approximate functional and check its accuracy.

Our strategy is thus:
\begin{enumerate}
\item Postulate simple functional forms capturing the relevant physics
  with a small number of parameters.
\item Use ab initio results to fix these parameters.
\item Validate the functional with different ab initio and
  experimental results.
\item Make interesting and verifiable physical predictions.
\end{enumerate}
The computational cost of minimizing the density functional is much
less than solving for many-body wavefunctions, and one may consider
substantially larger systems, untenable with ab initio methods.
This allows one to make direct contact with typical mesoscopic
experiments for example.  In this way, one may view the density
functional as a bridge between microscopic and mesoscopic physics.

As we have noted, although \DFT\ is exact in principle, for
non-trivial systems we must postulate a form for the functional.
Nevertheless, it provides a substantial improvement to the ad hoc
mean-field methods typically employed to study the properties of large
non-perturbative many-body systems.  Without a program for
systematically correcting the functional, the \DFT\ approach will not
be the final word.  However, judging from the success of the approach
in quantum chemistry, and from the results presented here, we expect
that without too much effort one should be able to obtain percent
level accuracy for a wide range of systems, which should be sufficient
for quite some time.

The qualitative success of the Eagles-Leggett~\cite{Eagles:1969-10,
  Leggett:1980uq} mean-field model describing the \BCS--\BEC\
crossover suggests a functional description of the unitary Fermi gas
in terms of quasi-particle fermionic states (see \eqref{eq:DF_BdG}).
As discussed in Sec.~\ref{cha:introduction}, although the \BdG\
approximation is quite successful, it is quantitatively inaccurate as
it describes \emph{all} interaction effects through the condensation
energy (pairing) alone, completely omitting the ``Hartree-Fock''
contribution which dominates the energetics.  To see this, consider
the typical local interaction
$g\op{a}^{\dagger}\op{b}^{\dagger}\op{b}\op{a}$ between species $a$
(spin up) and species $b$ (spin down).  The mean-field approximation
retains the pairing term
$g\braket{\op{a}^{\dagger}\op{b}^{\dagger}}\braket{\op{b}\op{a}} =
g\nu^{\dagger}\nu$ and the Hartree term
$g\braket{\op{a}^{\dagger}\op{a}}\braket{\op{b}^{\dagger}\op{b}} =
gn_a n_b$.  (The other quadratic Fock term
$\braket{\op{a}^{\dagger}\op{b}}\braket{\op{b}^{\dagger}\op{a}}$ has
zero expectation.) The problem arises upon renormalization: As
discussed below, the anomalous density $\nu$ is formally divergent,
and regularization requires taking the coupling $g \rightarrow 0$ to
keep the gap parameter $\Delta = -g\nu$ finite.  Since the densities
remain finite, the Hartree contribution $gn_a n_b \rightarrow 0$
vanishes.

In weak coupling, one can carefully take the zero-range limit while
summing ladders~\cite{AGD:1975, Fetter:1971fk}, obtaining the well
known form $a n_a n_b$ of the Hartree interaction, which is clearly
invalid in the unitary limit $\abs{a}\rightarrow \infty$.  In
particular, for the symmetric phase $n_a = n_b = n$, there is no
additional length scale, and so we must have a dependence $\sim
n^{5/3}$ as dictated by dimensional analysis.  This physics---the
dominant contribution to the energetics (see the discussion
below~(\ref{eq:delta_xi}))---is completely missing from the \BdG\
(mean-field) approach and is one of the main deficiencies we hope to
overcome within an improved \DFT\ description.

We shall first discuss an improved local \DFT\ for symmetric
systems $n_a = n_b = n$: the Superfluid Local Density Approximation
(\SLDA).  This is a generalization of the Kohn-Sham Local Density
Approximation (\LDA) to includes pairing effects and subsumes the
\BdG\ form, adding an $n^{5/3}$ Hartree interaction term.

We subsequently extent the \SLDA\ to study asymmetric systems $n_a
\neq n_b$ through the use of the Asymmetric \SLDA\ (\ASLDA) functional
that subsumes the \SLDA.  The approach of both these approximations is
to introduce as few parameters as possible that are consistent with the
scaling and symmetries of the problem, then to determine the
coefficients of these terms by matching to ab initio properties in the
thermodynamic limit.  The form of the functionals is described in
Sec.~\ref{sec:energy-dens-funct}, the fitting of the
parameters is discussed in Sec.~\ref{sec:fitting-functionals}, and
some physical applications are presented in
Sec.~\ref{sec:using-slda-aslda}.

\subsection{The Energy Density Functional}
\label{sec:energy-dens-funct}

We start with the most restrictive conditions of a cold ($T=0$)
symmetric ($n_{a}=n_{b}$, $m_{a}=m_{b}$) unitary
($\abs{a}=\infty$) Fermi gas.  As discussed in
Sec.~\ref{cha:introduction}, the only dimensionful scale in the
problem is the density $n$, so dimensional analysis provides
significant constraints on the form of the functional and
thermodynamic functions, allowing us to postulate a simple functional
form characterized by only three dimensionless parameters.  Relaxing
any of these conditions will introduce additional dimensionless
parameters.  In particular, we consider the dimensionless polarization
$p = (n_{a} - n_{b})/(n_{a} + n_{b})$ to formulate
\ASLDA~\cite{Bulgac:2006cv,Chevy:2006b}.  The generalized \ASLDA\
functional promotes the dimensionless parameters to dimensionless
functions of this asymmetry parameter $p$.
 
\subsubsection[Local Density Approximation (\Lda)]{Local Density Approximation (LDA)}
\label{sec:local-dens-appr}

In general, the energy functional might be a highly non-local and
extremely complicated object.  One major simplification is to assume
that the functional is local and can be represented by a function of
various types of densities.  This amounts to introducing the energy
density $\mathcal{E}$ which is a function (as opposed to a functional)
of the local densities and their derivatives (referred to as gradient
corrections):
\begin{equation}
  \label{eq:DFT}
  E_{KS} = \int\D^{3}\vec{r}\;
  \mathcal{E}_{KS}[n(\vec{r}), \tau(\vec{r}), \vec{\nabla}n(\vec{r}),\dotsc]
  + U(\vec{r}) n(\vec{r}) + \cdots,
\end{equation}
where $U(\vec{r})$ represents an external (trapping) potential.  This
local density approximation (\LDA) has met with remarkable success in
quantum chemistry applications \cite{Kohn:1999fk, Dreizler:1990lr,
  Parr:1989uq}.

The simplest function contains a single term $E\propto n^{5/3}$.
This---along with gradient corrections---has been explored
in~\cite{Papenbrock:2005fk, Rupak:2008fk}, and, while it can model the
energetics of the symmetric gas, it does not include information about
pairing correlations.  The extensions we describe here include both
kinetic terms and an anomalous pairing density.

\subsubsection{Densities and Currents}
The first task is to construct the densities and currents.  In the
\SLDA, we consider five types of densities: the standard particle
densities $n_{a}(\vec{r})\propto
\braket{a^{\dagger}(\vec{r})a(\vec{r})}$ and
$n_{b}(\vec{r})\propto\braket{b^{\dagger}(\vec{r})b(\vec{r})}$, the
kinetic densities $\tau_{a}(\vec{r}) \propto
\braket{a^{\dagger}(\vec{r})\Delta a(\vec{r})}$ and
$\tau_{b}(\vec{r})\propto \braket{b^{\dagger}(\vec{r})\Delta
  b(\vec{r})}$, and an anomalous density $\nu(\vec{r}) \propto
\braket{a(\vec{r}) b(\vec{r})}$.  When considering time dependence
(Sec.~\ref{cha:time-depend-superfl}), we must also include the
currents $\vec{j}_{a}(\vec{r})\propto
\braket{a^{\dagger}(\vec{r})\vec{\nabla}a(\vec{r})}$ and
$\vec{j}_{b}(\vec{r})\propto
\braket{b^{\dagger}(\vec{r})\vec{\nabla}b(\vec{r})}$ to restore
Galilean invariance as discussed in
Sec.~\ref{sec:galilean-invariance}.  In principle, these densities may
be non-local, but to simplify the functional we wish to consider only
local quantities.  The local form of the anomalous density $\nu$ leads
to \UV\ divergences that we must regularize as we discuss in
Sec.~\ref{sec:regularization}.

The formal analysis proceeds with a four-component formalism discussed
in Sec.~\ref{sec:superfl-single-part}, but the symmetries of the cold
atom systems allow everything to be expressed in terms of
two-component wavefunctions (see Appendix~\ref{cha:appendix})
\begin{equation}
  \psi_{n}(\vec{r}) = \begin{pmatrix}
    u_{n}(\vec{r})\\
    v_{n}(\vec{r})
  \end{pmatrix}
\end{equation}
with energy $E_{n}$.  The densities and currents are constructed from
these as 
\begin{subequations}
  \label{eq:Densities}
  \begin{gather}
    \begin{aligned}
      n_{a}(\vec{r}) &= \sum_{n}\abs{u_{n}(\vec{r})}^2f_{\beta}(E_{n}), &
      n_{b}(\vec{r}) &= \sum_{n}\abs{v_{n}(\vec{r})}^2f_{\beta}(-E_{n}),\\
      \tau_{a}(\vec{r}) &= \sum_{n}\abs{\nabla u_{n}(\vec{r})}^2f_{\beta}(E_{n}),&
      \tau_{b}(\vec{r}) &= \sum_{n}\abs{\nabla v_{n}(\vec{r})}^2f_{\beta}(-E_{n}),
    \end{aligned}\\
    \begin{aligned}
      \label{eq:nu_dens}
      \nu(\vec{r}) &= \frac{1}{2}\sum_{n} u_{n}(\vec{r})v_{n}^{*}(\vec{r})
      \Bigl(f_{\beta}(-E_{n}) - f_{\beta}(E_{n})\Bigr),\\
      \vec{j}_{a}(\vec{r}) &=
      \frac{\I}{2} 
      \sum_{n} \left[u^*_{n}(\vec{r}) \nabla u_{n}(\vec{r}) 
        - u_{n}(\vec{r}) \nabla u^*_{n}(\vec{r})\right]
      f_{\beta}(E_{n}),\\
      \vec{j}_{b}(\vec{r}) &=
      \frac{\I}{2} 
      \sum_{n} \left[v^*_{n}(\vec{r}) \nabla v_{n}(\vec{r})  
        - v_{n}(\vec{r}) \nabla v^*_{n}(\vec{r})\right]
      f_{\beta}(-E_{n}),
    \end{aligned}
  \end{gather}
\end{subequations}
where $f_{\beta}(E_{n})=1/(\exp(\beta E_n)+1)$ is the Fermi
distribution and $\beta=1/T$ is the inverse temperature. Even though
we shall only discuss the zero temperature limit of \SLDA\, it is
convenient for numerical purposes to introduce a very small
temperature (much smaller than any other energy scale in the system)
so that $\mathcal{E}(\mu)$ is a smooth function.

\subsubsection{Functional Form}
\label{sec:functional-form}


Our functionals generically include a kinetic term and a pairing term
of the form
\begin{equation}
  \mathcal{E} = \frac{\hbar^2}{m}\Biggl(\frac{\tau_{a} 
    + \tau_{b}}{2}\Biggr) +
  g\nu^{\dagger}\nu + \dotsb,
\end{equation}
along with additional density dependent terms, where all of the densities and
currents $n(\vec{r})$ etc. are functions of position but have no non-local
structure.  (Note that here and in many of the following formulae we suppress
the explicit dependence on position $\vec{r}$.)  In the superfluid, this local
approximation has formal difficulties since the anomalous density $\nu(\vec{r},
\vec{r}') \sim \sum u_{n}(\vec{r})v_{n}^{*}(\vec{r}') \sim \abs{\vec{r} -
  \vec{r}'}^{-1}$ diverges for small $\abs{\vec{r} - \vec{r}'}$ if the pairing
field is taken to be a multiplicative operator $\Delta(\vec{r})$.  The kinetic
energy densities $\tau_{a,b}(\vec{r})$ diverge as well.  A proper local
formulation thus requires regularization~\cite{BY:2002fk} as discussed in
Sec.~\ref{sec:regularization}.  We introduce an energy cutoff
$E_{c}$---$\nu_{c}(\vec{r}) \sim
\sum_{\abs{E}<E_c}u_{n}(\vec{r})v_{n}^{*}(\vec{r})$---and a cutoff dependent
effective interaction $g_{\text{eff}}$ such that
\begin{equation}
  \Delta = -g\nu = -g_{\text{eff}}\nu_{c}
\end{equation}
is finite and independent of the cutoff as $E_{c}\rightarrow \infty$.
Once this is done, we can write the functional as
\begin{equation}
  \mathcal{E} = \frac{\hbar^2}{m}\Biggl(\frac{\tau_{a}+\tau_{b}}{2}\Biggr) 
  - \Delta^{\dagger}\nu + \dotsb.
\end{equation}
Note that $\nu$ is still formally divergent, but will cancel with
a similar divergence in the kinetic piece such that the energy
density is finite.  The full forms of the local functionals considered
here are thus:
\begin{description}
\item[\textbf{Bogoliubov de-Gennes (BdG)~\cite{Gennes:1966gf}:}]
  \begin{equation}
    \label{eq:DF_BdG}
    \mathcal{E}_{\text{BdG}} = 
    \frac{\hbar^2\tau_{a}}{2m_{a}} +
    \frac{\hbar^2\tau_{b}}{2m_{b}} +
    g\nu^{\dagger}\nu.
  \end{equation}
  For homogeneous systems, this is equivalent to the Eagles-Leggett
  mean-field theory where the parameters here represent the bare
  parameters (elsewhere we shall only consider $m_a = m_b = m$) and
  the coupling constant is tuned to reproduce the vacuum two-body
  scattering length $a$.  Note the absence of a self-energy: all of
  the interaction effects are modelled through the pairing
  interaction.  One unphysical consequence is that the normal state is
  described as completely non-interacting in this model.  While this
  may capture some qualitative features of the theory, and provides a
  rigorous variational bound on the energy, it cannot be trusted for
  quantitative results beyond the rather poor variational upper bound.

\item[\textbf{SLDA:}]
  \begin{equation}
    \label{eq:DF_SLDA}
    \mathcal{E}_{\SLDA} =
    \frac{\hbar^2}{m}\left(
      \frac{\alpha}{2}(\tau_a + \tau_b) +
      \beta \frac{3}{10}(3\pi^2)^{2/3}(n_a + n_b)^{5/3}\right) +
    g\nu^{\dagger}\nu.
  \end{equation}
  This may be thought of as the unitary generalization of the symmetric \BdG\
  functional for symmetric matter $n_a = n_b = n_{+}/2$ to include a self-energy
  term $\smash{n_{+}^{5/3}}$ (whose form is fixed by simple dimensional
  analysis) and an effective mass $m_{\text{eff}} = m/\alpha$.  The three
  parameters here $\alpha$, $\beta$, and the pairing interaction $g$ must be
  fixed by matching to experiments or ab initio calculations as discussed in
  Sec.~\ref{sec:symm-superfl-state}.  Since $g$ is formally zero in the large
  coupling limit, we characterize it with a dimensionless constant $\gamma$ such
  that $g_{\text{eff}}^{-1} = (n_a + n_b)^{1/3}/\gamma -\Lambda$ where $\Lambda$ is
  the cutoff discussed in Sec.~\ref{sec:regularization}.
\item[\textbf{ASLDA:}]
\begin{equation}
    \label{eq:DF_ASLDA}
    \mathcal{E}_{\ASLDA} =
    \frac{\hbar^2}{m}\left(
      \alpha_{a}(n_{a},n_{b})\frac{\tau_{a}}{2} +
      \alpha_{b}(n_{a},n_{b})\frac{\tau_{b}}{2} +
      D(n_a,n_b)
    \right) +
    g\nu^{\dagger}\nu.
  \end{equation}
  Here we allow for polarization $n_a \neq n_b$ and so we must generalize
  the parameters such as the effective masses and self-interaction to
  be functions of the local polarization $p = (n_a - n_b)/(n_a +
  n_b)$.  Dimensional analysis restricts these $\alpha_{a,b}(\lambda
  n_{a},\lambda n_{b}) = \alpha(n_{a},n_{b})$ and $D(\lambda
  n_{a},\lambda n_{b}) = \lambda^{5/3}D(n_{a},n_{b})$ so that we need
  only to parametrize functions of the single variable $p$ as discussed
  in Sec.~\ref{sec:homogeneous-matter}.
\end{description}
To fully define these functionals, we must now regularize the pairing
interaction $g$ (Sec.~\ref{sec:regularization}) and then specify
the values and functional forms of the parameters and parametric
functions (Sec.~\ref{sec:fitting-functionals}).

\subsubsection{Regularization}
\label{sec:regularization}

As formulated, the local theory is ultraviolet divergent due
to the well known behaviour of the anomalous density:
\begin{equation}
  \nu(\vec{r}, \vec{r}') \sim \sum_{n}
  u_{n}(\vec{r})v_{n}^{*}(\vec{r}')
  \propto
  \frac{1}{\abs{\vec{r} - \vec{r}'}}.
\end{equation}
There are many ways of dealing with this.  For example, physical
potentials are always non-local, and the non-locality naturally
regulates the theory.  However, in the unitary gas, the non-local
(range of the interaction) is much smaller than any other length scale
in the system and the stability of the system (see
Sec.~\ref{cha:introduction}) indicates that the low-energy
large-distance physics should be independent of the short-range
details.  

As a result, one can choose any sort of regularization scheme that is
convenient and obtain the universal results with an appropriate
limiting procedure.  In the homogeneous case, one can use a variety of
techniques: some interesting choices include dimensional
regularization~\cite{Papenbrock:1998wb} and selective distribution
functions~\cite{Tan:2008kx}.  The most straightforward is
to use a momentum cutoff, but for inhomogeneous systems, momentum is
not a good quantum number.  Instead, an energy cutoff $E_{c}$
suffices.  All quantities---especially the divergent anomalous
density---can be computed from states with energies below this cutoff:
\begin{equation}
  \label{eq:regularized_nu}
  \nu_c = \sum_{\abs{E_{n}}<E_{c}}u_{n}v_{n}^{*}\frac{f_{\beta}(E_{n})
    - f_{\beta}(-E_n)}{2}.
\end{equation}
(To improve the behaviour, we actually use a smooth cutoff so that
discontinuities are not introduced when levels cross in and out of the
sum during the self-consistent iterations.)

To better understand the nature of these divergences, consider the
ultraviolet limit where the length scale is much smaller than any
other scale in the system.  In this limit, the semi-classical
Thomas-Fermi approximation may be applied locally.  The linear
divergences in both the symmetric combination of the kinetic energy
and in the anomalous density have the form
\begin{align}
  \label{eq:UV_divergence}
  \tau_{+}(k) = \tau_{a}(k) + \tau_{b}(k) &\rightarrow 
  \frac{2(m^*)^2\Delta^{\dagger}\Delta}{\hbar^4k^2}, &
  \nu(k) &\rightarrow  \frac{m^{*}\Delta}{\hbar^2k^2},
\end{align}
where the average effective mass $m^{*} = m/\alpha_{+} =
2m/(\alpha_{a} + \alpha_{b})$ enters explicitly through the equations
of motion.  From this it is clear that the combination
\begin{equation*}
  \label{eq:energy_p}
  \frac{\hbar^2\tau_{+}}{2m^{*}} - \Delta^{\dagger}\nu
  =
  \frac{\hbar^2}{m}\left(
    \frac{\alpha_{a}\tau_{a}}{2} + \frac{\alpha_{b}\tau_{b}}{2}
  \right)
  + g\nu^{\dagger}\nu
\end{equation*}
remains finite if we regularize the theory such that the gap
parameter remains finite for all values of the cutoff
\begin{equation}
  \Delta = -g_{\text{eff}}\nu_{c}. 
\end{equation}
When regularizing the \BdG\ equations~(\ref{eq:DF_BdG}), we hold fixed
the vacuum two-body scattering length,
\begin{equation}
  \label{eq:two-body-scattering}
  \frac{m}{4\pi\hbar^2 a}
  =
  \frac{1}{g}
  +
  \frac{1}{2}
  \dashint\frac{\D^{3}\vec{k}}{(2\pi)^3}
  \frac{1}{\dfrac{\hbar^2k^{2}}{2m} + \I 0^{+}}
\end{equation}
where $\dashint$ is the principal value integral.  This may be easily
derived from the pseudo-potential approach (see for
example~\cite{Blatt:1952-74-76,Huang:1987-pp230-238} or for higher
partial waves~\cite{Huang:1957,Lee:1957}).

In the other \DFT{}s~\eqref{eq:DF_SLDA} and~\eqref{eq:DF_ASLDA}, $g$
does not represent the physical interaction, but is simply another
parameter of the theory.  Thus, we define a similar regularization
scheme by introducing a finite function $\tilde{C}(n_{a},n_{b})$ that
must be fit in order to characterize the pairing interaction and
correlations.\footnote{We have changed notations slightly
  from~\cite{Bulgac;Forbes:2008-08} using $\tilde{C}(n_{a},n_{b}) =
  \alpha_{+}C(n_{a},n_{b})$ which simplifies the equations because, in
  the limit of infinite cutoff, $\Lambda$ is independent of any
  densities and functional parameters.}
\begin{equation}
  \tilde{C}(n_{a}, n_{b})
  =
  -\frac{\alpha_{+}\nu}{\Delta}
  +
  \frac{1}{2}
  \dashint\frac{\D^{3}\vec{k}}{(2\pi)^3}
  \frac{1}{\dfrac{\hbar^2k^{2}}{2m} 
    - \dfrac{\mu_{+}}{\alpha_{+}} + \I 0^{+}}
  = 
  \frac{\alpha_{+}}{g_{\text{eff}}}
  +
  \Lambda.
\end{equation}
This differs from (\ref{eq:two-body-scattering}) in two ways: 1) we
have included a factor of the effective mass parameter $\alpha_{+}$ to
ensure that the divergences~(\ref{eq:UV_divergence}) cancel and, 2) we
have shifted the pole of the integral by the average local chemical
potential $\mu_{+} = (\mu_a - V_a + \mu_b + V_b)/2$ to improve
convergence.  As pointed out in~\cite{BY:2002fk}, the shift does not
change the integral in the limit of infinite cutoff, but greatly
improves the convergence if a cutoff is used.  Given a fixed momentum
cutoff $k < k_c$, the integral $\Lambda$ in the second term can be
performed exactly
\begin{equation}
  \Lambda = 
  \frac{m}{\hbar^2}\frac{k_c}{2\pi^2}
  \left\{
    1 - \frac{k_0}{2k_c}
    \ln\frac{k_c + k_0}{k_c - k_0}
  \right\}
\end{equation}
where $\hbar^2k_{0}^2/(2m) = \mu_{+}/\alpha_{+}$ defines the location
of the pole.  In general, translational invariance is not preserved,
and so we must use the fixed energy cutoff $\abs{E(k)} < E_c$ that
enters~(\ref{eq:regularized_nu}) rather than a momentum cutoff
as the latter is not a good quantum number.  To relate the two we used
the local quasiparticle dispersion relationship:
\begin{subequations}
  \begin{align}
    \frac{\hbar^2}{2m}\alpha_{+}(\vec{r})k_0^2(\vec{r}) -
    \mu_{+}(\vec{r})
    &= 0, \\
    \frac{\hbar^2}{2m}\alpha_{+}(\vec{r})k_c^2(\vec{r}) -
    \mu_{+}(\vec{r}) 
    &= E_c.
  \end{align}
\end{subequations}
This defines a position-dependent momentum cutoff $k_c(\vec{r})$ and
effective coupling constant $g(\vec{r})$ that can be used to regulate
the anomalous density at any point in space:
\begin{subequations}
  \begin{align}
    \Lambda(\vec{r}) &= 
    \frac{m}{\hbar^2}
    \frac{k_c(\vec{r})}{2\pi^2}
    \left\{
      1 - \frac{k_0(\vec{r})}{2k_c(\vec{r})}
      \ln\frac{k_c(\vec{r}) + k_0(\vec{r})}{k_c(\vec{r}) - k_0(\vec{r})}
    \right\},\\
    \frac{\alpha_{+}(\vec{r})}
         {g_{\text{eff}}(\vec{r})} &=
    \tilde{C}\Bigl(n_a(\vec{r}), n_b(\vec{r})\Bigr)
    - 
    \Lambda(\vec{r}),\\
    \Delta(\vec{r}) &= -g_{\text{eff}}(\vec{r})\nu_{c}(\vec{r}).
  \end{align}
\end{subequations}
Varying the functional with respect to the occupation numbers (see
Appendix~\ref{sec:A-form-descr-dft} for a formal description) allows
us to derive the self-consistency conditions. Recall that the
functional has the form
\begin{equation}
  \alpha_{-}(n_{a}, n_{b})\frac{\hbar^2\tau_{-}}{2m}
  +
  \alpha_{+}(n_{a}, n_{b})\left(\frac{\hbar^2\tau_{+}}{2m} 
    + \frac{g_{\text{eff}}}{\alpha_+}\nu_{c}^\dagger\nu_{c}\right)
  + \frac{\hbar^2}{m}D(n_{a}, n_{b}),
\end{equation}
and that, in the limit of infinite cutoff, $\Lambda$ has no
dependence on the functional parameters so that\footnote{There is a
  small correction due to the residual density dependence
  of $\Lambda$ at finite cutoff but in practice this is
  insignificant.}
\begin{equation}
  \D\tilde{C} = \D\left(\frac{\alpha_{+}}{g_{\text{eff}}}\right)
  \qquad \implies \qquad
  \D\left(\frac{g_{\text{eff}}}{\alpha_{+}}\right) =
  -\left(\frac{g_{\text{eff}}}{\alpha_{+}}\right)^2\D\tilde{C}.
\end{equation}
Thus, we have the following equations:
\begin{equation*}
  \begin{pmatrix} 
    \op{K}_{a} - \mu_{a} + V_{a} & \Delta^{\dagger}\\
    \Delta & -\op{K}_{b} + \mu_{b} - V_{b}
  \end{pmatrix}
  \begin{pmatrix}
    u_{n}\\
    v_{n}
  \end{pmatrix}
  =
  E_{n}
  \begin{pmatrix}
    u_{n}\\
    v_{n}
  \end{pmatrix}
\end{equation*}
where
\begin{align*}
  \op{K}_{a}u &= -\frac{\hbar^2}{2m}
  \nabla_{i}\bigl(\alpha_{a}(n_{a},n_{b})\nabla_{i}u\bigr)\\
  \op{K}_{b}v &= -\frac{\hbar^2}{2m}
  \nabla_{i}\bigl(\alpha_{b}(n_{a},n_{b})\nabla_{i}v\bigr)\\
  V_{a} &= \pdiff{\alpha_{-}(n_{a},n_{b})}{n_{a}}
  \frac{\hbar^2\tau_{-}}{2m}
  +
  \pdiff{\alpha_{+}(n_{a},n_{b})}{n_{a}}
  \left(
    \frac{\hbar^2\tau_{+}}{2m} 
    - \frac{\Delta^{\dagger}\nu}{\alpha_{+}(n_{a},n_{b})}
  \right)\\
  & \quad 
  -\pdiff{\tilde{C}(n_{a},n_{b})}{n_{a}}\frac{\Delta^{\dagger}\Delta}{\alpha_{+}}
  +\frac{\hbar^2}{m}\pdiff{D(n_{a},n_{b})}{n_{a}} + U_{a}(\vec{r}),\\
  \alpha_{\pm}(n_{a},n_{b}) &= 
  \tfrac{1}{2}[\alpha_{a}(n_{a},n_{b})\pm\alpha_{b}(n_{a},n_{b})],\\
  \tau_{\pm} &= \tau_{a} \pm \tau_{b}.
\end{align*}
and similarly with $a \leftrightarrow b$.

\subsection{Determining the \Slda\ and \Aslda\ Energy Density Functionals}
\label{sec:fitting-functionals}

Unless one has perturbative control over the theory, one cannot in
general determine the correct functional from first principles.
Instead, the functional must be treated as a model incorporating the
most relevant physics for the application at hand.  As such, one must
determine some parameters in order to make predictions about other
properties of the system.  Here we use properties of homogeneous
matter in the thermodynamic limit to determine the parameters of our
functional, and then use the functional to compute the properties of
non-uniform systems such as trapped gases.  Our hope is that the
single particle states in the self-consistent approach will provide a
good description of the finite size (shell) effects missing in the
Thomas Fermi approximation.

Fortunately, the thermodynamic functions describing the unitary Fermi
gas are tightly constrained~\cite{Bulgac:2006cv}, and have both
calculational and experimental verification.  We shall now describe
how to use these constraints to determine the form of the
dimensionless parameters describing the functional.

\subsubsection{Homogeneous Matter}
\label{sec:homogeneous-matter}

A simple Thomas-Fermi calculation can be employed to describe states
of homogeneous matter by exploiting the translational invariance of the
system.  This allows us to fix all non-gradient terms in the
functional.  The only remaining term---the effective mass---must be
fixed by other means and we use the quasiparticle properties to
determine this coefficient.

\paragraph{Normal Phase}
The energy-density for the normal phase of homogeneous matter has the
form
\begin{subequations}
  \begin{align}
    \mathcal{E}[n_{a},n_{b}] &=
    \frac{\hbar^2}{m}\frac{\Bigl(6\pi^2 (n_a + n_b)\Bigr)^{5/3}}{20\pi^2}G(p),
    &
    p &= \frac{n_{a}-n_{b}}{n_{a} + n_{b}} \in[-1, 1].
  \end{align}
  where
  \begin{equation}
    \label{eq:G_x}
    G(p) = \alpha(p)\left(\frac{1+p}{2}\right)^{5/3} + 
    \alpha(-p)\left(\frac{1-p}{2}\right)^{5/3} + 2^{-2/3}\beta(p)
  \end{equation}
  and $\beta(p)$ is defined through
  \begin{equation}
    \label{eq:D}
    D(n_a, n_b) = \frac{\Bigl(6\pi^2(n_a + n_b)\Bigr)^{5/3}}{20\pi^2}
    2^{-2/3}\beta(p).
  \end{equation}
\end{subequations}
The function $G(p)$ will be the main function that enters our
numerical formulae.\footnote{In our previous
  calculations~\cite{BF:2008,Bulgac;Forbes:2008-08}, we used a more
  complicated parametrization: the present form $G(p)$ is just as good
  and much simpler and we advocate its use instead.}

We shall define the dimensionless function $G(p)$ by fitting a simple
even polynomial to the Monte-Carlo data tabulated for
$f[p(x)]$.\footnote{ $G(p)$ is related to the other dimensionless
  functions $f(x)$ and $g(x)$ discussed in the literature as:
    \begin{align*}
      G(p) &= \left(\frac{1+p}{2}g(p)\right)^{5/3} =
      \left(\frac{1+p}{2}\right)^{5/3} f(p), &
      x &= \frac{n_{b}}{n_{a}} \in [0,\infty].
    \end{align*}
    The function $g(x) = g[p(x)]$ introduced in~\cite{Bulgac:2006cv} has the
    necessary and sufficient requirement of convexity to satisfy the
    second law of thermodynamics; and the function $f(x) = f[p(x)]$ was
    introduced in~\cite{Cohen:2005ea} and has been tabulated using
    Monte-Carlo methods~\cite{LRGS:2006}.} From this,
the function $D(n_a, n_b)$ may be directly expressed in terms of the
inverse effective mass $\alpha(p)$, which may be independently parametrized:
\begin{equation*}
  D(n_a, n_b) = \frac{\Bigl(6\pi^2(n_a + n_b)\Bigr)^{5/3}}{20\pi^2}
  \left[
    G(p) 
    - \alpha(p)\left(\frac{1+p}{2}\right)^{5/3}
    - \alpha(-p)\left(\frac{1-p}{2}\right)^{5/3}
  \right].
\end{equation*}
 
\begin{figure}[htbp]
  \begin{center}
    \psfrag{sin(x)}{$\sin(x)$}
    \psfrag{cos(x)}{$\cos(x)$}
    \psfrag{x (radians)}{$x$ (radians)}
    \psfrag{y}{$y$}
    {\footnotesize                  
      \psfrag{1.0}{1.0} 
      \psfrag{0.5}{0.5} 
      \psfrag{0.0}{0.0}
      \psfrag{0.2}{0.2}
      \psfrag{0.4}{0.4}
      \psfrag{0.6}{0.6}
      \psfrag{0.8}{0.8}
      \psfrag{1.0}{1.0}
      \psfrag{1.2}{1.2}
      \psfrag{1.4}{1.4}
      \psfrag{1.6}{1.6}
      \psfrag{-1.0}{-1.0}
      \psfrag{-0.5}{-0.5}
      \psfrag{G_p}{$G(p)$}
      \psfrag{f=g^{5/3}}{$f(x)=g^{5/3}(x)$}
      \psfrag{x=n_b/n_a}{$x=n_b/n_a$}
      \psfrag{p=(n_a-n_b)/(n_a+n_b)}{$p=(n_a-n_b)/(n_a+n_b)$}
      \includegraphics[width=\textwidth]{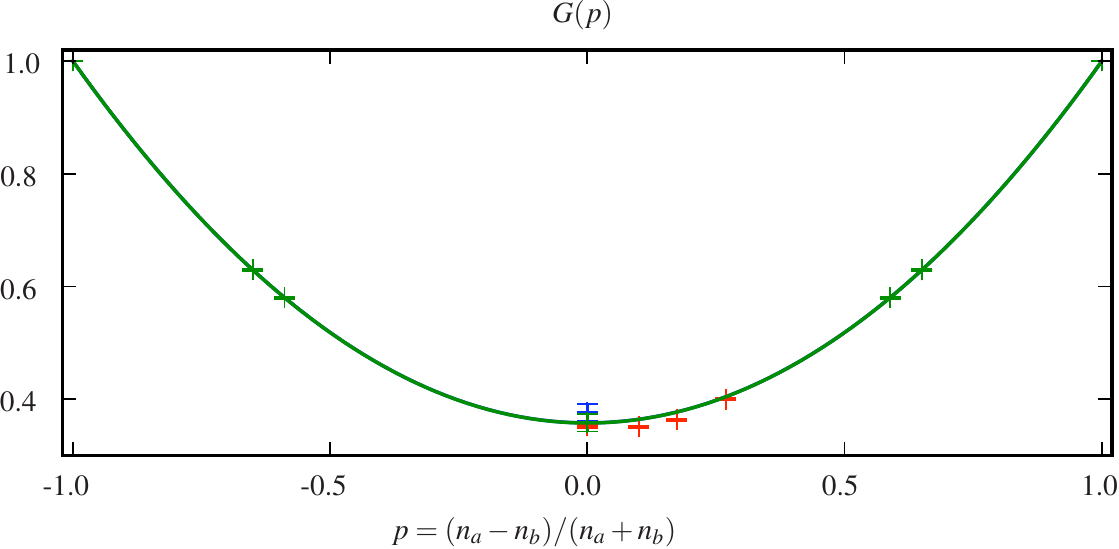}
      \includegraphics[width=\textwidth]{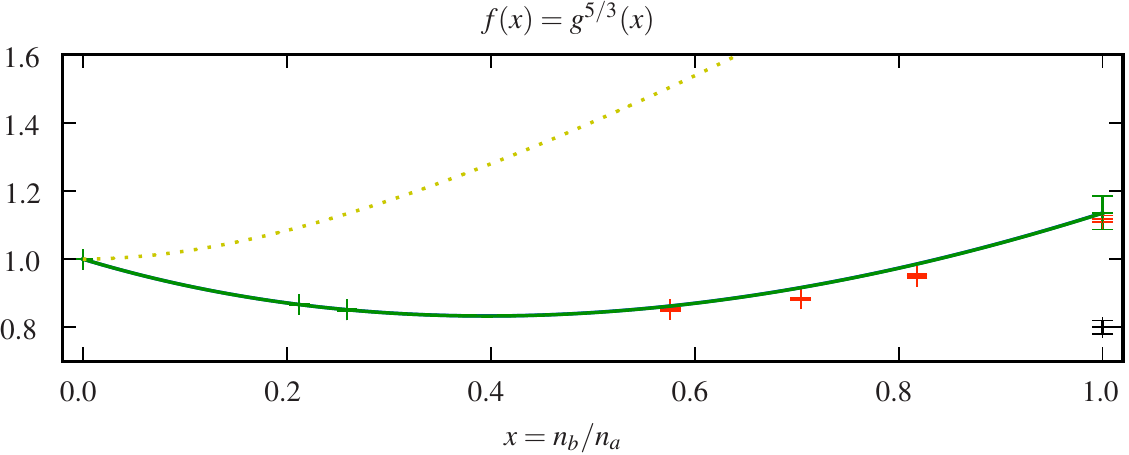}
    }
    \caption{
      \label{fig:giorgini}
      Monte Carlo data used to fit the function $G(p)$ (top) and in
      its raw form $f(x)=g^{5/2}(x)$ (bottom) representing the energy
      of the normal state with respect to the energy of the free
      system.  We excluded the red points from our fit because we
      suspect that they slightly contaminated by the superfluid state
      (and hence have a lower energy).  Fitting these close to the
      superfluid state would require a double hump structure in $G(p)$
      for which we do not yet see any physical motivation.  To anchor
      the solution in the superfluid phase, we include a datum
      $\beta_{p=0}$ extracted from the symmetric
      state~(\ref{eq:beta_0}).  The value here depends slightly on
      whether or not we also extract an effective mass, or hold
      $\alpha = 1$ constant.  Both fits are shown (but lie on top of
      each other).  The present fit is the simple two-parameter
      quadratic given in~(\ref{eq:G_p_fit}).  At the lower-right of
      the lower plot we have shown the values of $f_{x=1}$ for the
      superfluid state (black point).  Finally, for comparison, we
      have included the function $f(x)$ obtained using the standard
      mean-field (Eagles-Leggett) approximation as a dotted yellow
      line to show that it bears little resemblance to the physical
      curves.}
  \end{center}
\end{figure}

The function $G(p)$
describing the normal state has been well-constrained by Monte-Carlo
data~\cite{LRGS:2006} (see Fig.~\ref{fig:giorgini}).  As shown in
Fig.~\ref{fig:giorgini}, the function $G(p)$ is very
well parametrized by a simple quadratic polynomial:
\begin{equation}
  \label{eq:G_p_fit}
  G(p) = 0.357 + 0.642p^2.
\end{equation}

\paragraph{Symmetric Superfluid State $n_{a} = n_{b}$}
\label{sec:symm-superfl-state}

As suggested in~\cite{Bulgac:2007a}, by considering the calculated
properties of the fully paired symmetric superfluid, one may determine
the values of the functions $\alpha(p)$, $\tilde{C}(n_{a}, n_{b})$, and
$D(n_{a}, n_{b})$ at the point $p=0$ where the energy density
functional depends only on the symmetric combination of parameters
$n_a = n_b$ and $\tau_a = \tau_b$.  For any value of the inverse
effective mass $\alpha = \alpha_{p=0}$, one can uniquely determine the
self-energy $\beta = \beta_{p=0}$ and pairing interaction $\gamma$ by
requiring that the energy and gap satisfy
\begin{subequations}
  \label{eq:SymmConditions}
  \begin{align}
    \mathcal{E}_{SF} = \mathcal{E}(n,n) &= \xi \mathcal{E}_{FG} = 
    \xi\frac{\hbar^2}{m}\frac{(6\pi^2n)^{5/3}}{10\pi^2},\\
    \Delta &= \eta\varepsilon_{F} 
    = \eta\frac{\hbar^2}{m}\frac{(6\pi^2n)^{2/3}}{2}.
  \end{align}
\end{subequations}
The parameters $\xi$ and $\eta=\Delta/\varepsilon_{F}$ have been
calculated using several Monte-Carlo techniques~\cite{CCPS:2003,
  Carlson:2005kg, Carlson:private_comm2, Magierski:2009, BDM:2008}.  We
take the following values in our estimates~\cite{Carlson:2005kg,
  Carlson:private_comm2}:
\begin{align}
  \label{eq:xi_Delta0}
  \xi &= \frac{\mathcal{E}(n,n)}{\mathcal{E}_{FG}(n,n)} = 0.40(1),&
  \eta &= \frac{\Delta}{\varepsilon_{F}} = 0.504(24).
\end{align}
In order to determine the effective mass, we consider the
quasiparticle dispersion relationship~\cite{Carlson:2005kg}.  Within
our density functional, this has the form
\begin{equation}
  \label{eq:qp_disp}
  E_{qp}(k) = 
  \sqrt{\left(\frac{\hbar^2k^2}{2m_{\text{eff}}} - \mu_{\text{eff}}\right)^2 
    + \Delta^2}
\end{equation}
where $n_a + n_b = k_F^3/3\pi^2$ is the Fermi wave-vector and
$\mu_{\text{eff}}$ is the effective chemical potential.  It turns out
that $\mu_{\text{eff}}$ also depends on $\Delta/\varepsilon_{F}$, so the
quasiparticle dispersion relation is really a function of only two
parameters: the effective mass and $\Delta/\varepsilon_{F}$.

\begin{figure}[htbp]
  \begin{center}
    \psfrag{k2/kF2}{$k^2/k_{F}^2$}
    \psfrag{Eqp/eF}{$E_{\text{qp}}/\varepsilon_{F}$}
    {\footnotesize                  
      \psfrag{0.4}{0.4}
      \psfrag{0.6}{0.6}
      \psfrag{0.8}{0.8}
      \psfrag{1.0}{1.0}
      \psfrag{0}{0}
      \psfrag{0.5}{0.5}
      \psfrag{1}{1}
      \includegraphics[width=0.8\textwidth]{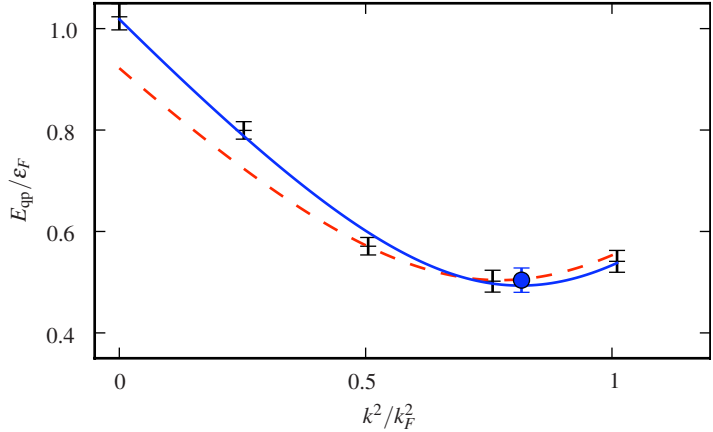}
    }
    \caption{
      \label{fig:qp}
      Fit of the Monte-Carlo data for the quasiparticle dispersions
      from~\cite{Carlson:2005kg} with the \BCS\
      form~(\ref{eq:qp_disp}). The solid blue curve is the full
      two-parameter fit including the mass as a parameter.  This is
      used to determine the effective mass of the fully paired
      symmetric matter $m_{\text{eff}}=0.91(1)$.  The dashed red curve
      is a one-parameter fit holding the mass fixed to $m=1$.  }
  \end{center}
\end{figure}

The fit to the Carlson-Reddy data~\cite{Carlson;Reddy:2008-04} is
shown in Fig.~\ref{fig:qp} and gives the following parameter
values:\footnote{We have performed a simple two-parameter non-linear
  least-squares fit which has a reduced $\chi_{\text{red}}^2 = 1.1$,
  indicating a very good fit.}
\begin{subequations}
  \label{eq:symmetric_fit}
  \begin{align}
    \alpha_{p=0} = m_{\text{eff}}^{-1}/m^{-1} &=
    \phantom{-}1.094(17), \label{eq:alpha_0}\\
    \beta_{p=0} &= -0.526(18), \label{eq:beta_0}\\
    \gamma^{-1} &= -0.0907(77),
    \label{eq:gammainv_0}\\
    \eta = \Delta/\varepsilon_{F} &= \phantom{-}0.493(12),
    \label{eq:Delta_0}\\
    \xi_{N} = \alpha + \beta &= \phantom{-}0.567(24).
    \label{eq:xi_N}
  \end{align}
\end{subequations}
where $\xi_{N}\mathcal{E}_{FG}$ is the energy of the interacting
normal state predicted by the functional.  Note that this agrees very
well with the value given by $G(p)$ in Fig.~\ref{fig:giorgini} (we
have used this parameter as an additional point in the fitting of
$G(p)$).

In principle, one should use some form of ab initio calculation
or experimental measurement for polarized systems to determine the
dependence of the parameters $\alpha$, $\beta$, and $\gamma$ on the
polarization $p=n_{b}/n_{a}$.  Unfortunately, the fermion sign problem
has made this difficult and there is presently insufficient quality
data to perform such a fit.  Instead, we simply fix
\begin{equation}
  \label{eq:gamma_p}
  \gamma(p) = \gamma_{p=0} = -11.11(94).
\end{equation}
If high quality data about polarized superfluid states become
available, one might consider promoting this parameter to a
polarization dependent function similarly to $\alpha(p)$ and $G(p)$.
This fixes the pairing interaction:
\begin{equation}
  \tilde{C}(n_a, n_b) 
  = \frac{m}{\hbar^2}\frac{\alpha_{+}(p)(n_a + n_b)^{1/3}}{\gamma(p)}.
\end{equation}

\paragraph{Effective Mass Parametrization: $\alpha(n_{a},n_{b})$}

As discussed above, the effective mass cannot be determined solely
from the properties of homogeneous matter.  It is also clear in \DFT's
developed perturbatively~\cite{Bhattacharyya:2005,
  Bhattacharyya;Furnstahl:2005-10} that the effective mass is
arbitrary.  In the \ASLDA, however, the only gradient terms that enter
the functional are the kinetic terms $\tau$ whose coefficients are the
effective masses.  To allay the need for additional gradient
corrections, one must provide a parametrization of the effective mass.
Fortunately, three values are well determined: In a fully polarized
system, the effective mass of the majority species remains unchanged,
$m_{p=1} = 1.0 m$, while in the minority species, the effective
``polaron'' mass $m_{p=-1}= 1.20 m$~\cite{CG:2008vn}.  We use this
value, but note that there are other estimates: Monte Carlo
calculations give $m_{-1} = 1.04(3)$~\cite{LRGS:2006} and $m_{-1} =
1.09(2)$~\cite{PG:2008}, and experiments measure $m_{-1}=1.06$ (no
error given)~\cite{shin-2008} and
$m_{-1}=1.17(10)$~\cite{Nascimbene:2009qy}. The third value for
symmetric matter $m_{0} = m/\alpha_{p=0}$ is determined
in~(\ref{eq:alpha_0}).

We now have three data-points constraining the effective mass
parametrization of $\alpha(p)$.  For numerical reasons, in order to
ensure that the effective potentials $V_{a,b}$ approach zero as the
density falls to zero, we impose the additional constraint that the
first and second derivatives of $\alpha(p)$ vanish at the end-points $p
= \pm 1$.  Taken together, this fixes a sixth order, two parameter polynomial
approximation for $\alpha(p)$:
\begin{equation}
  \label{eq:alpha_p_fit}
  \alpha(p) = 1.094 + 0.156 p (1 - 2p^2/3 + p^4/5) 
  - 0.532 p^2(1 - p^2 + p^4/3).
\end{equation}
\begin{figure}[ht]
  \begin{center}
    \psfrag{a=m/meff}{$\alpha = m/m_{\text{eff}}$}
    \psfrag{p=na-nb/na+nb}{$p=(n_a - n_b)/(n_a + n_b)$}
    {\footnotesize            
      \psfrag{0.8}{0.8}
      \psfrag{0.9}{0.9}
      \psfrag{1.0}{1.0}
      \psfrag{1.1}{1.1}
      \psfrag{1.2}{1.2}
      \psfrag{-1.0}{-1.0}
      \psfrag{-0.5}{-0.5}
      \psfrag{-1}{-1}
      \psfrag{0}{0}
      \psfrag{1}{1}
      \includegraphics[width=0.8\textwidth]{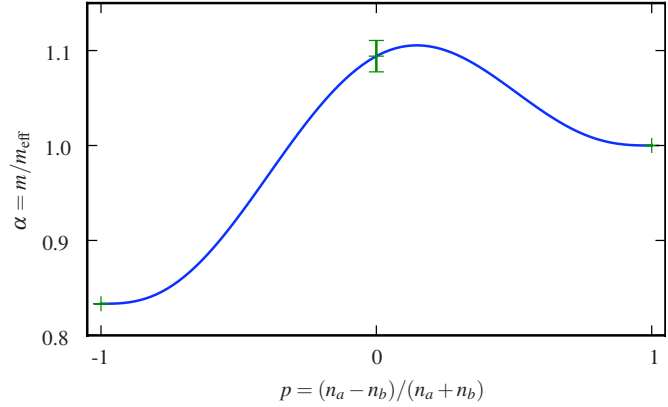}
    }
    \caption{
      \label{fig:alpha_p}
      Inverse effective mass $\alpha(p)=m/m_{\text{eff}}(p)$ as a
      function of the polarization $p=(n_a-n_b)/(n_a+n_b)$.  The
      functional fit is the polynomial~(\ref{eq:alpha_p_fit}).
    }
  \end{center}
\end{figure}

\subsubsection{Summary}
Here we summarize the complete definition of the \ASLDA\ functional.
The \SLDA\ functional follows by setting the local polarization
\begin{equation}
  p(\vec{r}) = \frac{n_{a}(\vec{r}) - n_{b}(\vec{r})}
                    {n_{a}(\vec{r}) + n_{b}(\vec{r})}
\end{equation}
to zero.  First, fitting the quasiparticle dispersion relationships,
gap and energy for the superfluid state gives the \SLDA\ parameters at
$p=0$:
\begin{align}
  \alpha_{p=0} &= 1.094(17), &
  \beta_{p=0} &= -0.526(18), &
  \gamma^{-1}_{p=0} &= -0.0907(77). \tag{\text{from} (\ref{eq:symmetric_fit})}
\end{align}
Using this derived effective mass, and the energy data for the normal
state from Monte Carlo data we obtain the following polynomial fits
defining the polarization dependence of the effective mass and
self-energy:
\begin{align}
  \alpha(p) &= 1.094 + 0.156 p \left(
    1 - \frac{2p^2}{3} + \frac{p^4}{5}\right) 
  - 0.532 p^2\left(1 - p^2 + \frac{p^4}{3}\right),
  \tag{\text{from (\ref{eq:alpha_p_fit})}}\\
  G(p) &= 0.357 + 0.642p^2,
  \tag{\text{from (\ref{eq:G_p_fit})}}\\
  \gamma(p) &= \gamma_{p=0} = -11.11(94).
  \tag{\text{from (\ref{eq:gamma_p})}}
\end{align}
These fix the specification of the functional parameters
\begin{gather*}
  \begin{aligned}
    \alpha_{a}(n_a, n_b) &= \alpha(p), &
    \alpha_{b}(n_a, n_b) &= \alpha(-p),&
    \tilde{C}(n_a, n_b) &= \frac{m}{\hbar^2}
    \frac{\alpha_{+}(p)(n_a + n_b)^{1/3}}{\gamma(p)},
  \end{aligned}\\
  D(n_a, n_b) = \frac{\Bigl(6\pi^2(n_a + n_b)\Bigr)^{5/3}}{20\pi^2}
  \left[
    G(p) 
    - \alpha(p)\left(\frac{1+p}{2}\right)^{5/3}
    - \alpha(-p)\left(\frac{1-p}{2}\right)^{5/3}
  \right],
\end{gather*}
in terms of the densities
\begin{gather*}
  \begin{aligned}
    n_{a}(\vec{r}) &= \sum_{\abs{E_{n}}<E_{c}}
    \abs{u_{n}(\vec{r})}^2f_{\beta}(E_{n}), &
    n_{b}(\vec{r}) &= \sum_{\abs{E_{n}}<E_{c}}
    \abs{v_{n}(\vec{r})}^2f_{\beta}(-E_{n}),\\
    \tau_{a}(\vec{r}) &= \sum_{\abs{E_{n}}<E_{c}}
    \abs{\nabla u_{n}(\vec{r})}^2f_{\beta}(E_{n}),&
    \tau_{b}(\vec{r}) &= \sum_{\abs{E_{n}}<E_{c}}
    \abs{\nabla v_{n}(\vec{r})}^2f_{\beta}(-E_{n}),
  \end{aligned}\\
  \begin{aligned}
    \nu(\vec{r}) &= \frac{1}{2}\sum_{\abs{E_{n}}<E_{c}}
    u_{n}(\vec{r})v_{n}^{*}(\vec{r})
    \Bigl(f_{\beta}(-E_{n}) - f_{\beta}(E_{n})\Bigr),\\
    \vec{j}_{a}(\vec{r}) &=
    \frac{\I}{2} 
    \sum_{\abs{E_{n}}<E_{c}} \left[u^*_{n}(\vec{r}) \nabla u_{n}(\vec{r}) 
      - u_{n}(\vec{r}) \nabla u^*_{n}(\vec{r})\right]
    f_{\beta}(E_{n}), \\
    \vec{j}_{b}(\vec{r}) &=
    \frac{\I}{2} 
    \sum_{\abs{E_{n}}<E_{c}} \left[v^*_{n}(\vec{r}) \nabla v_{n}(\vec{r})  
      - v_{n}(\vec{r}) \nabla v^*_{n}(\vec{r})\right]
    f_{\beta}(-E_{n}),
  \end{aligned}
\end{gather*}
in the form 
\begin{equation*}
  \mathcal{E}_{\ASLDA} =
  \frac{\hbar^2}{m}\left(
    \alpha_{a}(n_{a},n_{b})\frac{\tau_{a}}{2} +
    \alpha_{b}(n_{a},n_{b})\frac{\tau_{b}}{2} +
    D(n_a,n_b)
  \right) +
  g_{\text{eff}}\nu^{\dagger}\nu
\end{equation*}
together with the renormalization conditions
\begin{gather*}
  \begin{aligned}
    \Delta(\vec{r}) &= -g_{\text{eff}}(\vec{r})\nu_{c}(\vec{r}), &
    \frac{\alpha_{+}(\vec{r})}{g_{\text{eff}}(\vec{r})} =
    \tilde{C}(\vec{r})
    - 
    \Lambda(\vec{r})
  \end{aligned} \\
  \Lambda(\vec{r}) = \frac{m}{\hbar^2}\frac{k_c(\vec{r})}{2\pi^2}
  \left\{
    1 - \frac{k_0(\vec{r})}{2k_c(\vec{r})}
    \ln\frac{k_c(\vec{r}) + k_0(\vec{r})}{k_c(\vec{r}) - k_0(\vec{r})}
  \right\},\\
  \begin{aligned}
  \frac{\hbar^2}{2m}\alpha_{+}(\vec{r})k_0^2(\vec{r}) -
  \mu_{+}(\vec{r}) &= 0, &
  \frac{\hbar^2}{2m}\alpha_{+}(\vec{r})k_c^2(\vec{r}) -
  \mu_{+}(\vec{r}) &= E_c
  \end{aligned}
\end{gather*}
where $\alpha_{+} = (\alpha_{a} + \alpha_{b})/2$ and $\mu_{+} = (\mu_a
- V_a + \mu_b + V_b)/2$ is the average chemical potential defined
through the equations:
\begin{equation*}
  \begin{pmatrix} 
    \op{K}_{a} - \mu_{a} + V_{a} & \Delta^{\dagger}\\
    \Delta & -\op{K}_{b} + \mu_{b} - V_{b}
  \end{pmatrix}
  \begin{pmatrix}
    u_{n}\\
    v_{n}
  \end{pmatrix}
  =
  E_{n}
  \begin{pmatrix}
    u_{n}\\
    v_{n}
  \end{pmatrix}
\end{equation*}
where
\begin{align*}
  \op{K}_{a}u &= -\frac{\hbar^2}{2m}
  \nabla_{i}\bigl(\alpha_{a}(n_{a},n_{b})\nabla_{i}u\bigr)\\
  \op{K}_{b}v &= -\frac{\hbar^2}{2m}
  \nabla_{i}\bigl(\alpha_{b}(n_{a},n_{b})\nabla_{i}v\bigr)\\
  V_{a} &= \pdiff{\alpha_{-}(n_{a},n_{b})}{n_{a}}
  \frac{\hbar^2\tau_{-}}{2m}
  +
  \pdiff{\alpha_{+}(n_{a},n_{b})}{n_{a}}
  \left(
    \frac{\hbar^2\tau_{+}}{2m} 
    - \frac{\Delta^{\dagger}\nu}{\alpha_{+}(n_{a},n_{b})}
  \right)\\
  & \quad 
  -\pdiff{\tilde{C}(n_{a},n_{b})}{n_{a}}\frac{\Delta^{\dagger}\Delta}{\alpha_{+}}
  +\frac{\hbar^2}{m}\pdiff{D(n_{a},n_{b})}{n_{a}} + U_{a}(\vec{r}),\\
  \alpha_{\pm}(n_{a},n_{b}) &= 
  \tfrac{1}{2}[\alpha_{a}(n_{a},n_{b})\pm\alpha_{b}(n_{a},n_{b})],\\
  \tau_{\pm} &= \tau_{a} \pm \tau_{b}.
\end{align*}
and similarly with $a \leftrightarrow b$.

\subsection{Using the \Slda\ and \Aslda}
\label{sec:using-slda-aslda}

Once the form of the \DFT\ and its parameters have been fixed, the
function needs to be tested and applied.  Since we fit the parameters
using \QMC\ results for homogeneous matter, a non-trivial test is to
compare it with ab initio results in inhomogeneous situations.  This
will asses the accuracy of the approximation we have made in
neglecting gradient corrections beyond the kinetic term.  In
Sec.~\ref{sec:trapped-systems} we compare the predictions of the
\DFT{}s with \QMC\ calculations of trapped systems.  Next we show how
the functionals can be used to explore mesoscopic physics
inaccessible to \QMC\ analysis techniques: we consider the structure
of superfluid vortices in Sec.~\ref{sec:vortex-structure}, and the
prediction of a supersolid phase in the asymmetric case in
Sec.~\ref{sec:loff}.

\subsubsection{Trapped Systems}
\label{sec:trapped-systems}
\begin{figure}[b]
  \begin{center}
    \includegraphics[width=0.75\textwidth]{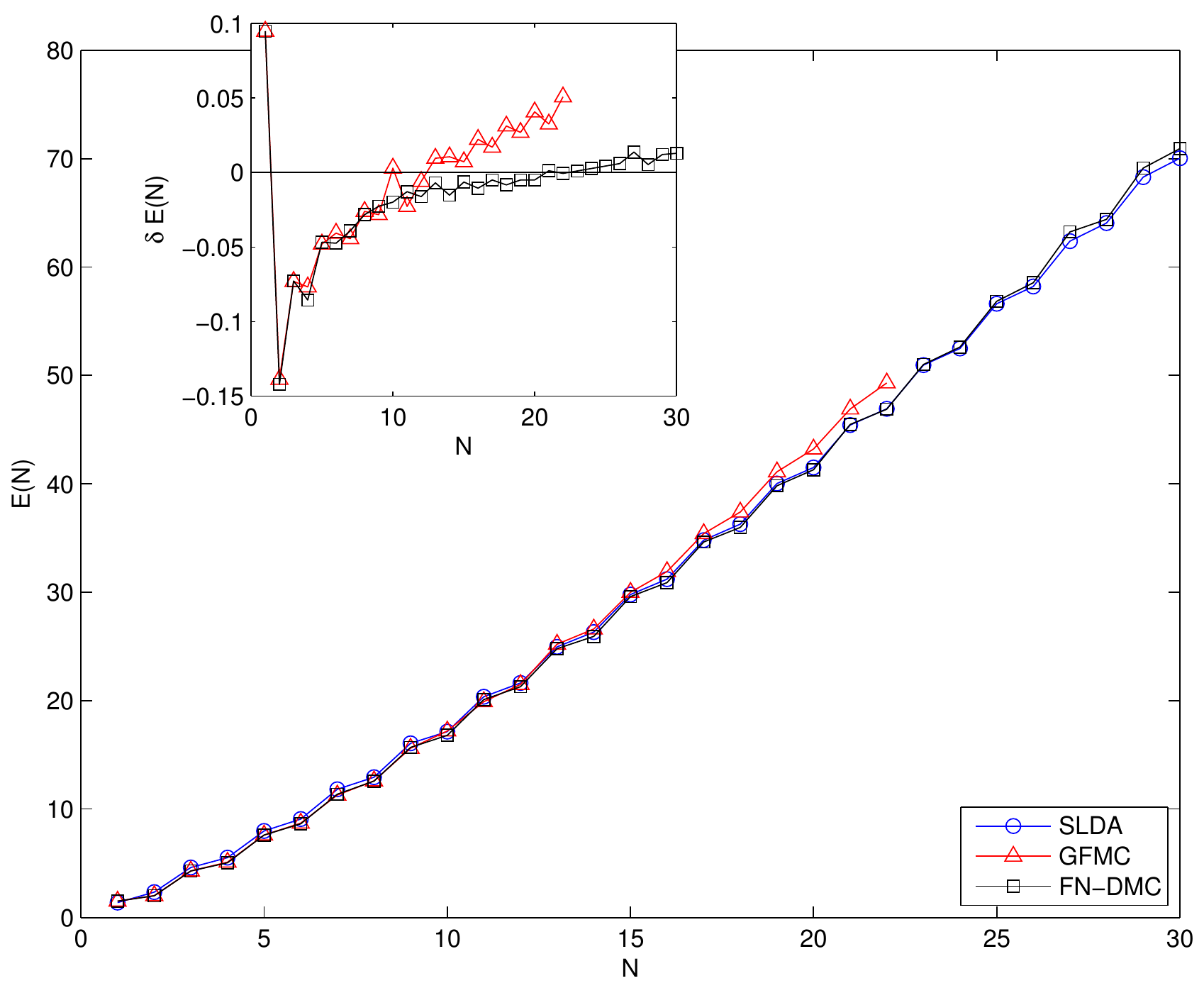}
    \caption{
      \label{fig:trap} 
      The comparison between the \GFMC~\cite{Chang;Bertsch:2007-08},
      \FNDMC~\cite{Blume;Stecher;Greene:2007-12} and \SLDA\ total
      energies $E(N)$.  The clear odd-even staggering of the energies
      is due to the onset of the pairing correlations. The inset shows
      the discrepancy between the \GFMC\ and \FNDMC\ and \SLDA\
      energies, $\delta E(N)=E_{MC}(N)/E_{\SLDA}(N)-1$, where
      $E_{MC}(N)$ stands for the energies obtained in \GFMC\ or
      \FNDMC\ respectively.}
  \end{center}
\end{figure}
\begin{table}[ht]
\begin{center}
\begin{tabular}[t]{r l l r}
  \toprule
  \multicolumn{4}{c}{Normal State}\\
  $(N_a, N_b)$ & $E_{FNDMC}$ & $E_{ASLDA}$ & (error)\\
  \midrule
  $( 3,  1)$ & $   6.6\pm  0.01$ & $6.687$ & $1.3\%$\\
  $( 4,  1)$ & $  8.93\pm  0.01$ & $8.962$ & $0.36\%$\\
  $( 5,  1)$ & $  12.1\pm   0.1$ & $12.22$ & $0.97\%$\\
  $( 5,  2)$ & $  13.3\pm   0.1$ & $13.54$ & $1.8\%$\\
  $( 6,  1)$ & $  15.8\pm   0.1$ & $15.65$ & $0.93\%$\\
  $( 7,  2)$ & $  19.9\pm   0.1$ & $20.11$ & $1.1\%$\\
  $( 7,  3)$ & $  20.8\pm   0.1$ & $21.23$ & $2.1\%$\\
  $( 7,  4)$ & $  21.9\pm   0.1$ & $22.42$ & $2.4\%$\\
  $( 8,  1)$ & $  22.5\pm   0.1$ & $22.53$ & $0.14\%$\\
  $( 9,  1)$ & $  25.9\pm   0.1$ & $25.97$ & $0.27\%$\\
  $( 9,  2)$ & $  26.6\pm   0.1$ & $26.73$ & $0.5\%$\\
  $( 9,  3)$ & $  27.2\pm   0.1$ & $27.55$ & $1.3\%$\\
  $( 9,  5)$ & $    30\pm   0.1$ & $30.77$ & $2.6\%$\\
  $(10,  1)$ & $  29.4\pm   0.1$ & $29.41$ & $0.034\%$\\
  $(10,  2)$ & $  29.9\pm   0.1$ & $30.05$ & $0.52\%$\\
  $(10,  6)$ & $    35\pm   0.1$ & $35.93$ & $2.7\%$\\
  $(20,  1)$ & $ 73.78\pm  0.01$ & $73.83$ & $0.061\%$\\
  $(20,  4)$ & $ 73.79\pm  0.01$ & $74.01$ & $0.3\%$\\
  $(20, 10)$ & $  81.7\pm   0.1$ & $82.57$ & $1.1\%$\\
  $(20, 20)$ & $ 109.7\pm   0.1$ & $113.8$ & $3.7\%$\\
  $(35,  4)$ & $   154\pm   0.1$ & $154.1$ & $0.078\%$\\
  $(35, 10)$ & $ 158.2\pm   0.1$ & $158.6$ & $0.27\%$\\
  $(35, 20)$ & $ 178.6\pm   0.1$ & $180.4$ & $1\%$\\
  \bottomrule
\end{tabular}
\hspace{1ex}
\begin{tabular}[t]{r l l r}
  \toprule
  \multicolumn{4}{c}{Superfluid State}\\
  $(N_a, N_b)$ & $E_{FNDMC}$ & $E_{ASLDA}$ & (error)\\
  \midrule
  $( 1,  1)$ & $ 2.002\pm     0$ & $2.302$ & $15\%$\\
  $( 2,  2)$ & $ 5.051\pm 0.009$ & $5.405$ & $7\%$\\
  $( 3,  3)$ & $ 8.639\pm  0.03$ & $8.939$ & $3.5\%$\\
  $( 4,  4)$ & $12.573\pm  0.03$ & $12.63$ & $0.48\%$\\
  $( 5,  5)$ & $16.806\pm  0.04$ & $16.19$ & $3.7\%$\\
  $( 6,  6)$ & $21.278\pm  0.05$ & $21.13$ & $0.69\%$\\
  $( 7,  7)$ & $25.923\pm  0.05$ & $25.31$ & $2.4\%$\\
  $( 8,  8)$ & $30.876\pm  0.06$ & $30.49$ & $1.2\%$\\
  $( 9,  9)$ & $35.971\pm  0.07$ & $34.87$ & $3.1\%$\\
  $(10, 10)$ & $41.302\pm  0.08$ & $40.54$ & $1.8\%$\\
  $(11, 11)$ & $46.889\pm  0.09$ & $45$ & $4\%$\\
  $(12, 12)$ & $52.624\pm   0.2$ & $51.23$ & $2.7\%$\\
  $(13, 13)$ & $58.545\pm  0.18$ & $56.25$ & $3.9\%$\\
  $(14, 14)$ & $64.388\pm  0.31$ & $62.52$ & $2.9\%$\\
  $(15, 15)$ & $70.927\pm   0.3$ & $68.72$ & $3.1\%$\\
  $( 1,  0)$ & $   1.5\pm   0.0$ & $ 1.5$  & $0\%$\\
  $( 2,  1)$ & $ 4.281\pm 0.004$ & $4.417$ & $3.2\%$\\
  $( 3,  2)$ & $  7.61\pm  0.01$ & $7.602$ & $0.1\%$\\
  $( 4,  3)$ & $11.362\pm  0.02$ & $11.31$ & $0.49\%$\\
  $( 7,  6)$ & $24.787\pm  0.09$ & $24.04$ & $3\%$\\
  $(11, 10)$ & $45.474\pm  0.15$ & $43.98$ & $3.3\%$\\
  $(15, 14)$ & $69.126\pm  0.31$ & $62.55$ & $9.5\%$\\
  \bottomrule
\end{tabular}
\end{center}
\caption{\label{tab:asymmetic_trap}
  Comparison between the \ASLDA\ density functional as described in
  this section and the \FNDMC\
  calculations~\cite{Blume;Stecher;Greene:2007-12,D.Blume:2008-05} for
  a harmonically trapped unitary gas at zero temperature.  The normal
  state energies are obtained by fixing $\Delta = 0$ in the
  functional: In the \FNDMC\ calculations, this is obtained by
  choosing a nodal ansatz without any pairing.  In the case of small
  asymmetry, the resulting ``normal states'' may be a somewhat artificial
  construct as there is no clear way of preparing a physical system in this
  ``normal state'' when the ground state is superfluid.
}
\end{table} 
The functional form of both the \SLDA\ and \ASLDA\ have been
completely fixed by considering only homogeneous matter.  Hence, a
non-trivial test of the theory is to compare the energy of trapped
systems with Monte Carlo calculations.  This was first done for the
\SLDA\ in~\cite{Bulgac:2007a} and the results are shown in
Fig.~\ref{fig:trap}.  Even for systems with only a few
particles---which have large gradients---the agreement is within 10\%.
This rapidly improves to the percent level as one move to larger
systems.

The agreement is somewhat remarkable.  In particular, we have included
no gradient corrections in the theory beyond the Kohn-Sham kinetic
energy.  These gradient corrections will contribute at some level, but
in the present system the coefficients are extremely tiny (the leading
gradient correction $\sim (\nabla n)^2/n$ should give corrections that
scale as $E \propto N^{2/3}$ for which there is no evidence in the
Monte Carlo data).  In any case, the agreement provides strong
evidence that the \SLDA\ captures the relevant energetics to provide a
quantitative model of the unitary Fermi gas.

We should point out that the gradient terms in the \SLDA\ are
completely characterized by the kinetic terms.  Thus, finite size
effects are highly sensitive to the inverse effective mass parameter
$\alpha$.  As mentioned in Sec.~\ref{sec:symm-superfl-state}, the
energy and gap can be fit with $\alpha=1$, but the resulting
parametrization demonstrates a marked systematic deviation from the
trap energies shown in Fig.~\ref{fig:trap}.  It is reassuring that the
agreement is restored when the effective mass is
chosen~(\ref{eq:alpha_0}) to reproduce the quasiparticle spectrum.

We have validated the \ASLDA\ in a similar manner for trapped
systems in Table~\ref{tab:asymmetic_trap}.  Again, the agreement is at
the few percent level in virtually all cases.  In general, the
formulation of the unitary \DFT\ has a remarkable ability to
capture the finite size effects in systems down to even a few
particles~\cite{FGG:2010}, lending credence to the approximation of
neglecting further gradients beyond the standard kinetic terms.  This
was somewhat anticipated since the kinetic terms completely
describe finite size (shell) effects in the non-interacting system,
but is non-trivial in the strongly interacting case of the unitary gas.

Note that the \BdG\ and \SLDA\ functionals have also been considered in larger
trapped systems~\cite{Baksmaty:2010, Pei:2010}.

\subsubsection{Vortex Structure}
\label{sec:vortex-structure}

\begin{figure}[ht]
  \begin{center}
    \includegraphics[width=0.5\textwidth]{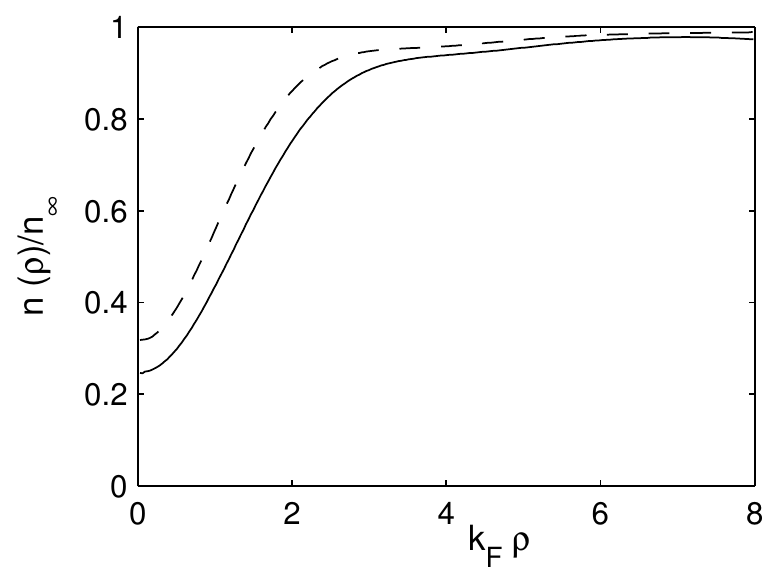}%
    \includegraphics[width=0.5\textwidth]{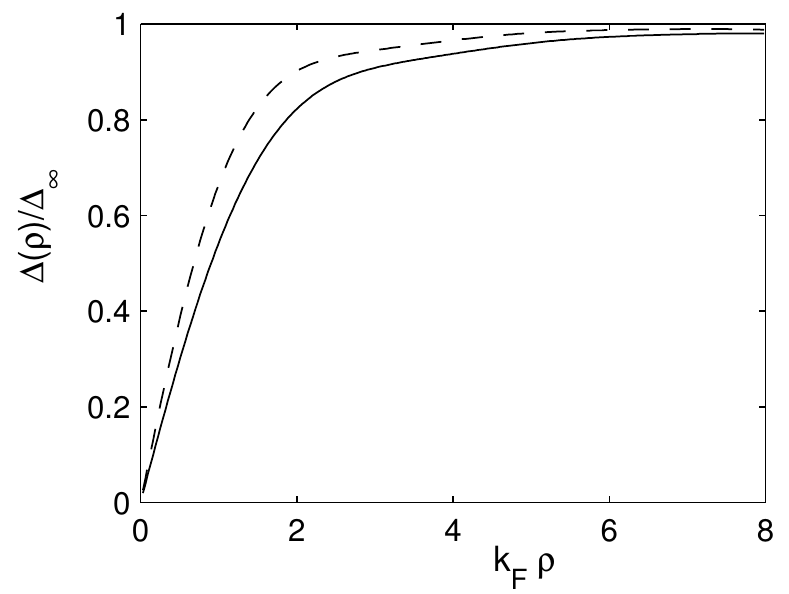}
    \caption{
      \label{fig:vortex}
      Density profile (left) and gap parameter (right)
      from~\cite{BY:2003} for a superfluid vortex in the symmetric
      $n_a = n_b$ unitary Fermi gas with unit circulation.  The solid
      curve corresponds to a parametrization of the \SLDA\ with no
      self-energy $\beta = 0$ but including an effective mass
      correction.  The dotted curve corresponds to a version with unit
      effective mass $\alpha=1$.  The other two parameters were fixed
      to reproduce the best approximation to energies of the normal
      and superfluid states known at the time: $\xi_{N} = 0.54$ and
      $\xi_{SF} = 0.44$.  The current parameter
      set~(\ref{eq:symmetric_fit}) should be preferred, but gives
      similar results.  Note: The solid curve does not have the
      required currents to restore Galilean invariance (see
      Sec.~\ref{sec:galilean-invariance}), but the effect should be
      small here. Since the dotted curve has no effective mass
      correction, Galilean corrections are not required.}
  \end{center}
\end{figure}

The first use of the \SLDA\ was to determine the structure of
superfluid vortices~\cite{BY:2003}.  In this work, two forms of \SLDA\
(slightly different parameter values) were considered, and the
solution for an axial symmetric vortex with unit circulation was
found. The method of solution uses a technique that properly treats
the infinite boundary conditions without truncating the physical space
and introducing finite-size artifacts (see~\cite{Belyaev:1987uq,
  Fayans:2000fk} for details). The profile for this vortex is shown in
Fig.~\ref{fig:vortex}.  In particular, it was predicted that the
vortices should have a significant density depletion in the
core---something that is not observed in the weak-coupling limit where
pairing is exponentially suppressed.  This predicted core depletion
allows for the direct imaging of vortices in rotating trapped
gasses~\cite{ZA-SSSK:2005lr}, providing direct evidence for
superfluidity in these systems.

\subsubsection[\Fflo/\Loff]{FFLO/LOFF}
\label{sec:loff}

\begin{figure}[tbp]
  \begin{center}
    \psfrag{x=n_b/n_a}{$x=n_{b}/n_{a}$}
    \psfrag{g(x)}{$g(x)$}
    \psfrag{0.9}{\footnotesize 0.9}
    \psfrag{1.0}{\footnotesize 1.0}
    \psfrag{1.1}{\footnotesize 1.1}
    \psfrag{0.0}{\footnotesize 0.0}
    \psfrag{0.2}{\footnotesize 0.2}
    \psfrag{0.4}{\footnotesize 0.4}
    \psfrag{0.6}{\footnotesize 0.6}
    \psfrag{0.8}{\footnotesize 0.8}
    \includegraphics[width=\textwidth]{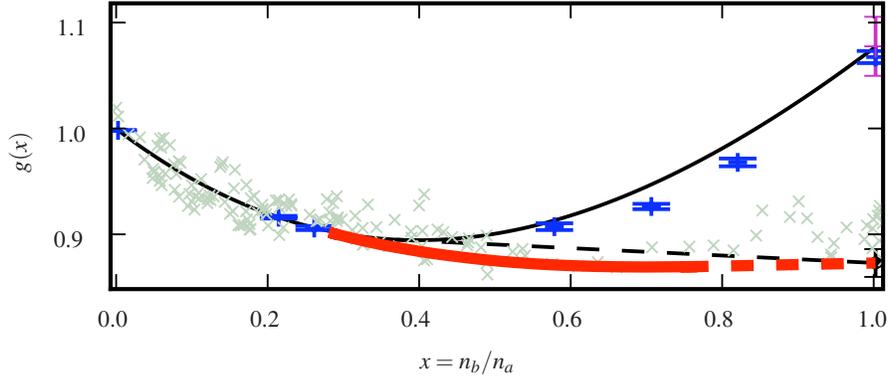}
    \caption{
      \label{fig:g_x_loff}
      The dimensionless convex function $g(x)$ \cite{Bulgac:2006cv}
      that defines the energy density $\mathcal{E}(n_a,n_b) =
      \tfrac{3}{5}\tfrac{\hbar^2}{2m}(6\pi^2)^{2/3}\left[n_{a}
        g(x)\right]^{5/3}$ as a function of the asymmetry $x=n_b/n_a$
      (this plot is very similar to Fig.~\ref{fig:giorgini}). The
      points with error-bars (blue online) are the Monte Carlo data
      from~\cite{PG:2008, CRLC:2007, LRGS:2006}.  The fully-paired
      solution $g(1)=(2\xi)^{3/5}$ is indicated to the bottom right,
      and the recent \MIT\ data~\cite{shin-2008} is shown (light
      $\times$) for comparison.  The phase separation discussed
      in~\cite{PG:2008, CRLC:2007, LRGS:2006, Bulgac:2006cv} is shown
      by the Maxwell construction (thin black dashed line) of the
      first-order transition.  The \LO\ state (thick red curve) has
      \emph{lower energy} than all pure states and phase separations
      previously discussed.  The Maxwell construction of the weakly
      first-order transition between the superfluid and \LOFF\ phase
      is shown by the thick dashed line (red).}
  \end{center}
\end{figure}
\begin{figure}[ht]
  \begin{center}
    \psfrag{Delta(z)}{$\Delta(z)/\Delta_{0}$}
    \psfrag{n(z)/n0}{$n(z)/n_{0}$}
    \psfrag{z/l0}{$z/l_{0}$}
    \psfrag{-1}{\footnotesize -1}
    \psfrag{0}{\footnotesize 0}
    \psfrag{1}{\footnotesize 1}
    \psfrag{0.0}{\footnotesize 0.0}
    \psfrag{0.5}{\footnotesize 0.5}
    \psfrag{1.0}{\footnotesize 1.0}
    \psfrag{-5}{\footnotesize -5}
    \psfrag{0}{\footnotesize 0}
    \psfrag{5}{\footnotesize 5}
    \includegraphics[width=\columnwidth]{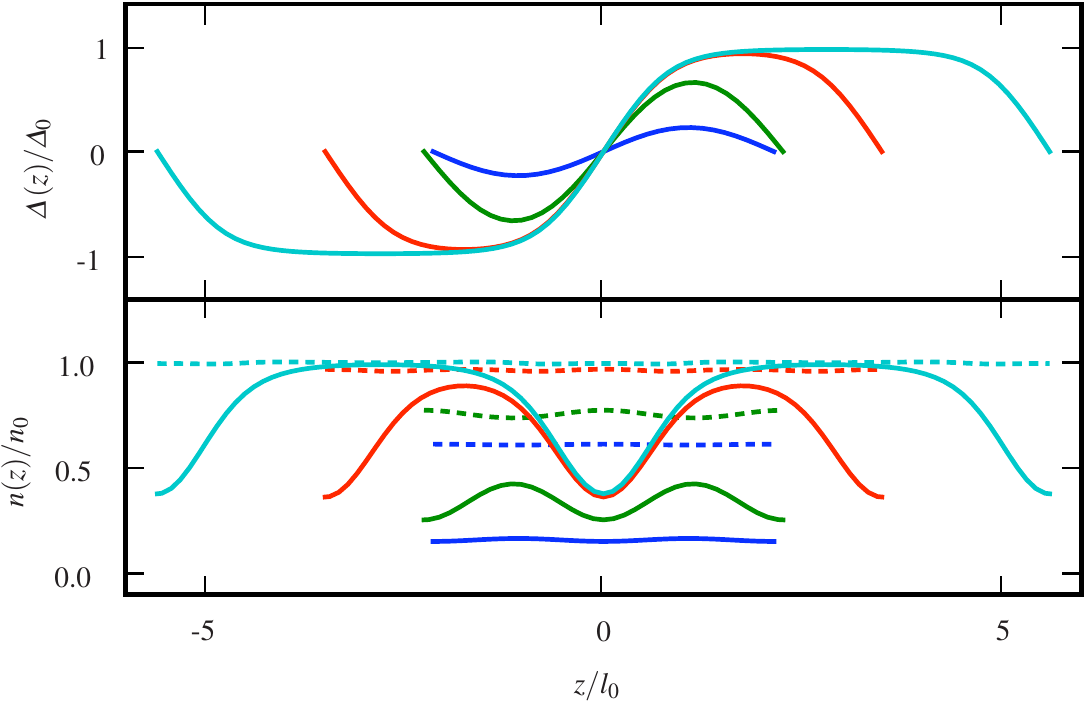}
    \caption{
      \label{fig:delta_x}
      A single \LO\ period showing the spatial dependence of the
      pairing field $\Delta(z)$ (top) and the number densities of the
      majority (dotted) and minority (solid) species (bottom) at the
      values.  Units are fixed so that $\mu_{-} = \mu_{a} - \mu_{b}$
      is held fixed as it is for trapped systems.  We normalize
      everything in terms of the density $n_{0} = n_{a} = n_{b}$,
      interparticle spacing $l_{0} = \smash{n_{0}}^{-1/3}$, and superfluid
      pairing gap $\Delta_{0}$ of the fully paired superfluid at the
      superfluid/\LO\ transition point close to the center of the
      cloud.  At the superfluid/\LO\ transition, the character of the
      solution is that of widely spaced domain walls (see for
      example~\cite{YY:2007}).  As one proceeds outward in the trap,
      the period and amplitude of the solution decreases until it is
      almost sinusoidal at the transition point to the interacting
      normal phase.}
  \end{center}
\end{figure}

The first application of the \ASLDA\ was to consider the energetic
stability of a Larkin-Ovchinnikov--Fulde-Ferrell
(\LOFF)~\cite{FF:1964, LO:1965} polarized superfluid
state~\cite{BF:2008}.  The density functional as constructed naturally
supports a strong first-order phase transition between the fully
paired superfluid state and the interacting normal state (dashed line
in Fig.~\ref{fig:g_x_loff}).  We seed the functional with a periodic
solution of the form shown in Fig.~\ref{fig:delta_x} with a node in
the gap $\Delta(z)$.  Allowing the system to relax to the optimal
period $L$ we find that this Larkin-Ovchinnikov type of solution has
significantly lower energy than the competing pure and mixed phases
over a large range of the phase diagram.

This is a qualitatively new prediction of the \ASLDA: such states are
only meta-stable in the \BdG.  The effect of the self-energy
corrections is to reduce the energy of these states to favor them over
the homogeneous phases.  It is interesting to note that the density
contrast of these states is comparable to the density contrast in
vortices (see Fig.~\ref{fig:vortex}).  Such states have yet to be
observed in experiments: this may be because the physical region in
which the \LO\ state is favoured exists only in a thin shell.  Also,
the one-dimensional structure discussed here will be unstable at any
finite temperature~\cite{Dewel:1979} (see
also~\cite{Radzihovsky:2009kx}) but might be stabilized in traps.
The ground state will most likely be some sort of three-dimensional
lattice structure (see for example~\cite{Bowers:2002xr}) which will
likely require a fairly large physical volume to exists without
significant frustration.  The ideal situation would be a very flat
trap tuned so that the \LO\ region occupies a large physical space at
the center of this trap, however, the construction of such traps
presently poses some experimental difficulties that we hope will be
overcome in the near future.



\section{Time-Dependent Superfluid Local Density Approximation}
\label{cha:time-depend-superfl}

\subsection{Time-Dependent Equations for the Quasiparticle Wave Functions} 

The equations for the time-dependent quasiparticle wave functions
$u_{n,\sigma}(\vec{r},t),\; v_{n,\sigma}(\vec{r},t)$ have the
time-dependent Bogoliubov-de Gennes form
\begin{equation}\label{eq:tddft}
  \I \hbar \frac{\partial}{\partial t}
  \begin{pmatrix}
    u_a \\ 
    u_b \\ 
    v_a \\ 
    v_b
  \end{pmatrix} = 
  \begin{pmatrix}
    h_a+U_a & 0 & 0 & \Delta \\
    0 & h_b+U_b  & -\Delta & 0 \\
    0 & -\Delta^*&  -h_a^*-U_a & 0 \\
    \Delta^*& 0 & 0& -h_b^*-U_b 
  \end{pmatrix}
  \begin{pmatrix}
    u_a \\ 
    u_b \\ 
    v_a \\ 
    v_b
  \end{pmatrix}.
\end{equation}
For the sake of simplicity, we have dropped the arguments
$(\vec{r},t)$ for all functions in these equations. Note also that the
external potentials $U_\sigma(\vec{r},t)$ are real. The only difference
with the static \SLDA\ in the structure of $h_\sigma(\vec{r},t)$ are
the contributions arising from the variation of the current density
correction to the kinetic energy density $\tilde{\tau}(\vec{r},t)$
(\ref{eq:kincur},\ref{eq:new_tau}), which are required by Galilean
invariance to be discussed in Sec.~\ref{sec:galilean-invariance}.  The
chemical potentials $\mu_{a,b}$, which can always be thought of as
external constraints, are implicitly included in
$U_\sigma(\vec{r},t)$.  The chemical potentials can also be removed by
a simple gauge transformation of the quasi-particle wave functions.
It is straightforward to show that these equations conserve the total
particle number for arbitrary time-dependent external fields and also
for arbitrary time variations of the coupling constant $g(t)$. As
expected however, in the presence of an external pairing field,
particle number is not conserved: particles can be exchanged with the
coupled system implied by the source of the external pairing field.

\subsection{Galilean Invariance}
\label{sec:galilean-invariance}

The functionals as expressed in Sec.~\ref{cha:dens-funct-theory} are
not manifestly covariant under Galilean transformation (a subset of
the general coordinate invariance discussed in~\cite{SW:2006} which
restrict the form of higher-order gradient terms).  To restore this
covariance, the currents currents $\vec{j}_{a}(\vec{r})$ and
$\vec{j}_{b}(\vec{r})$ described in~(\ref{eq:Densities}) must be
included.  These vanish in time-reversal invariant ground states, but
are crucial for discussing states that break time reversal and for the
general time-dependent analysis.  In nuclear physics Galilean
covariance have been considered for quite some time~\cite{Engel:1975,
  Dobaczewski:1995, Nesterenko:2008, Bender:2003}, and the
contribution of these currents is often essential for describing the
properties of excited states.

We start by expressing the Galilean invariance of the Lagrangian
density for a single Fermi species (see~\cite{SW:2006} for a more
general discussion)
\begin{equation}
  \mathcal{L} =
  \psi^{\dagger}\left(
    \I\hbar\partial_{t}  - 
    \frac{(-\I\hbar \vec{\nabla})^2}{2m}
  \right)\psi.
\end{equation}
This is invariant under the following Galilean transformation:
\begin{align}
  \psi(\vec{x}, t) &\rightarrow 
\exp\left [-\I\left ( \tfrac{1}{2}m \abs{\vec{v}}^2 t 
    + m\vec{v}\cdot\vec{x}\right)/\hbar\right ]
  \psi(\vec{x} + \vec{v}t, t) .
\end{align}
From this, we see that the currents and kinetic densities transform as
\begin{subequations}
  \label{eq:Galilean_j}
  \begin{align}
    \vec{j} = \tfrac{\I}{2}\psi^{\dagger}\vec{\nabla}\psi + \text{h.c.}
    &\rightarrow
    \vec{j} + m\vec{v}n,\\
    \tau = \tfrac{1}{2m}\vec{\nabla}\psi^{\dagger}\vec{\nabla}\psi 
    &\rightarrow
    \vec{\tau} + \vec{v}\cdot\vec{j} + \tfrac{1}{2}m\abs{\vec{v}}^2n.
  \end{align}
\end{subequations}
It follows directly that for a two-component system, the following
combinations are Galilean invariant:
\begin{align}
  \tilde{\tau}_{\sigma} &= \tau_{\sigma} 
  - \frac{\abs{\vec{j}_{\sigma}}^2}{2m_{\sigma} n_{\sigma}}, &
  &\frac{\vec{j}_{b}}{m_{b}n_{b}} - \frac{\vec{j}_{a}}{m_{a}n_{a}}.
\label{eq:kincur}
\end{align}
We would like to separate out the center of mass motion from the
intrinsic functional, so we introduce the total mass current and density:
\begin{align}
  \label{eq:jp_mp}
  \vec{j}_{+} &= \vec{j}_{a} + \vec{j}_{b}, &
  \rho_{+} &= m_{a}n_{a} + m_{b}n_{b}.
\end{align}
We may then write the functional in the following way:
\begin{equation}
  \mathcal{E} = \frac{\abs{\vec{j}_{+}}^2}{2\rho_{+}} 
  + \tilde{\mathcal{E}}.
\end{equation}
The first term captures the energy of the center of mass motion and
$\tilde{\mathcal{E}}$ describes the remaining intrinsic energy of
the system, and should be strictly Galilean invariant.

Excited states may be described by an extension of the \DFT\ method
commonly referred to as Time-Dependent Density Functional Theory
(\TDDFT)~\cite{Rajagopal:1973rt, Peuckert:1978fr, Runge:1984mz,
  tddft}.  This theory describes the evolution of the one-body number
density in the presence of an arbitrary one-body external field.  As
in the case of static \DFT, one can prove an existence
theorem~\cite{Rajagopal:1973rt, Peuckert:1978fr, Runge:1984mz}.  This
states that a functional exists from which one can determine the exact
time-dependent number density for a given quantum system, and can be
expressed in the form
\begin{multline}
  \label{eq:tdslda}
  \mathcal{S} = \frac{\I\hbar}{2}\int \D{t} \D^3r \sum_{n,\sigma}\left\{ 
    v_{n,\sigma}(\vec{r},t)\frac{\partial v^*_{n,\sigma}(\vec{r},t)}{\partial t}
    -v^*_{n,\sigma}(\vec{r},t)\frac{\partial
      v_{n,\sigma}(\vec{r},t)}{\partial t}\right\}\\
  -\int \D{t} \D^3r \Biggl\{\frac{\hbar^2}{2m}\sum_\sigma
    \tau_\sigma(\vec{r},t) 
    + \sum_\sigma U_\sigma(\vec{r},t) n_\sigma(\vec{r},t)\\
    + \mathcal{E}\left [ n_a(\vec{r},t),\tilde{\tau}_a(\vec{r},t),
      n_b(\vec{r},t),\tilde{\tau}_b(\vec{r},t)
      ,\abs{\nu(\vec{r},t)}^2,g(\vec{r}, t) \right ]  
  \Biggr\}.
\end{multline}
Here $\sigma=a,b$ labels the two fermion species. The existence
proof for superfluid systems is analogous to the proof for normal
systems~\cite{Rajagopal:1973rt, Peuckert:1978fr, Runge:1984mz}. Here
$U_\sigma(\vec{r},t)$ are arbitrary time-dependent one-body external
fields, which couple to the conserved number densities of the fermion
species $n_\sigma(\vec{r},t)$.  These external fields can represent
couplings to the laboratory environment, such as a trapping potential,
which can be used to manipulate and study these systems.

The direct coupling of an external gauge field to the electric charge
and magnetic moments of the particles can also be incorporated in a
straightforward manner, by the usual process of converting the global
particle number symmetry to a local symmetry by invoking the principle
of gauge invariance.  We can also couple an arbitrary time-dependent
external pairing field as well to represent interactions with another
superfluid system brought into the proximity of the system under
study.  As mentioned above, this will violate the conservation of
particle number as particles are now able to be exchanged with the
other system.  Finally, the last argument $g(\vec{r}, t)$ of the
interaction term $\mathcal{E}$ represents the possibility of varying
the coupling constants in space and time.  In particular, as discussed
in Sec.~\ref{sec:from-phys-probl}, by means of a Feshbach resonance
an external magnetic field can be used to directly control the
scattering length, providing yet another handle to manipulate and
study these systems.

In the functional $\mathcal{S}$ we have separated the kinetic energy
$\hbar^2 \sum_\sigma \tau_a(\vec{r},t)/2m$ from the interaction
energy in order to disentangle the dependence on the reference
frame. The interaction energy encoded in the functional ${\cal E}$
should be independent of the motion of the system as a whole.  By
default, the properties of the ground states of a physical system are
typically discussed in the center of mass reference frame. When the
system is excited by various external probes, inevitably currents
appear. In the \LDA\ it is natural to assume that the energy density
separates into the kinetic energy of center of mass (which depends
only on its local center of mass velocity and its corresponding mass)
and the internal energy (which should not depend on the local center
of mass velocity). The energy density $\mathcal{E}\left [
  n_a(\vec{r},t),\tilde{\tau}_a(\vec{r},t), n_b(\vec{r},t),\tilde{\tau}_b(\vec{r},t),\abs{\nu(\vec{r},t)}^2, g(\vec{r}, t)
\right ]$ is the same as in the static \SLDA, with the only difference
that the dependence on the modified kinetic energy density
$\tilde{\tau}_\sigma(\vec{r},t)$ now includes the current
densities~(\ref{eq:nu_dens}) to satisfy Galilean invariance
\begin{equation}
  \vec{j}_{\sigma}(\vec{r},t) =
  \frac{\I\hbar}{2}\sum_n \left[
    {\vec{\nabla}}v_{n,\sigma}(\vec{r},t) v^*_{n,\sigma}(\vec{r},t)
    - v_{n,\sigma}(\vec{r},t){\vec{\nabla}} v^*_{n,\sigma}(\vec{r},t)
  \right].
\end{equation}
Here we will describe a slightly different philosophy in implementing
the Galilean invariance than discussed at the beginning of this
section, which leads to a somewhat different definition of the
modified kinetic energy densities $\tilde{\tau}_\sigma(\vec{r},t)$
than those introduced in~(\ref{eq:kincur}) above. This ambiguity
illustrates the freedom one has in introducing currents and using no
other restriction except Galilean invariance.
  
Upon boosting the system to a frame with a velocity $\vec{V}$, the
current density changes $\vec{j}_\sigma(\vec{r},t)\rightarrow
\vec{j}_\sigma(\vec{r},t)+mn(\vec{r},t)\vec{V}$.  We introduce the
velocity of the local center of mass frame ($m_a=m_b=m$)
\begin{equation}
  \vec{V}(\vec{r},t)= \frac{ \vec{j}_{+}(\vec{r},t)}{\rho_{+}(\vec{r},t)},
\end{equation}
where we have introduced the total current $\vec{j}_{+}$ and density
$\rho_{+}$ from~(\ref{eq:jp_mp}).  Consequently, the following
combination of the kinetic energy density, current density and number
density
\begin{equation}
  \tilde{\tau}_\sigma(\vec{r},t)=\tau_\sigma(\vec{r},t)
  -\vec{j}_\sigma(\vec{r},t)\cdot\vec{V}(\vec{r},t)
  +\frac{mn_\sigma(\vec{r},t)\vec{V}^2(\vec{r},t)}{2}, \label{eq:new_tau}
\end{equation}
renders the energy density locally Galilean
invariant~\cite{Bulgac:2007a}.  $\tilde{\tau}_\sigma(\vec{r},t)$ is
therefore the internal kinetic energy density in the local center of
mass frame, which is different from the form of modified kinetic
energy introduced in~(\ref{eq:kincur}). The difference between the two
approaches to enforcing the Galilean invariance amounts to terms
proportional to $\abs{\vec{j}_{b}/m_{b}n_{b} -
  \vec{j}_{a}/m_{a}n_{a}}^2$, see~(\ref{eq:kincur}).
 
It is worth noticing that because the Galilean invariance is built in,
one of the famous relations in the Landau's Fermi liquid theory
linking the effective mass of the quasiparticles with the $p$-wave
interaction (denoted $F_1$) is automatically satisfied
(see~\cite{AGD:1975}).

Note also that if terms arise of the form $\vec{j}_a(\vec{r},t)\cdot
\vec{j}_b({\bf r},t)$, a new physical effect appears whereby the
local velocity of one species depends also on the velocity of the
other species.  In other words, the inverse mass becomes a tensor in
the spin (``isospin'') space. By including terms of the form
$\abs{(\vec{j}_a(\vec{r},t)\cdot\nabla n_\sigma(\vec{r},t))}^2$, the
effective mass becomes a tensor in real space.  This was discussed
in~\cite{Bulgac:1995} in connection with the construction of the
optimal local Schr\"odinger equation to represent a non-local
equation. In particular, it seems that, in order to describe some
rather subtle level orderings of the single-particle spectrum found in
the a non-local Schr\"odinger equation, one needs a tensor effective
mass in the local equations.  This is also related to the discussion of
superfluid mixtures, where it was observed long ago that one superfluid
can drag the other one without any dissipation: the Andreev-Bashkin
effect~\cite{Andreev:1975}. Similar effects arise when one considers
the terms induced by Galilean invariance~(\ref{eq:new_tau}) or
$\vec{j}_a(\vec{r},t)\cdot {\bf j}_b(\vec{r},t)$, when the
 presence of a current of one species induces a current of the
other species.

\subsection{The Excitation of the Pairing Higgs Mode}

We shall illustrate the power of the Time-Dependence \SLDA (\TD-\SLDA)
by examining the response of a superfluid unitary gas to the time
variation of the scattering length~\cite{Bulgac:2009}. This problem
has been studied extensively in the weak coupling regime when
$k_F\abs{a}\ll 1$ and $a<0$, see~\cite{Volkov:1974, Barankov:2004,
  Barankov:2006, Andreev:2004, Szymanska:2005, Tomadin:2008,
  Barankov:2006a, Warner:2005, Yi:2006, Nahum:2008, Teodorescu:2006,
  Robertson:2007, Yuzbashyan:2005a, Dzero:2007, Yuzbashyan:2005,
  Yuzbashyan:2006, Yuzbashyan:2008, Yuzbashyan:2006a, Dzero:2009}.
The initial state of the system will be the ground state, and at
subsequent times, the evolution will be adiabatic in the sense that no
entropy production is allowed. To some extent this is a rather strong
limitation of this time-dependent description of the quantum
evolution, a restriction which can be lifted if one would consider a
further extension of the formalism, the Stochastic
\TD-\SLDA~\cite{Bulgac:2010ul} which will not be discussed here.

\begin{figure}[h]
  \includegraphics[width=0.5\textwidth]{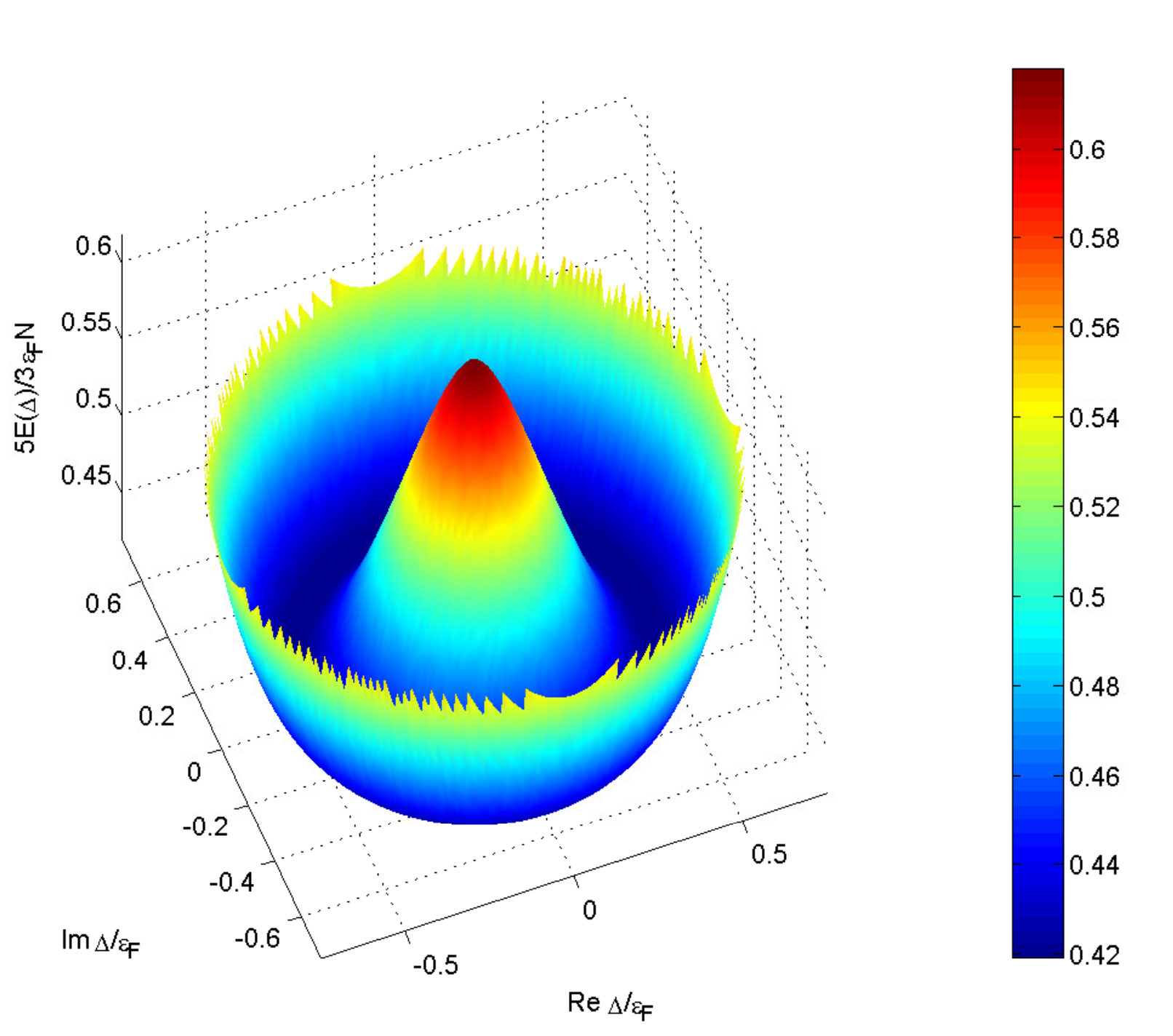}
  \includegraphics[width=0.5\textwidth]{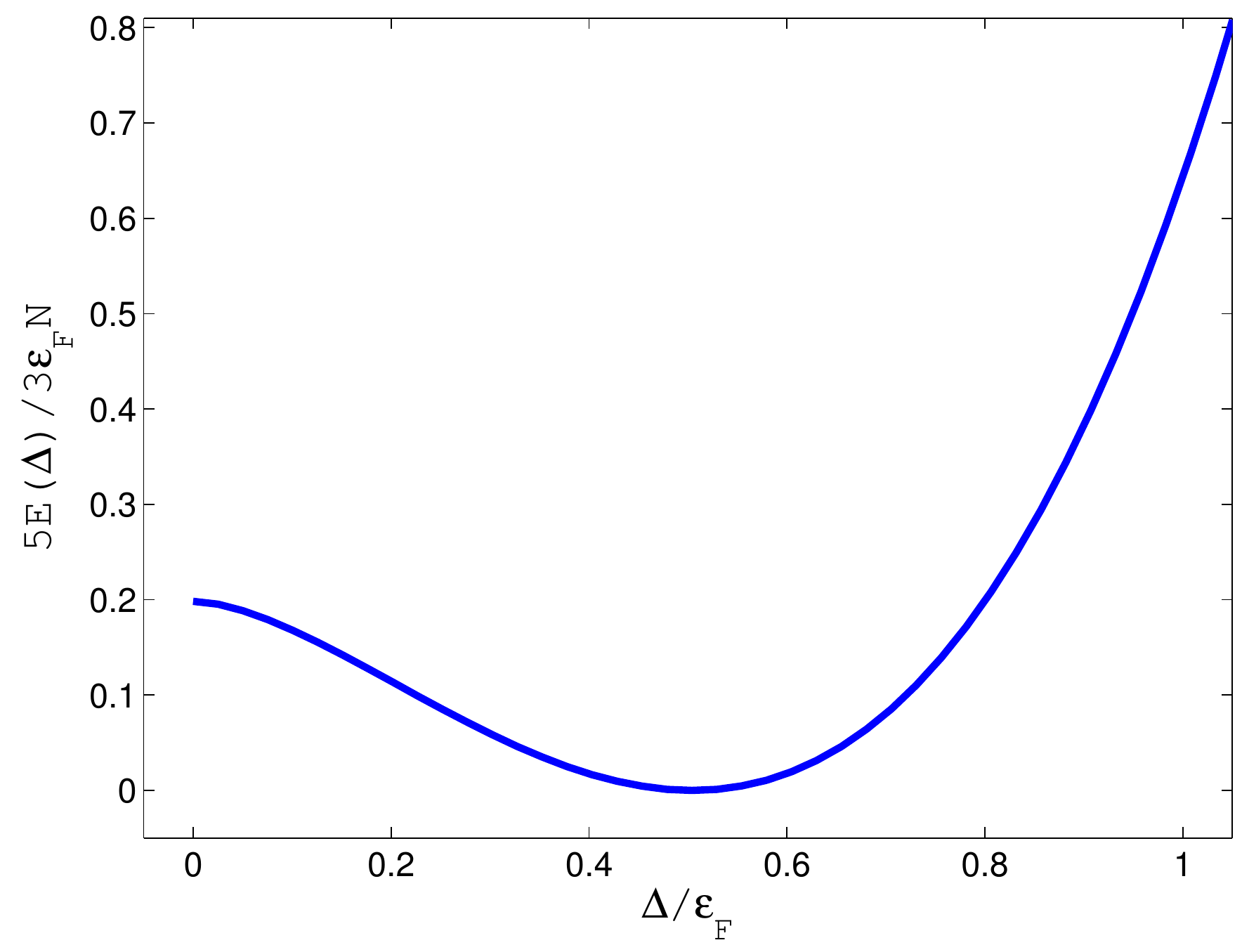}
  \caption{The profile of the energy of a unitary Fermi gas as a
    function of the pairing gap with respect to the energy of the
    ground state. One would na\"\i{}ely expect that this system if
    released from a point almost at the tip of the ``Mexican hat''
    will roll down along the radial direction, past the equilibrium
    value $\Delta_0= 0.5\varepsilon_F$ and oscillate indefinitely back
    and forth. \label{fig:mx_hat} }
\end{figure}

\begin{figure}[h]
  \includegraphics[width=\textwidth]{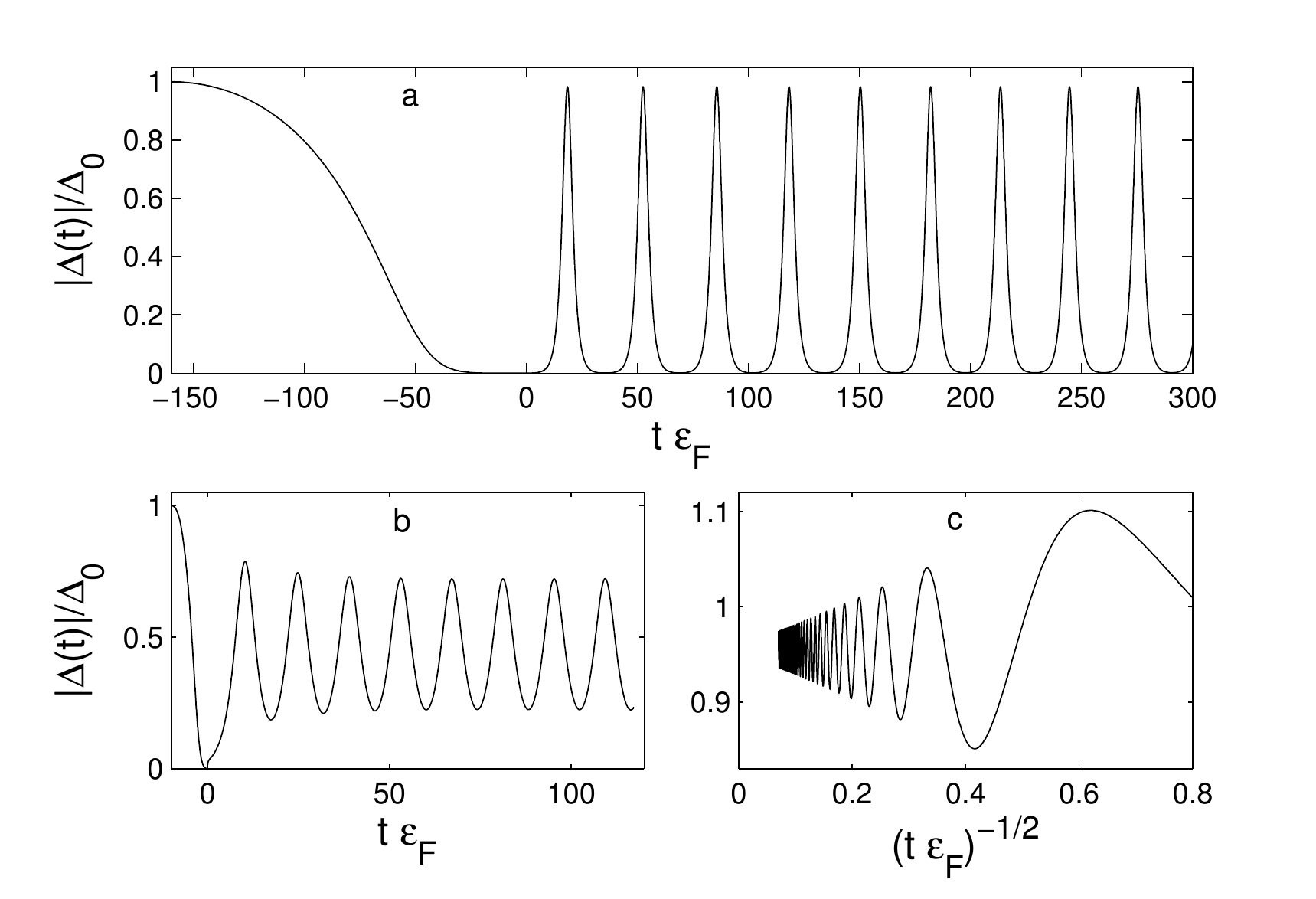}
  \caption{ The panels $a, b$ and $c$ display response of the
    homogeneous system to an initial switching time interval
    $t_0\varepsilon_F = 160, 10$ and 160 and values of the gap
    corresponding to $\gamma_s$ are $\gamma_s/\gamma =0.005, 0.005$
    and $0.5$ respectively, where $\gamma$ is the coupling constant
    controlling the magnitude of the pairing gap and $\Delta_0\approx
    0.5\varepsilon_F$ is the gap equilibrium value, both at unitarity.
    The panels $a$ and $b$ show that when the system is released from
    the neighborhood of the tip of the ``Mexican hat'' potential the
    pairing gap oscillates back and forth, but never past the
    equilibrium value $\Delta_0$. At the same time the system will
    rotate around the origin as the phase of the pairing field (not
    shown here) will monotonically evolve in time as well. However,
    when the system is released from an initial position closer to the
    minimum at $\Delta_0$ the oscillation is damped, $ \Delta (t) =
    \Delta_\infty+A\sin( 2\Delta_\infty t +\phi)/\sqrt{\Delta_\infty
      t}$, with a mean frequency $2\Delta_\infty$ and around a value
    smaller than the equilibrium $\Delta_\infty< \Delta_0$, a behavior
    which was first studied in~\cite{Volkov:1974} in the \BCS\
    limit, when the coupling is weak.
    \label{fig:modes} }
\end{figure}

Consider the following scenario~\cite{Bulgac:2009}: start with a
homogeneous unitary Fermi gas in its ground state.  At first slowly
reduce the coupling constant $\gamma$ from its unitary value to a very
small but still negative value.  If this change is slow enough, then
the system tracks the ground state into the ground state of the system
with an exponentially small pairing gap.  Now rapidly ramp the
coupling $\gamma$ back to its unitary value and let the system evolve.
This essentially looks at the evolution of an almost normal system
with the unitary \DFT. The behavior shown in Fig.~\ref{fig:modes} is
rather surprising.

Many approaches have been developed to describe the dynamics of a
fermionic superfluid at or near $ T=0$ including superfluid
hydrodynamics, a Landau-Ginzburg or Gross-Pitaevskii (\LGGP) like
description, and effective field theory, see~\cite{Babaev:2009,
  Babaev:2009b, Babaev:2009a, Nishida:2006, SW:2006, Rupak:2008fk,
  Rupak:2007, giorgini-2007, Khalatnikov:2000}. The common thread in
all these approaches is the desire to identify a significantly smaller
set of relevant degrees of freedom, and achieve an accurate
description of the phenomena within a reduced framework. As a rule,
when reducing the number of the degrees of freedom, one assumes that
the system evolves through states where local equilibrium is
maintained.  In this instance, one would na\"\i{}vely expect that the
system dynamics are governed by an effective ``Mexican hat''
potential, Fig.~\ref{fig:mx_hat}, representing the energy of the
system as a function of the complex pairing field.  The system is
brought adiabatically from the minimum of the potential to almost the
``tip of the Mexican hat'', and released with zero initial
``velocity''.  The na\"\i{}ve picture is that the system will ``roll''
down along the radial direction accelerating until it reaches the
minimum of the potential, pass through the minimum, and oscillate back
and forth along the ``radial'' direction without damping.

One might also inspect the \LGGP\ description of the dynamics
of the system using the nonlinear Schr\"{o}dinger equation
\begin{equation}
  \I\hbar \frac{\partial \Psi (\vec{r},t)}{\partial t}=
  -\frac{\hbar^2\Delta\Psi (\vec{r},t)}{4m}+
  \mathcal{U}(\abs{\Psi (\vec{r},t)}^2)\Psi (\vec{r},t).
\end{equation}
Since there are no spatial gradients in this system (we have changed
the coupling in a uniform manner so as not to break the translational
invariance), only the second term on the right hand side of this
equation contributes, and the solution is a simple monotonic evolution
of the condensate phase $\Psi (\vec{r},t)$ in time and the magnitude
of ``wave function'' $\Psi (\vec{r},t)$ remains constant. In lieu of a
better simple alternative, many authors have used this approach to
characterize dynamics of Fermi superfluids at essentially zero
temperatures, even though the \LGGP\ description is only justified
near the critical temperature.

Another approach is to use the zero-temperature limit of Landau's two
fluid hydrodynamics, which reduces to the following two equations at
zero temperature
\begin{align}
  \dot{n}(\vec{r},t) + \nabla\cdot [\vec{v}(\vec{r},t) n(\vec{r},t)]
  &= 0,&
  m\dot{\vec{v}}(\vec{r},t)+ \nabla \left\{
    \frac{m{\vec{v}}^2(\vec{r},t)}{2m} 
    +\mu[n(\vec{r},t)]\right \} &=0.
\end{align}
Here $\vec{v}(\vec{r},t)$ is the hydrodynamic velocity and
$\mu[n(\vec{r},t)]$ is the local thermodynamic potential. Since there
are no spatial gradients, these two equation simply predict that the
number density will remain constant and nothing else will happen.

\begin{figure}[h]
  \includegraphics[width=\textwidth]{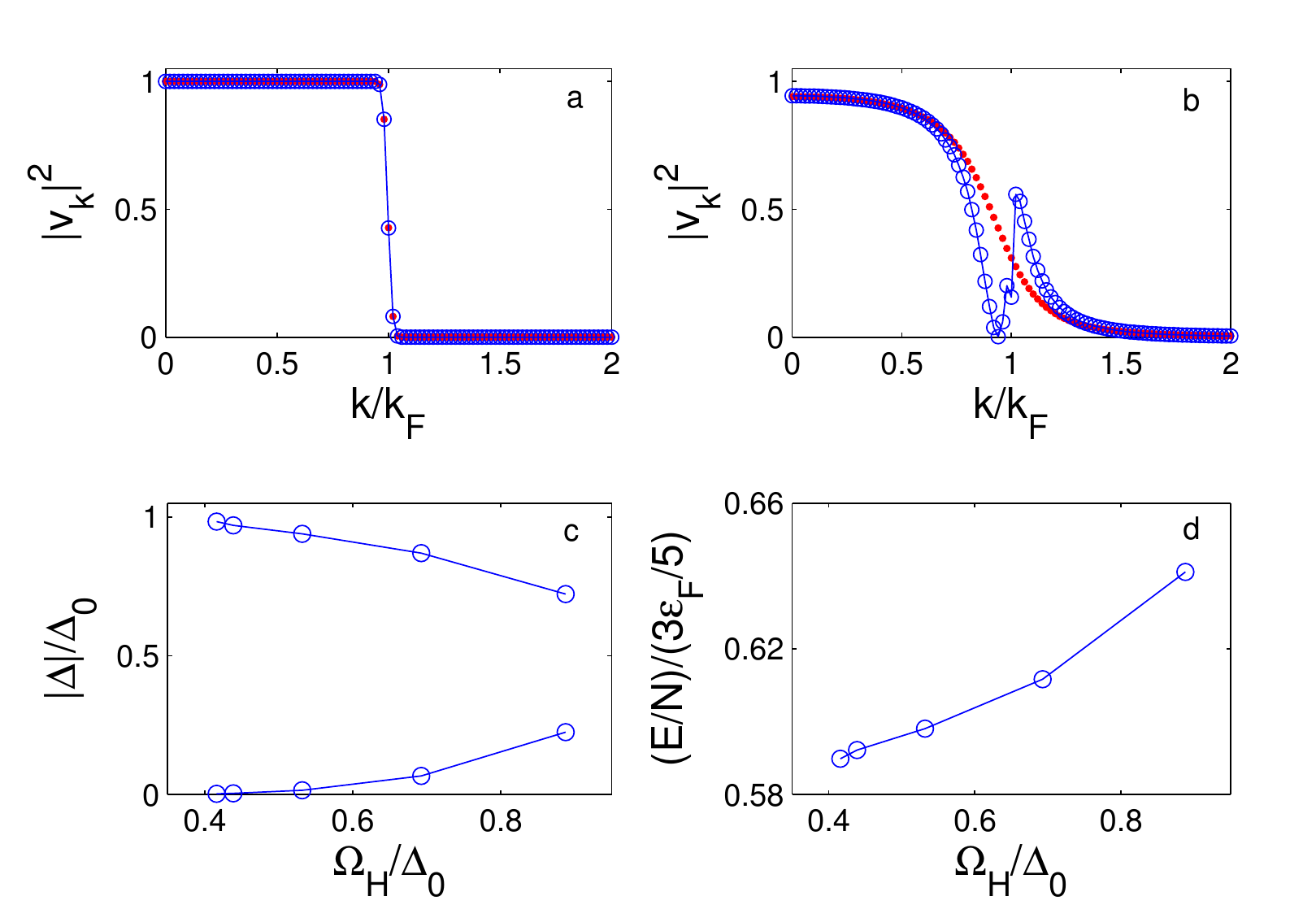}
  \caption{Panels $a$ and $b$  display the instantaneous occupation
    probabilities of the mode shown in upper panel of
    Fig.~\ref{fig:modes} corresponding to times $t>0$ when the pairing
    field is at its minimum and maximum values respectively with circles
    joined by a solid (blue with circles) line. With (red) dots we
    plotted the equilibrium occupation probabilities corresponding to
    the same instantaneous values of the pairing gap. In  panels $c$
    and $d$ we show the maximum and minimum values of the oscillating
    pairing field and the corresponding excitation energy as a function
    of the frequency of the Higgs-like modes, see Fig.~\ref{fig:modes}
    $a$ and $b$. 
    \label{fig:occup}  }
\end{figure}

Apart from the fact that the number density will remain constant and
spatially uniform, these three different na\"\i{}ve pictures lead to
drastically different predictions. The actual time evolution of the
system, shown in Fig.~\ref{fig:modes}, is qualitatively different from
any individual picture, but demonstrates a combination of the expected
features.  The pairing gap does increase from almost zero towards the
equilibrium value, and it oscillates, but it never crosses the minimum
equilibrium value $\Delta_{0}$ of the ``Mexican hat'' potential. At
the same time, the phase of the pairing gap increases monotonically
 in time and the number density is constant.

By preparing the initial state slightly differently one can excite
different types of these modes that have been dubbed ``Higgs'' modes of
the pairing field. One can vary the upper and the lower values between
which the pairing field oscillates, and also adjust the period of these
oscillations.  It is remarkable, however, that the frequencies of
these modes are always smaller than $2\Delta_0$, where
$\Delta_0$ is the equilibrium value of the pairing gap at unitarity,
even though the excitation energy is large. These are indeed very
collective excitations of the system, of extremely low frequency, but
with an excitation energy per particle significantly less than pairing
gap.

It is still an unresolved question of how these modes will eventually
decay and how the system will thermalize.  It is also instructive to
examine the time dependent population of the various single-particle
momentum states of these collective modes as shown in
Fig.~\ref{fig:occup}. When the value of the pairing gap is very small
the occupation probabilities are essentially those of a system in
equilibrium. However, when the system reaches a pairing gap
essentially equal to the equilibrium value $\Delta_0$, the occupation
probabilities are clearly very different from those in the ground
state, which clearly points to a non-equilibrium state. This aspect
should clarify why neither \LGGP\ nor quantum hydrodynamics are valid
as both assume local equilibrium is maintained.

\subsection{Generation and Dynamics of Vortices} 

A number of results concerning the generation and dynamics of vortices
in a unitary Fermi gas by an external time-dependent perturbation can
be found at ~\cite{Bulgac:2011b}.  As far as we are aware, this problem has
been studied in one paper for a pure 2D systems~\cite{Tonini:2006}. As
in the previous example, we do not yet consider entropy production in these
simulations.

In order to illustrate further the power of the \TD-\SLDA\ as well as
the limitations of traditional approaches such as superfluid
hydrodynamics or a \LGGP\ analysis, we now consider the
quantum dynamics of a stirred unitary Fermi gas~\cite{Bulgac:2011b}.  We
start with the gas in its ground state in a cylindrical trap, uniform
and with periodic boundary conditions in the third spatial
direction. We then subject the system to a time-dependent external
stirring field which breaks the cylindrical symmetry.  When
implemented numerically~\cite{Bulgac:2008}, if one places the system
on a spatial lattice with $N_s$ spatial lattice points in one
direction, one can show that the size of the problem scales as
$\propto N_s^5$. When the limitation of spatial homogeneity in the
$z$-direction is lifted the size of the problem scales as $\propto
N_s^6$, which as a rule requires an implementation on the largest
leadership class supercomputers available. For example, if $N_s=50$ an
efficient solution of the \TD-\SLDA\ equations becomes possible only
on the JaguarPF Cray XT5,\footnote{JaguarPF is a Cray XT5
  supercomputer with 224,256 processing cores, see
  \href{http://ww.nccs.gov}{http://ww.nccs.gov}.} which we are
currently utilizing to its full extent.

When homogeneity along the $z$-direction is enforced, the
quasiparticle wave functions have the structure $(u_n(x,y,t)\exp(\I
kz),v_n(x,y,t)\exp(\I kz))$ while self-energy $U(x,y,t)$ and the
pairing potential $\Delta(x,y,t)$ are translationally invariant along
$z$.  We adiabatically introduce a vertical rod into this ``soup can''
and stir the gas with a constant angular velocity. One can vary both the
stirring radius $R$ and stirring angular frequency $\omega$ to control
the speed $v_{\text{rod}}=R\omega$ of the rod. 

One might expect that if $v_{\text{rod}} \ll v_c$, where $v_c$ is the
critical velocity of a unitary Fermi gas, then the system will return
to its initial state after the stirring is turned off.  However, if
$v_{\text{rod}} > v_c$, then one might destroy the superfluid order,
resulting in a normal Fermi gas.  If $v_1< v_{\text{rod}} < v_c$,
where $v_1$ is some minimal stirring velocity, one expects that
vortices will be created.  Unfortunately, none of the simple theories
can shed much light on the outcome: superfluid hydrodynamics cannot
describe quantum vortices as there is no intrinsic quantization or
Planck's constant in its formulation: vortex quantization must be
imposed by hand, and there is nothing in principle to prevent decay of
a quantized vortex into two fractionally quantized vortices.  The time
dependent \LGGP\ approach will also fail to describe the
normal state and the transition from superfluid to normal state, as it
is formulated explicitly in terms of the order parameter alone, which
vanishes in the normal state.  Thus, it seems that the only viable
solution is to forgo a reduction in the degrees of freedom and deal
directly with the quasiparticles included in the \DFT.

\begin{figure}[t]
  \includegraphics[width=\textwidth]{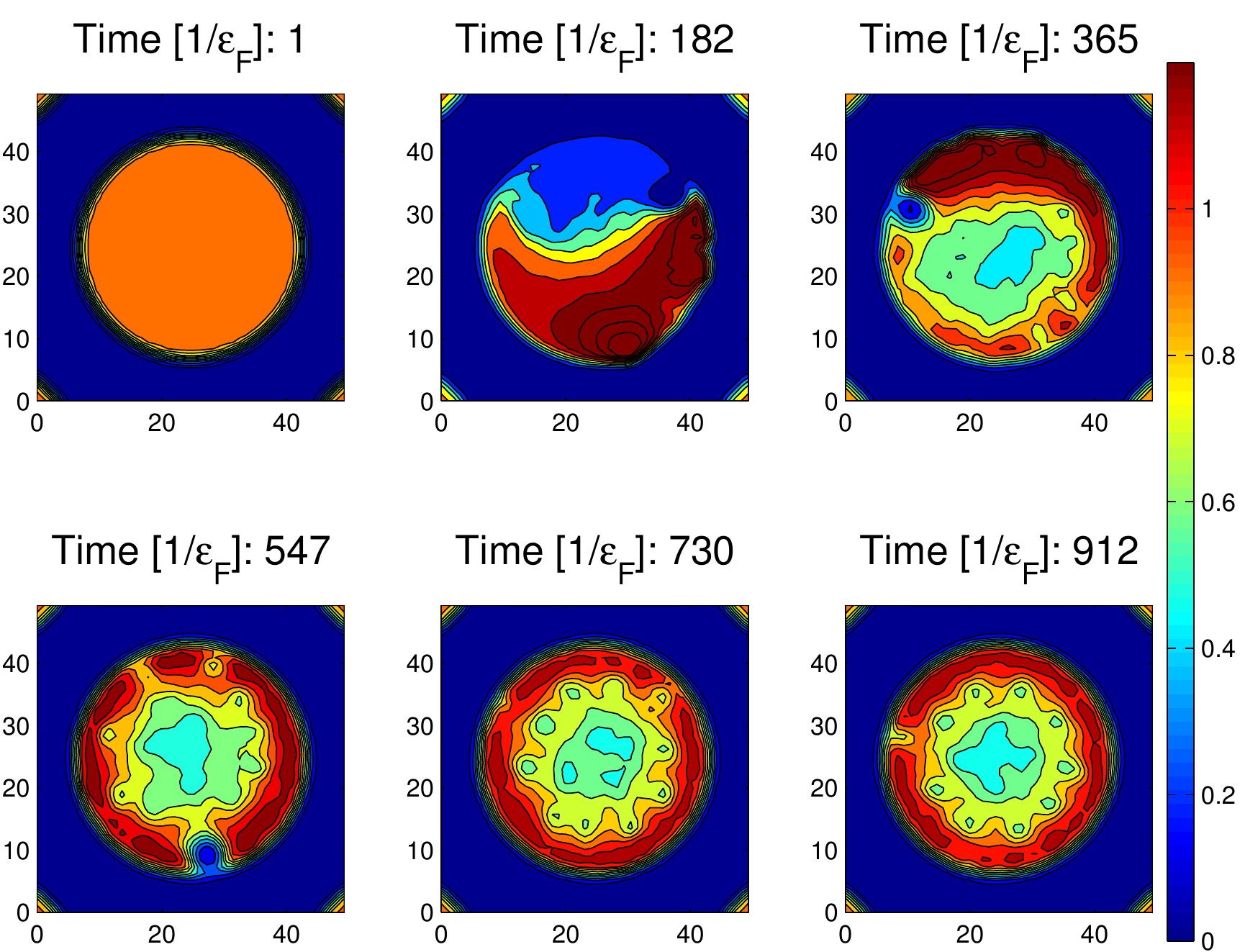}
  \caption{The contour density profiles of a unitary Fermi gas in a
    cylindrical container, stirred with a uniformly rotating rod,
    which is inserted and extracted adiabatically from the system. The
    position (and intensity) of the rod can be inferred as the deepest
    density depletion in the system, and it is actually visible only
    in panels 2-4. Initialy the gas shows an almost uniform density
    distribution, and subsequently it is gathered almost entirely in
    front of the stirring rod. The magnitude of the density is in
    units of the unperturbed central initial density of the cloud and 
    the colorbar on the right decodes the meaning of each color used.
    By the end of the simulation there are 13 vortices forming an
    almost perfect triangular Abrikosov lattice in this confined
    geometry.
    \label{fig:vortdens} }
\end{figure}

\begin{figure}[t]
  \includegraphics[width=\textwidth]{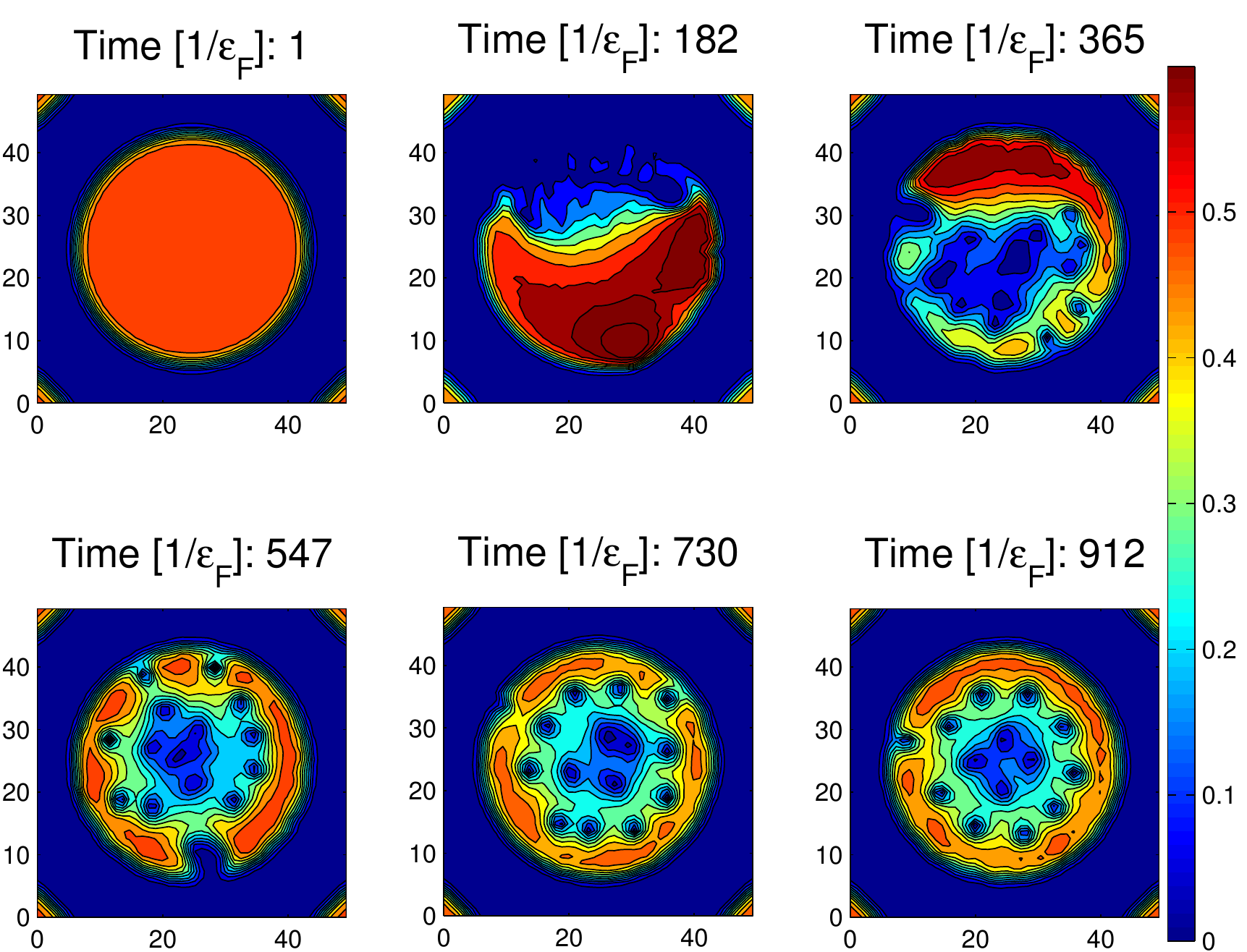}
  \caption{The corresponding contour profiles of the pairing field
    $\abs{\Delta(x,y,t)}$ a unitary Fermi gas in a cylindrical container,
    stirred with a uniformly rotating rod. A plot (not shown here) of
    the phase of the pairing field $\arg \Delta(x,y,t)$ shows that as
    one circles a vortex core the phase changes by $2\pi$, thus each
    of these vortices carries exactly $\hbar/2$ units of angular
    momentum per particle and both the number (normal) density and the
    pairing field are significantly depleted in the core of the vortex
    \cite{BY:2003}.
  \label{fig:vortdelt}  }
\end{figure}

A unitary Fermi gas is a special system in quite a number of ways: in
particular, it appears to have the highest critical velocity of all
known superfluids~\cite{Sensarma:2006, Combescot:2006}. On the \BCS\
side of the Feshbach resonance, if stirred fast enough, the system can
loose superfluidity by the breaking of the Cooper pairs $v_{qp}=\min
(E_k,k)$.  On the \BEC\ side of the Feshbach resonance, the dominant
mechanism for the loss of superfluidity is the excitation of the
Anderson-Bogoliubov sound modes $c=v_F\sqrt{\xi/3}$. In the a unitary
Fermi gas, these two different critical velocities appear to be
essentially identical, and exactly at unitarity one obtains
\begin{equation*}
  \label{eq:vcrit} 
  v_c = 
  \min(c,v_{qp})= v_F\min \left ( \sqrt{\frac{\xi}{3}}, 
    \sqrt{ \alpha \sqrt{(\overline{\beta}-\xi)^2+\eta^2} 
      +(\overline{\beta}-\xi) }  \right )
  \approx 0.365 v_F.
\end{equation*}
Since the amount of information one extracts in a \TD-\SLDA\
simulation of this type is very large, it is not sufficient to display
only a few pictures such as those in Figs.~\ref{fig:vortdens}
and~\ref{fig:vortdelt}.  We invite the interested reader to explore
some of the movies made of these simulations~\cite{Bulgac:2011b}. We shall
comment here only on a few selected aspects of these results, most of
which will be prepared for a publication at a later time.

Our expectation that, under gentle stirring, the unitary Fermi gas
will return to its initial superfluid state is supported by the
simulations~\cite{Bulgac:2011b}, and is in line with how one would expect a
superfluid to respond to such an external probe.  The other
expectation, that vigorous stirring can destroy the superfluid order
is also confirmed.  Within \TD-\SLDA, the dynamic generation of
vortices as well as formation of the celebrated Abrikosov vortex
lattice are also readily demonstrated. By varying shape, the number
and the stirring velocity we generated a plethora of quantized
vortices in this ``soup can'' of superfluid unitary Fermi
gas~\cite{Bulgac:2011b}.

While we expected to generate a relatively small number of
vortices at low stirring velocity, and that the
number of vortices will increase with more vigorous stirring, many of
the features of dynamic vortex generation are quite surprising. The
fact that this system is compressible results in surprisingly large
time-dependent variations of the local number density. Often the entire mass
of the system is gathered in front of the stirrer, leaving little
matter behind it:  The gas can occupy less than half of the available
volume, even though the volume excluded by the stirrer is quite
small. It also comes as a great surprise that the system does not loose
quantum coherence under such a violent perturbation.  Moreover, it
organizes itself in an almost perfect vortex lattice after the
stirring is turned off. Even more surprising is that the
system remains superfluid, \emph{even when stirred at supercritical
speeds!}  We have observed that the system forms a vortex lattice even
if stirred with speeds up to $v_s=0.60 v_F >v_c\approx 0.365 v_F$ (see
the case of 7 vortices with a large radius stirrer at
\cite{Bulgac:2011b}). We attribute this behavior to the fact that an increased
density of the cloud during the stirring process corresponds to an
increased local critical velocity, since the local Fermi velocity
increases as well accordingly.

These two cases of exciting and monitoring the unitary Fermi gas by
two drastically different methods illustrate both the power and
flexibility of this framework, as well as the richness of the
phenomena waiting to be fully explored.  One potential topic to be
explored by these techniques that has mesmerized the low temperature
community during the past few decades is quantum
turbulence~\cite{Vinen:2007, Vinen:2006, Tsubota:2008}.  Hopefully
this can also be replicated in experiments with cold atomic fermionic
gases. Due to the complexity of the full 3D time-dependent
Bogoliubov-de Gennes equations, this aspect has never been
theoretically addressed for fermionic systems. The \TD-\SLDA\ appears
as a framework of choice in this respect. In particular one can
address on a fully microscopic basis for vortex reconnection dynamics,
which is likely at the heart of quantum turbulence at zero
temperature, where dissipative processes are greatly inhibited.


\section{Concluding Remarks}

We have reviewed here three methods to describe the properties of
many-body systems starting from the bare Hamiltonian, and building a
practical framework for studying nontrivial properties of mesoscopic
systems and quantum dynamics.

The \QMC\ method is particularly suited to calculate properties of the
homogeneous phase of matter in an unbiased fashion.  It also can be
used for inhomogeneous systems, but is limited by system size and can
not handle large number of Fermions.  It is also generally plagued by
the infamous sign problem (except in exceptionally symmetric contexts)
and so far has not been used to describe systems in the time dependent
domain.

The complimentary approach of density functional theory (\DFT) through
the use of the \SLDA\ and \ASLDA\ can be applied to extend these
results to mesoscopic systems with larger number of particles and a
wide variety of geometries.  The time dependent \TDDFT\ (\TD-\SLDA)
extension brings these techniques to bear on time-dependent quantum
dynamics. The main difficulty with the \DFT\ is that there is no well
defined procedure to construct the functional. However, in the
particular case of a unitary Fermi gas, the form of the \SLDA\ and
\ASLDA\ functionals is sufficiently restricted by dimensional
analysis, \QMC\ results, and Galilean invariance as to be able to make
testable predictions with relatively small uncertainties.  This has
been validated with both ab initio theoretical and experimental results.

The next step is to use such tools to make predictions about the
properties of the unitary gas under various conditions: by changing
the geometry and even the Hamiltonian as a function of time, by
probing the system with a variety of external probes and exciting a
plethora of modes---both linear and nonlinear---and by studying both
the equilibrium and non-equilibrium dynamics. We have illustrated a
few of these applications, but it is clear that we have barely
scratched the surface of this subject.

The Fermi gas in the unitary regime proves to be an extraordinarily
rich physical system to study, not only because one can both
theoretically and experimentally address many of its properties with
both precision and accuracy, but because it has so many truly
unexpected phases and dynamical phenomena.

Many fascinating features of this systems are still waiting to be
revealed in experiments in their full glory, including: the pseudogap
phase, the supersolid \LOFF\ phase, $p$-wave
superfluidity~\cite{Bulgac:2006gh, Bulgac:2009a}, the Higgs
mode of the pairing field, the behavior and response to various
spatial and time varying trapping fields and probes, the dynamics of
vortices which opens a window to quantum turbulence both theoretically
and experimentally, the existence of supercritical superflow, and its
kinetic properties---in particular its viscosity.  One can safely
state that the most extraordinary features of the unitary gas are
still waiting to be demonstrated. 

Perhaps the most captivating part of the story of the unitary Fermi
gas is that it provides a link to an abundance of widespread fields of
physics, from optics and atomic physics, to condensed matter physics,
to nuclear physics and the physics of neutron stars, color
superconductivity in \QCD\ and dark matter, relativistic heavy ion
collisions, and the \AdS/\CFT\ approaches in quantum field theory.

We would like to thank our collaborators J.~E.~Drut, Y.~-L.~Luo, K.~J.~Roche,
G.~Wlaz\l{}owski, S.~Yoon, and Y.~Yu for their invaluable contributions during
various stages of the work reviewed here, some of which has not been published
yet. We thank D.~Blume, J.~Carlson, S.~Giorgini, L.~Luo, S.~Reddy, C.~Salomon,
Y.~Shin, and G.~E.~Thomas for providing their numerical results. Support is
acknowledged from the DOE under grants DE-FG02-97ER41014, DE-FC02-07ER41457, and
DE-FG02-00ER41132, from the Polish Ministry of Science under contracts No. N
N202 328234 and N N202 128439, and from the \textsc{ldrd} program at Los
Alamos. Calculations reported here have been in part performed on the UW Athena
cluster, on NERSC Franklin Cray XT4 supercomputer under grant B-AC02-05CH11231
and at the Interdisciplinary Centre for Mathematical and Computational Modelling
(ICM) at Warsaw University.  This document is unclassified with LANL release
number LA-UR 10-05509.


\choosebib{
  \bibliographystyle{apsrev}
  \bibliography{macros,master}
}{ 
  \bibliographystyle{apsrevM}
  \bibliography{master}
}


\section{Appendix}
\label{cha:appendix} 
\subsection{Formal Description of the \DFT}
\label{sec:A-form-descr-dft}

Here we present a somewhat formal derivation of the variational
property of the Kohn-Sham equations.  Consider a general free-energy
functional of the following form
\begin{equation}
  F = E(n_{A},n_{B},\cdots) + T\tr\left(\mat{\rho}\ln\mat{\rho}\right)
\end{equation}
where
\begin{align*}
  n_{A} &= \tr\left(\mat{\rho} \mat{A}^{T}\right),\\
  n_{B} &= \tr\left(\mat{\rho} \mat{B}^{T}\right),\\
  &\vdots\
\end{align*}
are the various densities, anomalous densities, etc. expressed
linearly in terms of the one-body density matrix $\mat{\rho}$.  By
varying the functional with respect to the density matrix $\mat{\rho}$
subject to the appropriate constraints on density matrix form
(discussed in section~\ref{sec:fermions}), one obtains a solution of
the form
\begin{equation}
  \label{eq:Formal_DFT}
  \mat{\rho} = f_{\beta}(\mat{H}[\mat{\rho}])
\end{equation}
where $f_{\beta}(E)$ is the appropriate thermal distribution for the
particles of interest, and $\mat{H}$ is a single-particle Hamiltonian that
depends on $\mat{\rho}$:
\begin{equation}
  \label{eq:Formal_H}
  \mat{H}[\mat{\rho}] = \pdiff{E}{n_{A}}\mat{A} 
  + \pdiff{E}{n_{B}}\mat{B} 
  + \cdots.
\end{equation}
The typical Kohn-Sham equations follow by diagonalizing the
self-consistency condition (\ref{eq:Formal_DFT}) with a set of
normalized Kohn-Sham eigenfunctions of definite energy:
\begin{equation}
  \label{eq:Formal_KS}
  \mat{H}\ket{n} = E_{n}\ket{n}.
\end{equation}
The density matrix is diagonal in this basis and expressed in terms of
the appropriate distribution functions $f_{\beta}(E)$:
\begin{equation}
  \mat{\rho} = \sum_{n} f_{\beta}(E_{n}) \ket{n}\bra{n}.
\end{equation}
All of the functionals considered in this chapter may be expressed in
this form.  Once the appropriate matrix structures $\mat{A}$,
$\mat{B}$ etc. are described, the form of the Kohn-Sham equations and
potentials follows directly from these expressions.

\subsubsection{Fermions}
\label{sec:fermions}
The only remaining complication is to impose the appropriate
constraint on the density matrix $\mat{\rho}$.  This ensures that the
appropriate statistics of the particles is enforces.  As we shall be
interested in Fermions, the relevant constraint on the density matrix
(dictated by the canonical commutation relationships) is
\begin{equation}
  \mat{\rho} + \mat{C}\mat{\rho}^{T}\mat{C} = \mat{1}
\end{equation}
where $\mat{C} = \mat{C}^{T}$ is the charge conjugation matrix:
\begin{equation}
  \mat{C}\ket{\psi} = \ket{\psi}^{*}.
\end{equation}
This follows from the anti-commutation relationship for fermions and
is discussed further in the Section~\ref{sec:superfl-single-part}.
The constrained minimization of the functional $F(\mat{\rho})$ results
in the standard Fermi distribution\footnote{Formally,
  this constraint can be implemented using a Lagrange multiplier, but
  it is much easier to see the results by letting $\rho = \mat{1}/2 +
  \mat{x} - \mat{C}\mat{x}^T\mat{C}$ where $\mat{x}$ is unconstrained,
  and then performing the variation with respect to $\mat{x}$.}
\begin{equation}
  \label{eq:Formal_Fermion}
  \mat{\rho} = \frac{1}
  {1+e^{\beta \left(\mat{H}[\mat{\rho}] - \mat{C}\mat{H}^{T}[\mat{\rho}]\mat{C}\right)}},
\end{equation}
which is the fermionic form of the self-consistency
condition~(\ref{eq:Formal_DFT}) for the density matrix $\mat{\rho}$.
In practise, one does not iterate the entire density matrix.  Instead,
one stores only the densities $n_{A}$, $n_{B}$, etc.
Through~(\ref{eq:Formal_H}), these define the Kohn-Sham Hamiltonian
$\mat{H}$, which is then diagonalized to form the new density matrix
and finally the new densities.  If, for example, symmetries allow the
Hamiltonian $\mat{H}$ to be block diagonalized, then one can construct
and accumulate the densities in parallel over each block.  Finally,
the densities represent far fewer parameters than the full density
matrix.  Thus, more sophisticated root-finding techniques such as
Broyden's
method~\cite{Baran;Bulgac;Forbes;Hagen;Nazarewicz...:2008-05} may be
efficiently employed: Applying these techniques to the full density
matrix would be significantly more expensive.

\subsection{Single Particle Hamiltonian}
\label{sec:superfl-single-part}

It is convenient to express these concepts in the language of second
quantization.  The Hamiltonian will appear as a quadratic
operator of the form
\begin{equation}
  \op{H}_{s} = \tfrac{1}{2}\op{\Psi}^{\dagger}\mat{\mathcal{H}}_{s}\op{\Psi}
\end{equation}
where $\op{\Psi}$ has several components and $\mat{\mathcal{H}}_{s}$
is a matrix.  The factor of $1/2$ accounts for the double counting to
be discussed below.  For a two component system, the most general
$\op{\Psi}$ that allows for all possible pairings has four components:

\begin{equation}
  \op{\Psi} = 
  \begin{pmatrix}
    \op{a}\\
    \op{b}\\
    \op{a}^{\dagger}\\
    \op{b}^{\dagger}
  \end{pmatrix}.
\end{equation}
In terms of components of the wavefunction, we will write
$\mat{\mathcal{H}}_{s} \psi = E\psi$ where:
\begin{equation}
  \psi = 
  \begin{pmatrix}
    u_{a}\\
    u_{b}\\
    v_{a}\\
    v_{b}
  \end{pmatrix}.
\end{equation}
The naming of these components is conventional (see for example
\cite{Gennes:1966gf}) and the functions $u$ and $v$ are typically
called ``Coherence Factors''.  Note that the convention is that
$v_{a,b}^{*}(\vec{r}, t)$ are the wavefunctions of the particles.
In this formulation the Hamiltonian has the form presented
in~\eqref{eq:tddft}:
\begin{equation}
  \mat{\mathcal{H}}_{s} = 
  \begin{pmatrix}
    h_a+U_a & 0 & 0 & \Delta \\
    0 & h_b+U_b  & -\Delta & 0 \\
    0 & -\Delta^*& -h_a^*-U_a & 0 \\
    \Delta^*& 0 & 0& -h_b^*-U_b 
  \end{pmatrix}
\end{equation}
 
\subsubsection{Four-component Formalism}

We shall start with this full four-component formalism but soon
utilized a reduction: If the superfluid pairing $\Delta \sim
\braket{\op{a}\op{b}}$ channel is attractive, then often the ``Fock''
channel is repulsive so we can take $\braket{\op{a}^\dagger\op{b}} =
0$.  In combination with the double-counting discussed below, this
will allow us to fully express the system in terms of two components.

The four-component formalism double counts the degrees of freedom:
$\op{\Psi}$ contains both $\op{a}$ and $\op{a}^{\dagger}$. This
degeneracy is described in terms of the charge conjugation matrix
$\mat{\mathcal{C}}$:
\begin{align}
  \op{\Psi}^{\dagger} &= \mat{\mathcal{C}}\op{\Psi} &
  &\text{where}&
  \mat{\mathcal{C}} &= \begin{pmatrix}
    \mat{0} & \mat{1}\\
    \mat{1} & \mat{0}
  \end{pmatrix}.
\end{align}
The operator $\op{\Psi}$ will satisfy the single-particle
Shr\"odinger equations
\begin{equation}
  \op{H}_{s}\ket{\Psi} = E\ket{\Psi}
\end{equation}
where the Hamiltonian
$\op{H}_{s}=\op{\Psi}^{\dagger}\mat{\mathcal{H}}_{s}\op{\Psi}$ can be
chosen to satisfy (the sign implements Fermi statistics)
\begin{equation}
  \label{eq:CHC}
  \mat{\mathcal{C}}\mat{\mathcal{H}}_{s}^{T}\mat{\mathcal{C}} 
  = -\mat{\mathcal{H}}_{s}.
\end{equation}
In this form, the charge conjugation symmetry ensures that the
eigenstates will appear in $\pm E$ pairs.\footnote{Suppose
  $\mat{\mathcal{H}}_{s}\psi = \epsilon\psi$.
  Applying~(\ref{eq:CHC}), using $\mathcal{C}^2 = \mat{1}$, and taking
  the transpose imply that $\psi^T\mathcal{C}^{T}\mat{\mathcal{H}}_{s}
  = -\epsilon\psi^{T}\mathcal{C}^{T}$.  Since left and right
  eigenvalues are the same, this implies that there is some other
  state such that $\mat{\mathcal{H}}_{s}\tilde{\psi} =
  -\epsilon\tilde{\psi}$.  For Hermitian Hamiltonians,
  $\mat{\mathcal{H}}_{s} = \mat{\mathcal{H}}_{s}^{\dagger}$, hence,
  the other state can be directly constructed as $\tilde{\psi} =
  \mathcal{C}\psi^{*}$.} Keeping only one set of pairs will ensure
that we do not double count.  Using this symmetry, we can formally
diagonalize the Hamiltonian by a unitary transformation
$\mat{\mathcal{U}}$ such that: 
\begin{equation}
  \mat{\mathcal{U}}^{\dagger}\mat{\mathcal{H}}_{s}\mat{\mathcal{U}} =
  \tfrac{1}{2}
  \begin{pmatrix}
    \mat{E} & \mat{0}\\
    \mat{0} & -\mat{E}
  \end{pmatrix}
\end{equation}
where $\mat{E} = \diag(E_{i})$ is diagonal.  The columns of the matrix
$\mat{\mathcal{U}}$ are the (ortho)normalized wave-functions and
describe the ``coherence'' factors.  To determine the correct
expressions for the densities in terms of the wavefunctions we form
them in the diagonal basis and then transform back to the original
basis using $\mat{\mathcal{U}}$.

Despite this formal degeneracy of eigenstates, we are not aware of a
general technique to block-diagonalize the original Hamiltonian in the
presence of non-zero terms of the form
$\braket{\op{a}^\dagger\op{b}}$, though perhaps the symmetry might be
incorporated into the eigensolver.
 
\subsubsection{Two-component Formalism}
\label{sec:two-comp-form}

If $\braket{\op{a}^\dagger\op{b}} = 0$, however, then the Hamiltonian
is naturally block diagonal:
\begin{align}
  \mat{\mathcal{H}}_{s} &= \frac{1}{2}
  \begin{pmatrix}
    \mat{H}_{s} & \mat{0}\\
    \mat{0} & -\mat{H}_{s}^{T}
  \end{pmatrix}, &
  \op{H}_{s} &= \op{\psi}^{\dagger}\mat{H}_{s}\op{\psi} + \text{const},
\end{align}
and one may consider only a single block in terms of the reduced set
of operators
\begin{equation}
  \op{\psi} = \begin{pmatrix}
    \op{a}\\
    \op{b}^{\dagger}
  \end{pmatrix}.
\end{equation}
This directly avoids any double counting issues.  This system may be
diagonalized:
\begin{equation}
  \mat{H}_{s}\mat{U} = \mat{U}\mat{E}.
\end{equation}
The matrix $\mat{U}$ defines the single ``quasi''-particle operators
$\op{\phi}$ as linear combination of the physical particle operators
contained in $\op{\psi}$:
\begin{equation}
  \op{\phi} = \mat{U}^{\dagger} \op{\psi}.
\end{equation}
The Hamiltonian is diagonal in this basis
\begin{equation}
  \op{H}_{s} = \op{\phi}^{\dagger}\cdot\mat{E}\cdot\op{\phi}
\end{equation}
and hence expectation values may be directly expressed
\begin{equation*}
  \braket{\op{\phi}\op{\phi}^{\dagger}} = \theta_{\beta}(\mat{E}) = 
  \begin{pmatrix}
    \theta_{\beta}(E_{0})\\
    & \theta_{\beta}(E_{1})\\
    && \ddots\\
    &&& \theta_{\beta}(E_{n})
  \end{pmatrix}
\end{equation*}
where $1-\theta_{\beta}(E) = f_{\beta}(E)$ is the appropriate
distribution function: For fermions we have
\begin{equation}
  \theta_{\beta}(E) = \frac{1}{1+e^{-\beta E}}.
\end{equation}
At $T=0$ this reduces to $\theta_{0}(E) = \theta(E)$ and is equivalent
to the zero-temperature property that negative energy states are
filled while positive energy states are empty.  This may be simply
transformed back into the original densities (on the diagonal) and
anomalous densities (off-diagonal):
\begin{align*}
  \mat{F}_{+} = \braket{\op{\psi}\op{\psi}^{\dagger}} =& 
  \begin{pmatrix}
    \braket{\op{a}\op{a}^{\dagger}} & \braket{\op{a}\op{b}}\\
    \braket{\op{b}^{\dagger}\op{a}^{\dagger}} & 
    \braket{\op{b}^{\dagger}\op{b}}
  \end{pmatrix}
  =
  \mat{U}\theta_{\beta}(\mat{E})\mat{U}^{\dagger},\\
  \mat{F}_{-}^{T} = \braket{\op{\psi}^{*}\op{\psi}^{T}} =&
  \begin{pmatrix}
    \braket{\op{a}^{\dagger}\op{a}} & 
    \braket{\op{a}^{\dagger}\op{b}^{\dagger}}\\
    \braket{\op{b}\op{a}} & 
    \braket{\op{b}\op{b}^{\dagger}}
  \end{pmatrix}
  =
  \mat{U}^{*}\theta_{\beta}(-\mat{E})\mat{U}^{T}.
\end{align*}
Fermi statistics demands $\mat{F}_{-} + \mat{F}_{+} = \mat{1}$ but we
may have to relax this requirement somewhat in order to regulate the
theory in terms of an energy cutoff $\theta_{c}(E)$.  The columns of
$\mat{U}_{n}$ of $\mat{U}$ correspond to the single-particle
``wavefunctions'' for the state with energy $E_{n}$.  We partition
these into two components sometimes referred to as ``coherence
factors''
\begin{equation}
  \mat{U}_{n} = \begin{pmatrix}
    \mat{u}_{n}\\
    \mat{v}_{n}
  \end{pmatrix}.
\end{equation}
The unitarity of $\mat{U}$ imposes the conditions that
\begin{subequations}
  \begin{align}
    \mat{u}_{m}^{\dagger}\mat{u}_{n} + \mat{v}_{m}^{\dagger}\mat{v}_{n}
    &=\delta_{mn},\\
    \sum_{n}\mat{u}_{n}\mat{u}_{n}^{\dagger} = 
    \sum_{n}\mat{v}_{n}\mat{v}_{n}^{\dagger} &= \mat{1},\\
    \sum_{n}\mat{u}_{n}\mat{v}_{n}^{\dagger} = 
    \sum_{n}\mat{v}_{n}\mat{u}_{n}^{\dagger} &= \mat{0}.
  \end{align}
\end{subequations}
From this we may read off the expressions for the densities
\begin{subequations}
  \begin{align}
    \mat{n}_{a} = \braket{\op{a}^{\dagger}\op{a}} =&
    \sum_{n} \mat{u}_{n}^{*}\mat{u}_{n}^{T} \theta_{\beta}(-E_{n}),\\
    \mat{n}_{b} = \braket{\op{b}^{\dagger}\op{b}} =&
    \sum_{n} \mat{v}_{n}\mat{v}_{n}^{\dagger} \theta_{\beta}(E_{n}),\\
    \mat{\nu} = \braket{\op{a}\op{b}} =&
    \sum_{n} \mat{u}_{n}\mat{v}_{n}^{\dagger} \theta_{\beta}(E_{n})\\
    =& -\sum_{n} \mat{u}_{n}\mat{v}_{n}^{\dagger} \theta_{\beta}(-E_{n})\\
    =& \sum_{n} \mat{u}_{n}\mat{v}_{n}^{\dagger}
    \frac{\theta_{\beta}(E_{n})-\theta_{\beta}(-E_{n})}{2}.
  \end{align}
\end{subequations}
The last form for $\mat{\nu}$ must be used if the regulator is
implemented such that $\theta_{c}(E) + \theta_{c}(-E)\neq 1$, in
particular, if $\theta_{c}(E)=0$ for $\abs{E}>E_{c}$.  Note that these
expressions are basis independent, e.g. in position space:
\begin{equation}
  n_{a}(\vec{r},\vec{r}') =
  \sum_{n} u_{n}(\vec{r})^{*}u_{n}(\vec{r}')^{T} \theta_{\beta}(-E_{n}).
\end{equation}
The energy $E_{n}$ here is the energy determined by solving these
equations and will contain both positive and negative energies.


\end{document}